\documentclass[11pt,a4paper]{article}
\pdfoutput=1
\usepackage{jheppub}
\usepackage{float}

\usepackage{multirow, graphicx,amssymb,url,mathrsfs,amsmath}
\usepackage{wrapfig,boxedminipage,setspace,epsfig}
\usepackage{amsxtra,amstext,latexsym,dsfont,amsfonts}
\usepackage{color,eucal}
\usepackage[dvipsnames]{xcolor}
\usepackage{float}
\usepackage{slashed,comment}
\usepackage{kotex}
\usepackage{contour}
\usepackage{tikz}
\usetikzlibrary{calc,patterns,angles,quotes}
\usetikzlibrary{decorations.pathreplacing,decorations.markings,snakes}
\usepackage{graphicx}
\usepackage{subcaption}
\usepackage{caption}
\usepackage{tabularx, array}

\usepackage{amsthm}
\usepackage{mathrsfs}
\usepackage{amssymb}
\usepackage{pifont}

\newcolumntype{L}[1]{>{\raggedright\arraybackslash}p{#1}}
\newcolumntype{C}[1]{>{\centering\arraybackslash}p{#1}}
\newcolumntype{R}[1]{>{\raggedleft\arraybackslash}p{#1}}

\newcommand{\td}{\text{d}}

\renewcommand{\l}{\lambda}

\newcommand{\bra}[1]{\mbox{$\langle #1 |$}}
\newcommand{\ket}[1]{\mbox{$| #1 \rangle$}}

\newcommand{\brac}[1]{\langle #1 \rangle}

\newcommand{\Tr}{{\rm Tr}\,}

\renewcommand{\Im}{{ \rm Im}}

\newcommand{\mco}{\mathcal{O}}
\newcommand{\mcu}{\mathcal{U}}

\newcommand{\mcc}{\mathcal{C}}
\newcommand{\mcv}{\mathcal{V}}
\newcommand{\red}{\textcolor{red}}

\newcommand{\mcD}{{\mathcal{D}}}

\usepackage{xparse}
\ExplSyntaxOn
\NewDocumentCommand{\HS}{m}
 {
  \seq_set_split:Nnn \l_tmpa_seq { ~ } { #1 }
  \seq_map_inline:Nn \l_tmpa_seq { \contour{green}{##1} ~ } \unskip
 }
\ExplSyntaxOff

\newcommand\VJ[1]{\textcolor{blue}{#1}}

\title{Quantum Signatures of Chaos  from Free Probability
}

\author[a]{Hugo A. Camargo,}
\author[a]{Yichao Fu,}
\author[a]{Viktor Jahnke,}
\author[a,g]{Keun-Young Kim,}
\author[a]{Kuntal Pal}

\emailAdd{hugo.camargo@gist.ac.kr}
\emailAdd{yichao.fu@gm.gist.ac.kr}
\emailAdd{viktorjahnke@gist.ac.kr}
\emailAdd{fortoe@gist.ac.kr}
\emailAdd{kuntalpal@gist.ac.kr }

\affiliation[a]{Department of Physics and Photon Science, Gwangju Institute of Science and Technology,\\123 Cheomdan-gwagiro, Gwangju 61005, Korea}
\affiliation[g]{Research Center for Photon Science Technology, Gwangju Institute of Science and Technology, 123 Cheomdan-gwagiro, Gwangju 61005, Korea}

\abstract{
A classical dynamical system can be viewed as a probability space equipped with a measure-preserving time evolution map, admitting a purely algebraic formulation in terms of the algebra of bounded functions on the phase space. Similarly, a quantum dynamical system can be formulated using an algebra of bounded operators in a non-commutative probability space equipped with a time evolution map. Chaos, in either setting, can be characterized by statistical independence between observables at \( t = 0 \) and \( t \to \infty \), leading to the vanishing of cumulants involving these observables. In the quantum case, the notion of independence is replaced by free independence, which only emerges in the thermodynamic limit (asymptotic freeness). In this work, we propose a definition of quantum chaos based on asymptotic freeness and investigate its emergence in quantum many-body systems including the mixed-field Ising model with a random magnetic field, a higher spin version of the same model, and the SYK model. The hallmark of asymptotic freeness is the emergence of the free convolution prediction for the spectrum of operators of the form \( A(0) + B(t) \), implying the vanishing of all free cumulants between \( A(0) \) and \( B(t) \) in the thermodynamic limit for an infinite-temperature thermal state. We systematically investigate the spectral properties of \( A(0) + B(t) \) in the above-mentioned models, show that fluctuations on top of the free convolution prediction follow universal Wigner-Dyson statistics, and discuss the connection with quantum chaos. Finally, we argue that free probability theory provides a rigorous framework for understanding quantum chaos, offering a unifying perspective that connects many different manifestations of it.

}

\begin{document}
\maketitle

%%%%%%%%%%%%%%%%%%%%%%%%%%%%%%%%%%%%%%%%%%%%%%%%%%%%%%%%%%%%%%%%%%%%%%%%%%%%%%%%%%%%%%%%%%%%%%%%%%%%%%%%%%%%%%%%%%%%%%%%%%%%%%%%%%%%%%%%%%%%%%%%%%%%%%%%%%%%%%%%%%%%%%%%%%%
\section{Introduction  }\label{sec:intro}

Understanding how macroscopic statistical behavior emerges from microscopic dynamics remains a fundamental open problem in physics. In classical systems, chaotic dynamics play a crucial role in driving systems toward thermal equilibrium, bridging deterministic laws and statistical mechanics~\cite{krylov1979works}. The quantum counterpart -- quantum chaos -- is equally significant, though its precise characterization remains elusive. Various notions of quantum chaotic behavior have been proposed, each with broad implications across multiple domains. The Eigenstate Thermalization Hypothesis (ETH)\cite{ETH-Deutsch,Srednicki:1994mfb} provides a statistical framework to explain how quantum chaotic systems reach thermal equilibrium\cite{DAlessio:2015qtq}, with implications for condensed matter physics and quantum statistical mechanics. Furthermore, out-of-time-order correlators (OTOCs) play a key role in the interplay between quantum chaos and the scrambling of quantum information in many-body systems~\cite{Xu:2022vko}, with important implications for quantum communication~\cite{Schuster:2021uvg}.

Recently, quantum chaos has also played an important role in the context of the AdS/CFT correspondence. The first holographic computations of OTOCs~\cite{Shenker:2013pqa, Shenker:2013yza, Roberts:2014isa, Shenker:2014cwa} led to the derivation of the chaos bound~\cite{Maldacena:2015waa}, which established a precise criterion for determining whether a quantum system admits a dual gravitational description within Einstein gravity. This insight catalyzed the development of simple holographic models, most notably the Sachdev--Ye--Kitaev (SYK) model~\cite{SachdevYeModel, Kitaev2015}, which, in certain limits, is dual to Jackiw--Teitelboim (JT) gravity~\cite{Teitelboim:1983ux,Jackiw:1984je}. Beyond OTOCs, another hallmark of quantum chaos -- spectral statistics governed by random matrix theory~\cite{BGS} -- has provided crucial insights into the nature of quantum gravity. In particular, it has illuminated the role of Euclidean wormholes in the gravitational path-integral~\cite{Saad:2018bqo} and reinforced the idea that semiclassical gravity in AdS should be understood as a mesoscopic approximation to full quantum gravity~\cite{Pelliconi:2024aqw}.

Despite its significance across various domains, quantum chaos remains poorly understood. Several diagnostics have been proposed, including spectral statistics governed by random matrix theory~\cite{BGS}, ETH~\cite{DAlessio:2015qtq}, OTOCs~\cite{Shenker:2014cwa}, Krylov complexity\cite{Parker:2018yvk}, and spread complexity~\cite{Balasubramanian:2022tpr}. While these probes offer valuable insights, their interconnections are not well established.\footnote{For previous work in this direction, see~\cite{Cotler:2017jue}, which relates averaged OTOCs to the spectral form factor, and~\cite{rozenbaum2019universal}, which discusses the spectral statistics of Lyapunovian operators. Other works in this direction include~\cite{Ma:2019ocx,Lau:2022med}. These topics are reviewed in Appendices~\ref{app-averageOTOC} and~\ref{app-Lyapunovian}, respectively.} It remains unclear whether an overarching theoretical framework exists to unify these different manifestations of quantum chaos or at least provide a systematic classification of them. In many respects, our current understanding of quantum chaos is largely phenomenological, relying on conjectures and heuristic arguments often inspired by semiclassical intuition. For instance, periodic orbit theory has been successfully employed to derive random matrix behavior from classical periodic orbits~\cite{Berry1985}. However, this approach inherently assumes the presence of well-defined classical trajectories and thus breaks down in systems without a classical limit. Moreover, the mathematically rigorous framework for characterizing chaos -- quantum ergodic theory (see~\cite{Gesteau:2023rrx, Ouseph:2023juq} for recent reviews) -- remains largely disconnected from the aforementioned diagnostics of quantum chaos. Developing a mathematically precise and conceptually unified theory of quantum chaos is, therefore, a crucial open challenge, one that could provide a systematic framework for addressing many outstanding questions in the field.

In this work, we propose a quantum ergodic theory based on free probability theory~\cite{mingo2017free, voiculescu1992free}. This approach is motivated by the fact that any quantum dynamical system can be viewed as a non-commutative probability space equipped with a time evolution map. Since free probability theory studies precisely such non-commutative probability spaces, it is natural to expect its relevance in quantum-mechanical systems. By analogy with classical ergodic theory, chaos can be characterized by how much operators in the future, $B(t)$, become statistically independent of operators in the past, $A(0)$. In free probability theory, a fundamental notion of statistical independence is {\it freeness}, meaning that all mixed free cumulants involving the corresponding random variables vanish. In fact, various manifestations of quantum chaos, such as the ETH and the vanishing of OTOCs, can be understood as different avatars of freeness~\cite{Jindal:2024zcg}.  Strictly speaking, freeness only emerges in the thermodynamic limit, where the rank $\mathcal{D}$ of the operators in a quantum many-body system tends to infinity. In this context, one speaks of {\it asymptotic freeness}, indicating that freeness holds as $\mathcal{D} \to \infty$. At finite $\mathcal{D}$, a hallmark of asymptotic freeness is that the spectrum of combinations of asymptotically free operators follows the predictions of free probability theory~\cite{Chen:2024zfj}. An important first application of free probability theory to quantum chaos was its role in developing generalized versions of the ETH to describe the behavior of OTOCs~\cite{Pappalardi:2022aaz}. This perspective has since been explored in subsequent works~\cite{PhysRevX.15.011031, Vallini:2024bwp, Pappalardi:2023nsj} (see also \cite{Wang:2022ots} for a discussion of the connection between the gravitational replica trick and free probability).

Building on the work of~\cite{Chen:2024zfj}, we investigate the spectral properties of operators of the form $A(0) + B(t)$ in various quantum many-body systems, including the mixed-field Ising model with a random magnetic field, a higher-spin variant of this model, and the SYK model. We derive the corresponding results from free probability theory and compare them across chaotic, near-integrable, and integrable dynamics, demonstrating that operator spectral statistics not only provide a robust probe of quantum chaos, but are also connected to the late-time vanishing of mixed cumulants in the thermodynamic limit, encompassing connected two-point functions, OTOCs, and their higher-order counterparts.  Additionally, we analyze fluctuations beyond the free probability predictions and show that they follow universal Wigner-Dyson statistics. Finally, we discuss how free probability theory can be employed in AdS/CFT to shed light on a novel manifestation of quantum chaotic behavior known as ``BPS chaos''~\cite{Chen:2024oqv}.

This work is organized as follows. In Section~\ref{sec-classicalErgodicTheory}, we review classical ergodic theory. In Section~\ref{sec-QuantumErgodicTheory} and \ref{sec-operatorStatistics}, we discuss quantum ergodic theory and briefly introduce some fundamental aspects of free probability theory that will be used in this paper. We explain why asymptotic freeness should be incorporated into the quantum ergodic hierarchy and explore its connection to operator spectral statistics through the predictions of free convolution. In Sections~\ref{Mixed_Ising},  \ref{sec-higherSpinModel}, and \ref{sec-SYK}, we analyze the spectral properties of time-evolved operators in three quantum many-body systems: the mixed-field Ising model with a random magnetic field,  a higher-spin variant of the mixed-field Ising model, and the SYK model, and compare the predictions of these statistics obtained from the free probability theory. Finally, in Section~\ref{sec-discussion}, we summarize our results and outline the directions for future research. Some technical details and reviews of previous works that are relevant to the present work are deferred to the appendices.

\section{Brief Review of Classical Ergodic Theory} \label{sec-classicalErgodicTheory}
In this section, we provide an overview of classical ergodic theory, a rigorous mathematical framework for studying both abstract and concrete dynamical systems \cite{walters1982introduction}. Recent reviews on this topic include~\cite{Gesteau:2023rrx, Ouseph:2023juq}.

\paragraph{Classical probability space.} A probability space is a triple $(\Omega, \Sigma, \mu)$, where $\Omega$ is a set (the sample space) whose elements represent all possible configurations (states) of the system. The $\sigma$-algebra $\Sigma$ consists of measurable subsets of $\Omega$, representing all possible events. The probability measure $\mu: \Sigma \to [0,1]$ assigns probabilities to events in a countably additive manner, satisfying $\mu(\Omega) = 1$.

Two events \( A, B \in \Sigma \) are said to be \textit{statistically independent} if \( \mu(A \cap B) = \mu(A) \mu(B) \). The degree of deviation from independence can be quantified by the connected correlation function
\begin{equation}
 F_2(A, B) = \mu(A \cap B) - \mu(A) \mu(B), 
\end{equation}
which measures the difference between the joint probability and the product of the individual probabilities.

\paragraph{Classical dynamical system.} A dynamical system is a probability space $(\Omega, \Sigma, \mu)$ equipped with a time-evolution map $T_t: \Omega \to \Omega$, where $t\in \mathbb{R}$. The focus is on invertible, measure-preserving maps, where $\mu(A_t) = \mu(A)$ for any $A \subset \Omega$. Here, $A_t = T_t(A)$ denotes the image of the set $A$ under the map $T_t$.

Physically, \( \Omega \) represents the phase space of the system. Given a point \( x \in \Omega \), the time evolution map \( T_t \) takes \( x \) to \( x_t = T_t(x) \), defining an orbit in phase space. For conservative systems, the orbits remain confined in a constant-energy hypersurface $\Omega_E \subset \Omega$, defined by the relation $E=H(q_i,p_i)$. The measure \( \mu \) specifies how averages are taken over phase space, representing the statistical state of the system. For instance, for a system described by generalized coordinates \( (q_i, p_i) \) with Hamiltonian \( H(q_i, p_i) \), the measure corresponding to a microcanonical ensemble is given by $\textrm{d}\mu = \delta(H(q_i,p_i)-E)\prod_{i=1}^{N} \textrm{d}q_i \, \textrm{d}p_i/Z(E)$, with $Z(E)=\int \delta(H(q_i,p_i)-E)\,\prod_i dq_i\,dp_i$.

\subsection{Classical ergodic hierarchy}
In classical ergodic theory, dynamical systems are characterized by the asymptotic behavior of connected correlation functions at large times. Intuitively, non-integrable dynamics progressively erases correlations between past and future events, rendering them statistically independent as time evolves \cite{berkovitz2006ergodic}.

\paragraph{Ergodic systems.} A classical dynamical system $(\Omega,\Sigma,\mu,T_t)$ is said to be ergodic if, for any $A,B \subset \Sigma$,
\begin{equation} \label{eq-ergodicity1}
    \lim_{T \rightarrow \infty} \frac{1}{T}\int_{0}^{T} F_2(A_t,B) \textrm{d}t =0\,.
\end{equation}
This implies that, on average, past and future events become statistically independent in ergodic systems. 

The definition of ergodicity given in (\ref{eq-ergodicity1}) is equivalent to the following definition:
 a classical dynamical system is ergodic if, for every $T_t-$invariant set $A \subset \Sigma$ (i.e. $T_t(A)=A$) has either \(\mu(A) = 0\) or \(\mu(A) = 1\)\footnote{See Section 1.6 of \cite{walters1982introduction} for an extensive discussion of equivalent notions of ergodicity.}. For an ergodic system, the only subsets of $\Sigma$ that are invariant under $T_t$ are the whole phase space $\Omega$ and the sets of zero measure. This implies that a typical trajectory cannot be confined to a restricted region of the phase space but instead explores the entire phase space, except for a subset of measure zero.\footnote{As explained in \cite{bohigas1984chaotic}, the strict form of Boltzmann's ergodic hypothesis can never hold, since a trajectory that passes through every point in \(\Omega\) would intersect itself, which is impossible.} In particular, this implies the equality between time averages and phase space averages. We will discuss this in more detail in Section \ref{sec-algebraicPerspective} after introducing functions in phase space.

 Ergodicity alone is not sufficient for a system to reach equilibrium. The system must also satisfy a stronger property, called mixing, which we will discuss next.

\paragraph{Strongly mixing systems.}
A classical dynamical system \( (\Omega, \Sigma, \mu, T_t) \) is said to be (strongly) \( 2 \)-mixing if, for any two measurable sets \( A, B \in \Sigma \), we have:
\begin{equation}
    \lim_{t \to \infty} \mu(A \cap B_t) = \mu(A) \mu(B)\,.
\end{equation}
The intuition behind this is that, under mixing dynamics, events in the future ($B_t)$ become statistically independent of events in the past ($A)$.
Similarly, a dynamical system is said to be (strongly) \( 3 \)-mixing if, for any three measurable sets \( A, B, C \in \Sigma \), we have:
\begin{equation} \label{eq-3mixing}
  \lim_{s \to \infty}  \lim_{t \to \infty} \mu(A \cap B_t \cap C_{t+s}) = \mu(A) \mu(B) \mu(C)\,,
\end{equation}
implying that the events $A$, $B_t$ and $C_{t+s}$ become mutually independent as $t,s \to \infty$\footnote{Here, we assume that the order of taking the limits does not affect the result.}. 
The degree of 3-mixing as a function of the time separation between the events can be quantified by the following connected correlation function:
\begin{eqnarray} \label{eq-F3}
    F_3(A,B_t,C_{t+s})&=&\mu(A \cap B_t \cap C_{t+s})-\mu(A)\mu(B)\mu(C)\,,
\end{eqnarray}
The notion of $n$-mixing generalizes the property (\ref{eq-3mixing}) to $ n $-fold intersections, and can be quantified using higher-order connected correlation functions defined similarly to (\ref{eq-F3}).

The concept of mixing refers to the idea that, under time evolution, any subset \( A \subset \Sigma \) is uniformly dispersed over the phase space \( \Omega \). A useful analogy is a tablespoon of milk added to a cup of coffee. The mixing process can be visualized as the effect of stirring the beverage with a spoon, creating a flow that gradually blends the two liquids. Suppose that the beverage initially consists of 80\% coffee and 20\% milk. At first, the milk concentration is highly non-uniform: it is much higher within the region in the cup where it was initially added and nearly zero in other parts of the cup. However, if the dynamics are mixing, the stirring action causes the milk to spread evenly throughout the cup. Over time, as \( t \to \infty \), the milk concentration becomes uniform, reaching 20\% in every region of the cup, reflecting the initial ratio of milk to coffee.

Let $B$ represent the initial milk portion and let $A$ denote a region of the cup where we measure the milk concentration. A system is said to be mixing if $\lim_{t \rightarrow \infty}F_2(A,B_t)=0$, which implies
\begin{equation} \label{eq-mixingExample}
    \lim_{t \rightarrow \infty} \frac{\mu(A \cap B_t)}{\mu(A)}=\mu(B) = \frac{\mu(B_t)}{\mu(\Omega)}\,,
\end{equation}
where we have used the fact that the flow is measure-preserving, so \( \mu(B) = \mu(B_t) \), and the total measure of the phase space is normalized, \( \mu(\Omega) = 1 \).  The left-hand side of \eqref{eq-mixingExample} represents the fraction of \( A \) occupied by \( B_t \), which corresponds to the milk concentration within the region \( A \). The right-hand side represents the fraction of \( \Omega \) occupied by \( B \), which corresponds to the total concentration of milk in the whole cup. As \( t \to \infty \), mixing ensures that the local concentration in any region \( A \) converges to the global average concentration.

\paragraph{K-systems (\cite{walters1982introduction}).}  
A classical dynamical system \((\Omega, \Sigma, \mu, T_t)\) is called a Kolmogorov system (or K-system) if there exists a \(\sigma\)-subalgebra \(\Sigma_0 \subset \Sigma\) such that:  
\begin{enumerate}
    \item \(T_t(\Sigma_0) \subset \Sigma\) for all \(t \in \mathbb{R}\);  
    \item \(\bigvee_{t \in \mathbb{R}} T_t(\Sigma_0) = \Sigma\), where \(\bigvee_{t \in \mathbb{R}} T_t(\Sigma_0)\) is the \(\sigma\)-algebra generated by all sets \(T_t(\Sigma_0)\);  
    \item \(\bigwedge_{t \in \mathbb{R}} T_t(\Sigma_0) = (\emptyset, \Omega)\), where \(\bigwedge_{t \in \mathbb{R}} T_t(\Sigma_0)\) is the intersection of all \(\sigma\)-algebras \(T_t(\Sigma_0)\).  
\end{enumerate}  
K-systems exhibit a sensitive dependence on initial conditions. Specifically, trajectories that start close to one another diverge exponentially on average over time, a behavior quantified by Lyapunov exponents. As a result, K-systems exhibit a stronger form of mixing known as \emph{K-mixing}. In fact, a dynamical system is a K-system if and only if it is K-mixing (see Section 4.9 of~\cite{walters1982introduction}).

\paragraph{K-mixing (\cite{walters1982introduction}).}  
Let \((\Omega, \Sigma, \mu, T_t)\) be a dynamical system on a Lebesgue space. For \(B \in \Sigma\) and \(t > 0\), let \(\Sigma(B_t)\) denote the smallest \(\sigma\)-algebra containing all sets \(T_s(B)\) for \(s \geq t\). Then, \((\Omega, \Sigma, \mu, T_t)\) is K-mixing if and only if, for all \(A, B \in \Sigma\):  
\begin{equation}
    \lim_{t \to \infty} \sup_{C \in \Sigma(B_t)} \Big| \mu(A \cap C) - \mu(A)\mu(C) \Big| = 0.
\end{equation}  
This condition ensures that correlations between any event in the present (\(A\) at \(t=0\)) and any event in the distant future (\(C\), even if \(C\) is generated by the time-evolved set \(B_t\)) are statistically independent. This implies that the system becomes not only statistically independent of the past, but also that its evolution is ``forgetful'' of any structure in the initial state after enough time. This loss of information about initial conditions can be characterized by the so-called {\it Kolmogorov--Sinai entropy}, which is positive in K-systems~\cite{bohigas1984chaotic}.

\paragraph{The ergodic hierarchy.} 
Classical dynamical systems can be classified based on their mixing properties. This classification is known as the \emph{ergodic hierarchy}, which organizes systems according to progressively stronger notions of mixing behavior. For physical systems, the hierarchy includes:
\begin{equation*}
    \text{K-mixing systems} \subset \text{Strongly mixing systems} \subset \text{Ergodic systems}.
\end{equation*}
Other significant classes of dynamical systems include Bernoulli systems, which represent the most chaotic systems at the top of the ergodic hierarchy, and weakly mixing systems, which fall between strongly mixing and ergodic systems. Additionally, Anosov systems, although not part of the ergodic hierarchy, are defined on manifolds where trajectories split into stable, unstable, and neutral directions, exhibiting exponential contraction or expansion in the stable and unstable directions.

Although these systems are mathematically rich and provide deep insights into the dynamics of chaos, they are not directly relevant to the discussions in this paper. Thus, we omit a detailed analysis of these classes and focus on the aspects of the ergodic hierarchy most pertinent to our work. For a detailed discussion of the classical ergodic hierarchy, its connection with randomness, along with a more extensive list of dynamical systems, we refer the reader to~\cite{berkovitz2006ergodic,Gesteau:2023rrx, Ouseph:2023juq}.

\subsection{Mixing and the Vanishing of Classical Cumulants}
In classical probability theory, two random variables are statistically independent if their classical cumulant vanishes. In a dynamical system, since future events become statistically independent of past events, the corresponding cumulants must vanish. Specifically, 2-mixing is equivalent to the vanishing of the classical cumulant:
\begin{equation}
    \kappa_2(A,B_t) = \mu(A \cap B_t) - \mu(A)\mu(B)\,,
\end{equation}
as \( t \to \infty \) for all \( A, B \subset \Sigma \). Similarly, 3-mixing (along with 2-mixing) implies the statistical independence of the three events \( A \), \( B_t \), and \( C_{t+s} \), which leads to the vanishing of the third classical cumulant:
\begin{multline}
       \kappa_3(A,B_t,C_{t+s}) = \mu(A \cap B_t \cap C_{t+s}) - \mu(A)\mu(B_t \cap C_{t+s})\\ 
       - \mu(B)\mu(A \cap C_{t+s}) - \mu(C)\mu(A \cap B_t) + 2\mu(A)\mu(B_t)\mu(C)\,,
\end{multline}
in the limit \( t,s \to \infty \) for all \( A, B, C \subset \Sigma \). Here, we use the fact that \( \mu(B_t) = \mu(B) \) and \( \mu(C_{t+s}) = \mu(C) \). More generally, \( n \)-mixing implies the vanishing of the \( n \)-th cumulant \( \kappa_n \), which involves \( n \) subsets of \( \Sigma \). The formula for the \( n \)-th cumulant can be derived using partition-based combinatorial techniques. For a pedagogical review, see Appendix \ref{sec-appA}.

In the classical case, thinking about mixing in terms of the vanishing of classical cumulants is not strictly necessary. However, as we argue in the following, this perspective becomes crucial when considering quantum systems, where the concept of statistical independence is replaced by the notion of free independence, which relies on the vanishing of certain mixed free cumulants involving non-crossing partitions.

\subsection{Functions in Phase Space and the Algebraic Perspective} \label{sec-algebraicPerspective}
Given a classical probability space \((\Omega, \Sigma, \mu)\), one can define real-valued random variables \(f: \Omega \to \mathbb{R}\) as measurable functions that map the sample space to real values. The set of such measurable functions forms an algebra. A notable example is the Banach algebra \(L^{\infty}(\Omega, \Sigma, \mu)\) (abbreviated as \(L^{\infty}\)), consisting of essentially bounded functions. The norm of a function in this algebra is defined as \(||f||_{L^{\infty}} = \sup |f|\), and is equipped with a trace functional \(\varphi: L^{\infty} \to \mathbb{R}\), given by  
\[
\varphi(f) =\langle f \rangle = \int_{x \in \Omega} f(x) \, \textrm{d}\mu,
\]
which represents the expectation value of \(f\).

Ergodicity and mixing can be discussed naturally in terms of measurable functions. First, a measure-preserving flow satisfies \(\int_{x \in \Omega} f(x) \, \textrm{d}\mu = \int_{x \in \Omega} f(T_t(x)) \, \textrm{d}\mu\). Given \(f, g \in L^\infty\), one can define the connected correlation function (cumulant) between them. E.g., the second-order cumulant can be written as:
\begin{equation}
    \kappa_2(f,g;t) = \int_{x \in \Omega} f(x) \, g(T_t(x)) \, \textrm{d}\mu - \int_{x \in \Omega} f(x) \, \textrm{d}\mu \int_{x \in \Omega} g(x) \, \textrm{d}\mu,
\end{equation}
with similar definitions for higher-order cumulants. Ergodicity can then be expressed as:\[ \lim_{T \to \infty} \frac{1}{T} \int_{0}^{T} \kappa_2(f,g;t) \, \textrm{d}t = 0.\]
Additionally, the Birkhoff Ergodic Theorem (see Sec. 1.6 of~\cite{walters1982introduction}) implies that time averages and phase space averages are equal:
\begin{equation}
   \lim_{T \to \infty} \frac{1}{T} \int_{0}^{T} f(T_t(x)) \, \textrm{d}t = \int_{x \in \Omega} f \, \textrm{d}\mu,
\end{equation}
a relation often used in statistical mechanics. The strongly mixing condition can be written as:
\[
\lim_{t \to \infty} \kappa_2(f,g;t) = 0,
\]
with a similar definition for \(K\)-mixing.

The fact that measurable functions in phase space form an algebra \( \mathcal{A} \) allows one to abstract away the sample space and define a probability space in a purely algebraic manner, as discussed in~\cite{Tao2010}. An algebraic probability space is a pair \( (\mathcal{A}, \tau) \), where:
\begin{itemize}
    \item \( \mathcal{A} \) is a unital commutative real algebra,  
    \item \( \varphi: \mathcal{A} \to \mathbb{R} \) is a linear functional satisfying \( \varphi(\mathbb{I}) = 1 \) and \( \varphi(f^2) \geq 0 \) for all \( f \in \mathcal{A} \),  
    \item Every element \( f \in \mathcal{A} \) is bounded, meaning that \( \sup_{k \geq 1} \varphi(f^{2k})^{1/2k} < \infty \).
\end{itemize}

This algebraic perspective allows for the generalization of the definition of dynamical systems to the quantum case by allowing \( \mathcal{A} \) to be a non-commutative algebra. The definition of a quantum dynamical system is then given in terms of a non-commutative probability space \( (\mathcal{A}_{NC}, \tau) \), equipped with a time evolution map \( T_t: \mathcal{A}_{NC} \rightarrow \mathcal{A}_{NC} \). Here \( NC \) stands for non-commutative.

\section{Quantum Ergodic Theory} \label{sec-QuantumErgodicTheory}
In this section, we present a mathematical framework for quantum dynamical systems, highlighting the fundamental role of non-commutative probability spaces. Additionally, we provide a brief overview of key aspects of free probability theory that are relevant to this work. For a more comprehensive discussion of free probability theory, we refer the reader to \cite{speicher2019lecture, mingo2017free,nica2006lectures, voiculescu1992free}.

\paragraph{Non-commutative probability space}  
A non-commutative probability space consists of a unital $*$-algebra\footnote{An algebra is a vector space equipped with a multiplication operation that is distributive over vector addition and compatible with scalar multiplication. A unital algebra contains a multiplicative identity, denoted by 1. Finally, a $*$-algebra $\mathcal{A}$ is an algebra over $\mathbb{C}$ endowed with an involution $*$ that satisfies the following properties:  
\[
(za)^* = \bar{z}a^*\,, \quad (a+b)^* = a^* + b^*\,, \quad (ab)^* = b^*a^*\,, \quad (a^*)^* = a\,, \quad 1^* = 1
\]  
for all $a,b \in \mathcal{A}$ and $z \in \mathbb{C}$, where $\bar{z}$ denotes the complex conjugate of $z$.} $\mathcal{A}$, together with a functional $\varphi: \mathcal{A} \rightarrow \mathbb{C}$ that satisfies the following properties for all $a \in \mathcal{A}$: 
\begin{itemize}  
    \item \textbf{$*$-linearity:} $\varphi(a^*) = \overline{\varphi(a)}$;  
    \item \textbf{Unitality:} $\varphi(\mathbb{I}) = 1$;  
    \item \textbf{Positivity:} $\varphi(aa^*) \geq 0$;  
    \item \textbf{Faithfulness:} $\varphi(aa^*) = 0 \iff a = 0$.  
\end{itemize}
The functional is referred to as a state in $\mathcal{A}$ and provides the expectation values of non-commutative random variables, which are elements of $\mathcal{A}$. One of the most familiar examples of a non-commutative probability space is a quantum system, which we define below.

\paragraph{Quantum Dynamical System}  
A quantum dynamical system is a non-commutative probability space $(\mathcal{A}, \varphi)$ equipped with a strongly continuous time evolution map $T_t: \mathcal{A} \to \mathcal{A}$, where $T_t$ represents the evolution of observables in the system over time. Typically, $\varphi$ is a normal state, and $\mathcal{A}$ is a von Neumann algebra\footnote{A \textit{von Neumann algebra} is a unital $*$-algebra of bounded operators on a Hilbert space that is closed under the weak operator topology. This means that given a sequence of operators $a_{n}$ in the algebra, the expectation values \( \langle \psi_1 | a_n | \psi_2 \rangle \) converge to \(\langle \psi_1 | a | \psi_2 \rangle \) in the limit $n\rightarrow \infty$ for all vectors \( |\psi_1\rangle, |\psi_2\rangle \) in the Hilbert space. In this case, the sequence of operators $a_{n}$ in the algebra is said to converge to $a$ under the weak operator topology. 
}. A \textit{normal state} on a von Neumann algebra is a type of state for which the expectation values of operators are continuous with respect to a topology called the ultraweak topology. Discussing the precise definition of the ultraweak topology is beyond the scope of this paper. We refer the reader to Section 20 of \cite{conway2000course} for further details. In summary, this condition ensures well-behaved convergence of expectation values, taken with the state, as the sequence of operators approaches its limit. For an operator $ a \in \mathcal{A} $, its expectation value is given by $\varphi(a)$. In what follows, we will also denote it by \(\langle a \rangle\). We denote time-evolved operators as $a_t= T_t(a)$ or  $a(t)$. Given the system's Hamiltonian, such operators can be computed as
$a_t=U^{\dagger} a\,U$, where $U=e^{-itH}$.

Similarly to the classical case, chaos can be quantified using cumulants involving operators at different times. The simplest cumulant that one can consider is a connected two-point function involving two operators $a,b \in \mathcal{A}$:
\begin{equation} \label{eq-quantumF2}
    F_2(a,b;t)= \varphi(a \, b_t)-\varphi(a)\varphi(b_t)\,.
\end{equation}
The quantum version of the classical ergodic hierarchy can be proposed on the basis of the properties of $F_2(a,b;t)$. 

\paragraph{Quantum ergodic system} A quantum dynamical system $(\mathcal{A},\varphi,T_t)$ is said to be ergodic if, for all $a,b \in \mathcal{A}$:
\begin{equation} \label{eq-Qergodic}
  \lim_{T \rightarrow \infty} \int_{0}^{T}F_2(a,b;t)~\textrm{d}t =0.
\end{equation}
In the context of modular flows, it can be demonstrated that the quantum ergodicity condition \eqref{eq-Qergodic} cannot be satisfied by type I algebras, whereas it may hold for type II$_1$ algebras in a tracial state or for type III algebras~\cite{Ouseph:2023juq}.

\begin{figure}
    \centering
    \includegraphics[width=0.9\linewidth]{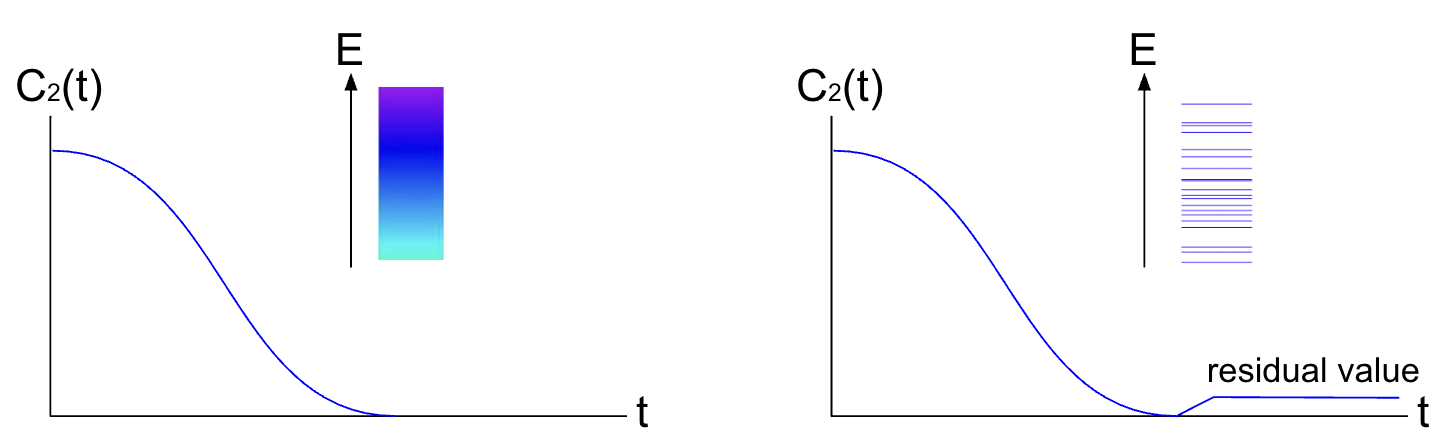}
    \caption{Typical behavior of a two-point function for strongly non-integrable quantum mechanical systems, showing the continuous spectrum (left panel) and discrete spectrum (right panel). The continuous spectrum is associated with type III algebras, which typically arise in the strict thermodynamic limit of many-body quantum systems, where the correlator vanishes. In contrast, the discrete spectrum corresponds to finite-dimensional Hilbert spaces, where the correlator does not vanish. The residual value of the correlator decreases as one increases the number of degrees of freedom.}
    \label{fig:MixingTypeIII}
\end{figure}

\paragraph{Quantum mixing system} A quantum dynamical system $(\mathcal{A},\varphi,T_t)$ is said to be quantum (strongly) 2-mixing if, for all $a,b \in \mathcal{A}$:
\begin{equation}
  \lim_{t \rightarrow \infty} F_2(a,b;t) =0.
\end{equation}
Mixing is expected to occur in systems with a continuous spectrum whose algebra is type II or III. This is consistent with the fact that in quantum mechanical systems with type I algebras, the spectrum of energies is discrete, and therefore two-point functions are periodic or quasi-periodic functions of time, and cannot decay to zero~\cite{Berry2001}.
Indeed, in the context of modular flows, it can be shown that quantum strong 2-mixing implies that the algebra is of type III~\cite{Ouseph:2023juq}. This is illustrated schematically in Figure \ref{fig:MixingTypeIII}. 

\paragraph{Algebraic Quantum K-system \cite{Narnhofer:2001}}  
A quantum dynamical system \((\mathcal{A}, \varphi, T_t)\) with a time-invariant state \(\varphi\) is called an \emph{algebraic quantum K-system} if there exists a subalgebra \(\mathcal{A}_0 \subset \mathcal{A}\) such that, for all \(t \geq 0\):
\begin{equation}
    T_t(\mathcal{A}_0) \supset \mathcal{A}_0, \quad
    \bigcup_{t \geq 0} T_t(\mathcal{A}_0) = \mathcal{A}, \quad
    \text{and} \quad
    \bigcap_{t \geq 0} T_{-t}(\mathcal{A}_0) = \mathbb{C} \, \mathbb{I}.
\end{equation}
Here, \(\mathbb{C} \,\mathbb{I}\) denotes the scalar multiples of the identity. Quantum K-systems exhibit \emph{quantum \(n\)-mixing} (see Theorem 2.3 in~\cite{Narnhofer:2001}), meaning that for any finite collection of operators \(A_1, A_2, \ldots, A_n \in \mathcal{A}\), the following holds :
\begin{equation} \label{eq:quantum-n-mixing}
    \lim_{\substack{t_{i+1} - t_i \to \infty \\ i=1,\ldots,n-1}} 
    \varphi\big( A_1(t_1) A_2(t_2) \cdots A_n(t_n) \big) 
    = \prod_{i=1}^n \varphi(A_i),
\end{equation}
where \(A_i(t) := T_t(A_i)\).

\begin{comment}
\paragraph{Quantum K-mixing system \VJ{[Replace by K-system]}}
A quantum dynamical system $(\mathcal{A}, \varphi, T_t)$ is said to be quantum K-mixing if, for all subalgebras $\mathcal{B} \subset \mathcal{A}$ and all $b \in \mathcal{B}$:
\begin{equation} \lim_{t \rightarrow \infty} \sup_{a \in \mathcal{A}_{(t,\infty)}} |F_2(a,b;t)| = 0\,, \end{equation}
where $\mathcal{A}_{(t,\infty)}= \{ T_s(A) \mid A \in \mathcal{A}, s > t \}$ denotes the time-band algebra associated with the interval $(t,\infty)$.
 That condition implies that correlations with the entire future decay to zero. Since the K-mixing condition is stronger than 2-mixing, it is likewise expected to occur only for type III algebras. 
\end{comment}

Since classical K-systems are typically characterized by Lyapunov exponents, it is reasonable to expect that quantum K-systems should likewise be characterized by a quantum analogue of the Lyapunov exponent. Indeed, for certain systems, the squared commutator $C_4(t)=-\langle [a(0), b(t)]^2 \rangle$ -- motivated by a semiclassical analogy with Poisson brackets~\cite{larkin1969sov} -- has been observed to exhibit exponential growth within a specific time window (see for example~\cite{Kobrin:2020xms,Craps:2019rbj}). This growth, however, cannot persist indefinitely, as quantum interference effects eventually disrupt the classical picture~\cite{Richter:2022sik}. Notably, this definition of a quantum Lyapunov exponent differs from its classical counterpart~\cite{Rozenbaum:2016mmv}, as it can lead to exponential behavior even in some classically integrable systems, a phenomenon known as saddle-dominated scrambling. This issue can be mitigated by considering logarithmic OTOCs~\cite{Trunin:2023xmw}. In the next subsection, we will discuss how the vanishing of the OTOCs, or equivalently the saturation of the squared commutator $C_4(t) = -\langle [a(0), b(t)]^2 \rangle$ to a constant value, typically associated with chaotic dynamics, can be understood as a consequence of a concept known as asymptotic freeness, a form of statistical independence in quantum systems.

\begin{figure}[h!]
    \centering
    \includegraphics[width=0.8\linewidth]{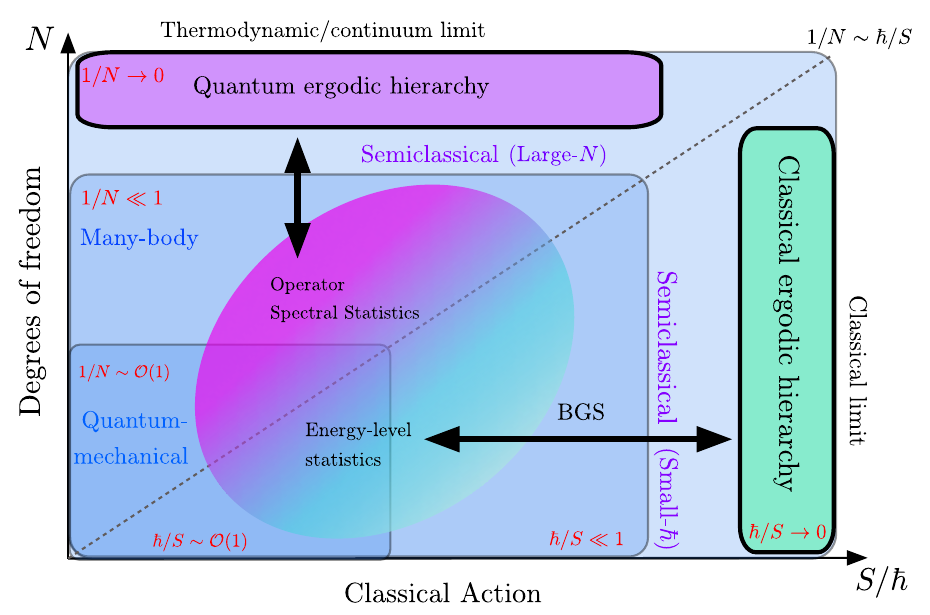}
    \caption{Illustration of the position of the classical and quantum ergodic hierarchies in relation to the number of degrees of freedom $N$, and classical action $S/\hbar$. The classical ergodic hierarchy deals with classical dynamical systems, which in the diagram exist in the classical limit $\hbar/S\rightarrow 0$ (right). In contrast, the quantum ergodic hierarchy deals with non-commuting variables and exists only in the strict thermodynamic/continuum limit $1/N\rightarrow 0$ (top). As probes of quantum chaos in quantum systems with $1/N>0$, energy-level statistics (arising from the BGS conjecture regarding quantum systems with chaotic classical analogues) and operator spectral statistics do not directly overlap with the quantum ergodic hierarchy. There are different ways in which one can reach an infinite number of degrees of freedom: by considering a lattice model with $L$ sites and taking $L \to \infty$; by taking the continuum limit, where the lattice spacing vanishes, as in QFTs; or by considering a gauge theory matrix/vector model with $N_c \to \infty$, where $N_c$ is the rank of the gauge group. Note that the parameter $1/N$ plays a role analogous to $\hbar$ in the semi-classical analysis. Diagram inspired by Figure 2 of~\cite{Richter:2022sik}. }
    \label{fig:DiagramErgodicHierarchies}
\end{figure}

\paragraph{Quantum Ergodic Hierarchy}
Analogous to the classical case, one can define a quantum ergodic hierarchy for quantum systems, which includes:
\begin{equation*}
    \text{Quantum K-systems} \subset \text{Quantum Mixing systems} \subset \text{Quantum Ergodic systems}.
\end{equation*}
However, it is important to note that the weakest form of chaos in this hierarchy, namely ergodicity, requires the algebra of observables to be of type II, while stronger notions of chaos may necessitate an algebra of type III \cite{Ouseph:2023juq}. In contrast, finite-dimensional quantum systems exhibit an algebra of type I. As a result, there is no quantum chaos in the sense of mixing for finite-dimensional systems. This is why Michael Berry famously stated, ``There is no quantum chaos, only quantum chaology" in his well-known paper~\cite{berry1989quantum}. However, under quantization, classically chaotic systems (specifically K-mixing systems) retain memory of their classical chaotic nature, exhibiting correlations in their spectra that resemble those found in random matrix theory~\cite{BGS}. In the thermodynamic limit, the algebra can change from type I to type II or III~\cite{Witten:2021jzq}, allowing the emergence of quantum ergodicity and mixing. Algebras of type III also arise in quantum field theory and gauge theories at large \(N\)~\cite{Witten:2021jzq}. For such systems, with an infinite number of degrees of freedom, one can truly talk about quantum chaos in a way that is parallel to classical chaos because the large-$N$ limit provides a sort of semiclassical limit~\cite{Richter:2022sik}. See Fig.~\ref{fig:DiagramErgodicHierarchies} for a comparison with the classical ergodic hierarchy in relation to number of degrees of freedom and ``quantumness'' of the classical action.\footnote{For a collection of works that appeared in the physics and mathematics literature on the types of systems that appear in the quantum ergodic hierarchy and related to the discussion presented above, see \cite{narnhofer1989mixing, narnhofer1989quantum, Narnhofer:2001, zaslavsky1981stochasticity, peres1984ergodicity, castagnino2009towards, vikram2023dynamical, benatti1998statistics}. Also, for recent progress in understanding quantum ergodic hierarchy in the context of unitary quantum circuits, a solvable model for many-body quantum chaos, we refer to \cite{Aravinda:2021sxc, Aravinda:2023vsg}.}

The conditions of ergodicity, mixing, and K-mixing all relate to the degree to which operators in the future become statistically independent from those in the past. In the context of free probability theory, there is a concept of statistical independence tailored for non-commutative probability spaces, known as freeness. In subsection \ref{subsec-freeness}, we introduce this concept and explore its potential application within the quantum ergodic hierarchy.

\paragraph{State Dependence.} 
The notions of chaos discussed above are defined with respect to a given quantum state, represented by the map \( \varphi: \mathcal{A} \to \mathbb{C} \). In this work, we fix the state to be an infinite temperature thermofield double (TFD) state and compute the expectation values of operators \( a \in \mathcal{A} \) as \( \varphi(a) = \langle \text{TFD} | a | \text{TFD} \rangle \). For operators acting on a single copy of the system, this reduces to a thermal expectation value. This choice is natural, as we aim to work with a state that is well-defined in the thermodynamic limit, and this is always the case for the TFD state~\cite{Witten:2021jzq}. In the case of finite-dimensional systems in which the operators $a_\mathcal{D}$ are $\mathcal{D} \times \mathcal{D}$ matrices, we are going to be interested in the expectation value defined as $\varphi(a) = \lim_{\mathcal{D} \rightarrow \infty} \langle \text{TFD} | a_\mathcal{D} | \text{TFD} \rangle $.\footnote{The TFD state is formally defined as the canonical purification of the thermal (Gibbs) density operator $\rho_{\beta}=e^{-\beta H}/\textrm{Tr}(e^{-\beta H})$, namely $\vert \textrm{TFD} \rangle\equiv\vert \sqrt{\rho_{\beta}}\rangle$. In the infinite temperature limit, corresponding to $\beta=0$, the TFD state becomes a maximally-entangled state obtained as a superposition of energy eigenstates over the doubled Hilbert space, weighted by the square root of the dimension of the Hilbert space.}

\subsection{Statistical Independence in Quantum Mechanics: Freeness} \label{subsec-freeness}
Given a non-commutative probability space \((\mathcal{A}, \varphi)\), two random variables \(a, b \in \mathcal{A}\) are said to be \emph{free} if, for all integers \(n \geq 1\) and for all polynomials \(P_1, Q_1, \dots, P_n, Q_n\) in one variable, we have~\cite{mingo2017free, speicher2019lecture}
\begin{equation} \label{eq-DefFreeness}
       \varphi\big((P_1(a)-\varphi(P_1(a)))(Q_1(b)-\varphi(Q_1(b))) \cdots (P_n(a)-\varphi(P_n(a)))(Q_n(b)-\varphi(Q_n(b)))\big) = 0\,.
\end{equation}
For polynomials of the form \(P(x)=Q(x)=x\), and denoting \(\varphi(\cdot)=\langle \cdot \rangle\), the above condition implies relations of the form:
\begin{align}\label{eq-Producs}
    &\langle (a-\langle a\rangle)(b-\langle b\rangle)\rangle=0\,,\nonumber\\
    &\langle (a-\langle a\rangle)(b-\langle b\rangle)(a-\langle a\rangle)\rangle=0\,,\nonumber \\
    &\langle (a-\langle a\rangle)(b-\langle b\rangle)(a-\langle a\rangle)(b-\langle b\rangle)\rangle=0\,,
\end{align}
that lead to factorization properties such as
\begin{equation}\label{eq-Factorization}
   \langle a b\rangle=\langle a\rangle \langle b\rangle\,,~~~~ \langle a b a\rangle=\langle a^2\rangle \langle b\rangle\,,~~~~ \langle a b a b\rangle=\langle a^2\rangle \langle b\rangle^2+\langle a\rangle^2 \langle b^2\rangle-\langle a\rangle^2 \langle b\rangle^2.
\end{equation}
Therefore, freeness between \(a\) and \(b\) can be understood as a notion of independence, as it implies that joint moments involving these variables factorize into products of the individual moments of \(a\) and \(b\).

\paragraph{Asymptotic Freeness of Large Random Matrices.}  
Starting from this section, when referring to operators that can be represented as finite-dimensional matrices (or finite-dimensional random matrices), we use capital letters such as $A$ and $B$, while for general statements about generic random variables we use lower-case letters such as $a$ and $b$.

In the study of large random matrices, it is often useful to consider the behavior of sequences of matrices as their dimension \( \mathcal{D} \) approaches infinity. A key concept in this context is the convergence of empirical eigenvalue distributions, which provides a way to analyze the asymptotic spectral properties of these matrices. One important phenomenon that arises in this limit is \emph{asymptotic freeness}, meaning that certain large random matrices behave as free elements in a non-commutative probability space. For example, independent realizations of Gaussian Unitary Ensemble (GUE) matrices are asymptotically free with respect to the expectation value~\cite{mingo2017free}: 
\begin{equation} \label{eq-expectationValueRM}
    \varphi(A) = \lim_{\mathcal{D} \rightarrow \infty} \varphi_{\mathcal{D}}(A)\,,~~~\text{where}~~\varphi_{\mathcal{D}}(A) = \mathbb{E}\left( \frac{\Tr A}{\mathcal{D}} \right),
\end{equation}
where \(\mathbb{E}\) denotes the ensemble average. 
Denoting the elements of the operator $A$ as $A_{ij}$, the above map for Gaussian 
random matrix ensembles is
$\varphi(A) = \lim_{\mathcal{D} \rightarrow \infty} \frac{1}{\mathcal{D}} \sum_{i=1}^{\mathcal{D}}\mathbb{E}\left(A_{ii} \right)$.

A fundamental result in this framework states that Haar-distributed unitary conjugation induces asymptotic freeness. Specifically, let \( A_\mathcal{D} \) and \( B_\mathcal{D} \) be two sequences of deterministic \( \mathcal{D} \times \mathcal{D} \) matrices whose empirical eigenvalue distributions converge as \( \mathcal{D} \to \infty \). If \( U_\mathcal{D} \) is an \( \mathcal{D} \times \mathcal{D} \) Haar-distributed unitary random matrix, then:  
\begin{equation} \label{eq-fundamentalResult}
    A_\mathcal{D},~~\text{and}~~ U_\mathcal{D} B_\mathcal{D} U_\mathcal{D}^\dagger \quad \text{are asymptotically free as } \mathcal{D} \to \infty,
\end{equation}  
with respect to the expectation value given in \eqref{eq-expectationValueRM}. For further details, see~\cite{mingo2017free}.

In the next subsection, we examine how freeness influences the spectral properties of the sum of two free operators.

\subsection{Asymptotic freeness and  the vanishing of OTOCs}\label{subsec-asymp freeness otocs}

Consider a quantum dynamical system \( (\mathcal{A}, \varphi, T_t) \), where the time evolution map \( T_t: \mathcal{A} \to \mathcal{A} \) is given by \( A(t) = U^\dagger A U \), with the unitary time evolution operator \( U = e^{-iHt} \),  \( H \) being the system's Hamiltonian. With a focus on many-body quantum systems, we will treat random variables in the algebra as operators represented by matrices of rank \( \mathcal{D} \).
The quantum state of interest is an infinite temperature TFD state, and the quantum expectation value of a given operator $A \in \mathcal{A}$ is therefore computed as
\begin{equation} \label{eq-TFDstate}
 \varphi(A) = \langle \text{TFD} | A | \text{TFD}\rangle \big|_{\beta=0}\,.
 \end{equation}
For an operator acting on a single copy of the system, this is equivalent to an infinite temperature thermal expectation value $\varphi(A)=\Tr(A)/\mathcal{D}$. If the model includes random couplings, the expectation value should also account for an average over the couplings, $\varphi(A)=\mathbb{E}  \left( \Tr(A)/\mathcal{D} \right)$, similar to the procedure in Eq. \eqref{eq-expectationValueRM}.

Under these conditions, we expect that for two operators $A, B \in \mathcal{A}$, if the dynamics is chaotic, the time-evolved operator $B(t)$ becomes asymptotically free from $A(0)$ with respect to the state \eqref{eq-TFDstate} at sufficiently large times. This expectation is motivated by the fact that in a chaotic system, the time evolution operator $e^{-iHt}$ can be approximated by a Haar-random unitary, i.e.,\footnote{Note that when the quantum dynamics modeled by a random unitary matrix from the Haar measure of the unitary groups, since there is no notion of time-translation symmetry, energy is not conserved in such a dynamics. 
A more physical approach, which is closer to the dynamics generated by a generic chaotic quantum system with energy conservation, would be to consider 
unitary evolution $U=e^{-i t H}$  generated 
random matrix Hamiltonian $H$ drawn from some ensemble of (time-independent) Hamiltonians.  The time evolution operators form an ensemble $\mathcal{E}(t)$, 
which, under time evolution, would gradually evolve into a complicated unitary ensemble,  though not exactly Haar random, in the long-time limit. 
In \cite{Cotler:2017jue} the time evolution by an ensemble of  GUE Hamiltonians was considered, and its approach to the Haar random unitary was quantified by examining when an approximate $k$-design is formed by computing the so-called frame potentials. Interestingly, it was found that time evolution by the GUE Hamiltonian does form an approximate $k$-design at an intermediate time scale; however, subsequently it then deviates from being an approximate $k$-design at later times.}  
\[
e^{-iHt} \approx U_{\text{Haar}}.
\]  
From the fundamental result in \eqref{eq-fundamentalResult}, we know that two deterministic matrices $A$ and $B$ are expected to be free when one of them is conjugated by a Haar-random unitary. That is, for two matrices \( A \) and \( B \) of rank \( \mathcal{D} \), when \( U \) is a Haar-random unitary matrix of size \( \mathcal{D} \times \mathcal{D} \), the conjugated matrix \( U_{\text{Haar}}^\dagger B U_{\text{Haar}} \) becomes asymptotically free from \( A \) with respect to the state $\varphi(A)=\mathbb{E}  \left( \Tr( A)/\mathcal{D} \right)$ in the limit \( \mathcal{D} \to \infty \).
 
The asymptotic freeness between \( A(0) \) and \( B(t) \) implies that all mixed cumulants involving these operators vanish in the limit \( \mathcal{D} \to \infty \). In particular, if we consider operators with zero expectation values, \( \langle A \rangle = 0 \) and \( \langle B(t) \rangle = 0 \), this leads to the vanishing of the following cumulants:  
\begin{equation*} 
   \langle A(0) B(t) \rangle \,,~~~ \langle A(0) B(t) A(0) B(t) \rangle\,,~~~ \langle A(0) B(t) A(0) B(t) A(0) B(t) \rangle
\end{equation*}
for sufficiently large times, as well as cumulants involving an odd number of operators.  The vanishing of the two-point function \( \langle A(0) B(t) \rangle \) is associated with thermalization, while the vanishing of the four-point OTOC \( \langle A(0) B(t) A(0) B(t) \rangle \), as well as its higher-order counterparts, is linked to the scrambling of quantum information\footnote{It is worth clarifying that in the literature on OTOCs, the term {\it scrambling} is sometimes used to describe the early-time exponential behavior of OTOCs in certain systems, such as the SYK model. In other cases -- particularly in the context of quantum information theory -- scrambling refers to the vanishing of OTOCs, regardless of whether this decay follows an exponential trend. See, for instance,~\cite{Hosur:2015ylk, Huang:2017fng}. Here, we adopt the latter definition, using the term {\it scrambling} to refer to the vanishing of OTOCs. }.

The asymptotic freeness between two operators \( A \) and \( B(t) \) can alternatively be understood as arising from the fact that, for sufficiently large times, these operators not only fail to commute but also exhibit a large commutator. Notably, this captures the same effect as the squared commutator, \( C(t) = -\langle [A, B(t)]^2 \rangle \), thereby making the connection with scrambling and operator growth particularly transparent. However, asymptotic freeness is an even stronger condition, as it implies not only that the squared commutator is large but also that all its higher-order generalizations, \( C_{(k)}(t) = -\langle [A, B(t)]^{2k} \rangle \), are large as well.

\subsection{Incorporating Freeness into the Quantum Ergodic Hierarchy}  
The previous discussions motivate the inclusion of freeness into the quantum ergodic hierarchy. We propose the following definition:

\paragraph{F-systems.} 
A quantum dynamical system \( (\mathcal{A}, \varphi, T_t) \) is said to be an \textit{F-system} or to possess the \textit{F-property} if, for any operators \( A, B \in \mathcal{A} \), the operators \( A \) and \( B(t) \) are free in the limit \( t \to \infty \). This implies the vanishing of all mixed free cumulants involving \( A \) and \( B(t) \):
\begin{equation}
    \lim_{t \rightarrow \infty}\kappa_n(A,B(t),A,B(t),\ldots,A,B(t))=0\,.
\end{equation}
We also refer to the F-property as \emph{pairwise freeness}.

Note that the F-property implies quantum 2-mixing, but it does not imply quantum $n$-mixing. Consider $n$ Heisenberg operators $A_1(t_1), A_2(t_2), \ldots, A_n(t_n)$, and take the limit $t_{i+1} - t_i \to \infty$ for all $i = 1, \ldots, n-1$. The F-property implies that, for any $i$, $A_{i+1}(t_{i+1})$ is free from $A_i(t_i)$ as $t_{i+1} - t_i \to \infty$, which in turn implies that $A_1(t_1), A_2(t_2), \ldots, A_n(t_n)$ are pairwise free. Pairwise freeness, however, does not imply mutual freeness, which is necessary for quantum $n$-mixing \eqref{eq:quantum-n-mixing}. It is interesting to compare the F-property with the notion of a quantum K-system, which implies quantum $n$-mixing (see Theorem 2.3 in~\cite{Narnhofer:2001}), but does not imply the vanishing of out-of-time-order correlators (see Example 2.2 in~\cite{Narnhofer:2001}). In particular, this shows that quantum K-property, as a mathematical condition, is not enough for the emergence of freeness. Therefore, while we expect that physical examples of quantum K-systems may also display the F-property, and vice versa, the mathematical definitions of these properties are not equivalent; neither condition is strictly stronger or weaker than the other.

In Figure \ref{fig:FreenessErgodicHierarchy}, we illustrate the quantum and classical ergodic hierarchies and their consequences for quantum systems with finite dimensional Hilbert spaces.

\begin{comment}
    This motivates the definition of a stronger notion of freeness, which we call the \emph{strong F-property}.

\VJ{\paragraph{Strong F-property.} 
A quantum dynamical system \( (\mathcal{A}, \varphi, T_t) \) is said to possess the strong F-property if, for any \( A \in \mathcal{A} \), the operator \( A \) is uniformly free from the entire asymptotic future:
\begin{equation}
    \lim_{t \rightarrow \infty} \sup_{B_t \in \mathcal{A}_{(t,\infty)}} |\kappa_n(A,B_t,A,B_t,\ldots,A,B_t)| = 0\,, \qquad \text{for all }  n \in \mathbb{N}.
\end{equation}
The strong F-property implies K-mixing and expresses the complete asymptotic independence of the present from the entire future -- not only in terms of the second cumulant, as in K-mixing, but also with respect to all higher-order cumulants.
Figure~\ref{fig:FreenessErgodicHierarchy} shows the position of F-property as well as strong F-property in the quantum ergodic hierarchy.}

\end{comment}

\subsection{Asymptotic freeness produced by time evolution}\label{sec:free_time_evol}

In this section, we use the formulas provided in Appendix \ref{app-averageOTOC} to gain intuition about the time scale when the mixed free cumulants of two deterministic operators vanish 
under time evolution, so that the operators approach freeness\footnote{In the context of Anosov flows, the exponential decay of correlation functions during the approach to equilibrium is governed by the so-called Pollicott--Ruelle resonances (see \cite{2017pollicottruelleresonances} for a review). One also expects these resonances to control the approach to equilibrium of OTOCs~ \cite{Polchinski:2015cea}. Since freeness emerges when correlation functions and OTOCs involving centered operators vanish, we expect the time scale for the onset of freeness to also be governed by the Pollicott–Ruelle resonances in systems in which such concepts are well-defined. We thank Zhenbin Yang for discussions on this point.}. 

To start with, first we consider the case of the time evolution operator modeled by a random unitary matrix that is drawn from the Haar measure on the unitary group.  
It is known that the operators $A$ and $B_U=U^{\dagger}B U$ are free when $U$ is a unitary matrix belonging to the Haar measure of the unitary group (where $A$ and $B$ are two deterministic $\mcD \times \mcD$ matrices which have convergent densities in the limit $\mcD \rightarrow \infty$).  
To see this from the explicit calculation of the moments, first consider  the two-point function between the operators $A$ and $B_U=U^{\dagger}B U$, which can be written as 
\begin{equation}
\big<AB_U\big>= \langle A\rangle\langle B_U\rangle~.
\end{equation}
This expression is the same one would expect when the operators $A$ and $B_U$ are free.  Next, considering the $4$-point OTOC one can show that \cite{Roberts:2016hpo, Yoshida:2017non}
\begin{equation}
	\langle AB_UAB_U \rangle= \frac{\mcD^2}{\mcD^2-1} \Bigg[\langle A^2 \rangle \langle B \rangle ^2+\langle A \rangle^2\langle B^2 \rangle-\langle A \rangle^2\langle B \rangle^2-\frac{1}{\mcD^2}\langle A^2 \rangle \langle B^2\rangle\Bigg]~.
\end{equation}
Clearly, in the limit $\mcD \rightarrow \infty$, the final term in the above expression vanishes, and it reduces to the one we get when the operators $A$ and $B_U$ are free.  One can follow a similar procedure to show that the higher-order correlators also reduce to the ones predicted by the free probability theory in the $\mcD \rightarrow \infty$ limit. 

To understand the effect of time-evolution by a generic Hamiltonian on two free variables, let us consider the following $2$-point correlation function between 
two deterministic matrices $A$ and $B$, $\brac{A e^{i H t} B  e^{-i H t}}$, where $H$ is a generic Hamiltonian, not necessarily belonging to some random matrix ensemble.  
Now, consider rotating one of the two operators by a random rotation by a unitary matrix $U$ belonging to the Haar measure of the unitary group, and compute the average of the correlation function over the group. A simple calculation shows that, 
\begin{equation}
	\brac{A e^{i H t} B_U  e^{-i H t}}_U=\frac{1}{\mathcal{D}}\int  \text{d}U~ \text{Tr}\Big[A e^{i H t} B_U  e^{-i H t}\Big] = \brac{A}\brac{B_U}~,
\end{equation}
i.e., the average two-point correlation factorises, and is independent of time, irrespective of the Hamiltonian. Similarly, computing the following four-point function, 
\begin{align}
\brac{Ae^{i H t} B_Ue^{-i H t}Ae^{i H t}B_Ue^{-i H t}}_U&=	\frac{1}{\mathcal{D}}\int  \text{d}U~ \text{Tr}\Big[Ae^{i H t} B_Ue^{-i H t}Ae^{i H t}B_Ue^{-i H t}\Big]  \\&=\frac{\mcD^2}{\mcD^2-1} \Bigg[\langle A^2 \rangle \langle B_U \rangle ^2+\langle A \rangle^2\langle B_U^2 \rangle-\langle A \rangle^2\langle B_U \rangle^2-\frac{1}{\mcD^2}\langle A^2 \rangle \langle B_U^2\rangle\Bigg]~,
\end{align}
we again see that it becomes time independent, and in the large $\mcD$ limit, $\mcD \rightarrow \infty$ matches with the free probability expression one gets when the variables $A$ and $B_U$ are free. Therefore, time evolution by a generic Hamiltonian does not change the mutual freeness of two initially free variables. This suggests that chaos should be understood as a mechanism that drives initially non-free operators toward asymptotic freeness, whereas initially free variables remain free under time evolution.

\paragraph{Time evolution generated by random matrices and the spectral form factor.}
Next, consider the quantum dynamics generated by a random matrix, and the correlation function between the operators $A$ and $B(t)=e^{iHt} B e^{-iHt}$. 
As we have discussed, for a random matrix  Hamiltonian drawn from the GUE, the two-point function can be written in terms of the spectral form factor as (see eq. \eqref{2pt_SFF}) 
\begin{equation}\label{2pt_GUE_SFF}
	\begin{split}
	\brac{AB(t)}_{\text{GUE}}=\int \text{d}U~\brac{A_U e^{i H t} B_U e^{-i H t}}_{\text{GUE}}
	=\brac{AB}\frac{\mathcal{R}_2(t)-1}{\mcD^2-1}+\brac{A}\brac{B(t)} \frac{\mcD^2-\mathcal{R}_2(t)}{\mcD^2-1}~.
\end{split}
\end{equation}
 We want to see from this formula and the known time-evolution pattern of the spectral form factor when the above correlation function goes to  $\brac{A}\brac{B(t)} $ consistent with the free probability prediction. First, note that, as expected, at the start of the evolution at $t=0$, when $\mathcal{R}_2(t)=\mcD^2$, the second term above vanishes, and the first one is equal to $\brac{AB}$. For $t>0$, which of the two terms dominates depends on the ratio, $r_f=(\mathcal{R}_2(t)-1)/(\mcD^2-\mathcal{R}_2(t))$. For instance, before reaching the dip, when $\mathcal{R}_2(t) $ crosses a value of $\mcD$, the above ratio is $\mathcal{O}(\mcD^{-1})$ so that the second term above dominates in the limit $\mcD \rightarrow \infty$ and the two-point correlation function matches approximately with the free probability prediction.  Furthermore, at dip time, $t_d \approx \sqrt{\mcD}$,  and $\mathcal{R}_2(t) \approx \sqrt{\mcD}$ \cite{Cotler:2017jue}, so that the ratio goes to zero faster ($\mathcal{O}(\mcD^{-3/2})$) 
in the limit $\mcD \rightarrow \infty$. However, for times larger than $t > t_d$, when the spectral form factor shows a ramp, the decay is slower, even though it still goes to zero 
when the limit $\mcD \rightarrow \infty$ is considered. Finally, when the spectral form factor reaches the plateau, at time $t_P>2\mcD$, $\mathcal{R}_2(t)$ reaches a constant value 
$\mathcal{O}(\mcD)$, and the ratio $r_f$ vanishes as $\mcD^{-1}$ in the large $\mcD$ limit for all subsequent times.  To summarise, for time evolution generated by the a random matrix Hamiltonian drawn from GUE, for times scales larger than at least the time when the spectral form factor takes a value $\mathcal{O}(\mcD)$, the $2$-point correlation 
function of two deterministic matrices factorises and matches with the free probability prediction in the large $\mcD$ limit. 

One can, in principle, use a similar procedure to the one reviewed in the Appendix. \ref{app-averageOTOC} to obtain the four-point OTOC by employing the formula for the integral over the Haar random unitary group. However, due to the fact that this integral gives a large number of terms, it is inconvenient to write the exact expression of the OTOC. Rather than following this approach, in the following, we employ techniques from free probability theory to compute the behavior of the OTOC valid in the large $\mathcal{D}$ limit.

\textbf{OTOCs for random matrices from free probability.}

\begin{figure}[h!]
    \centering
    \includegraphics[width=0.4\linewidth]{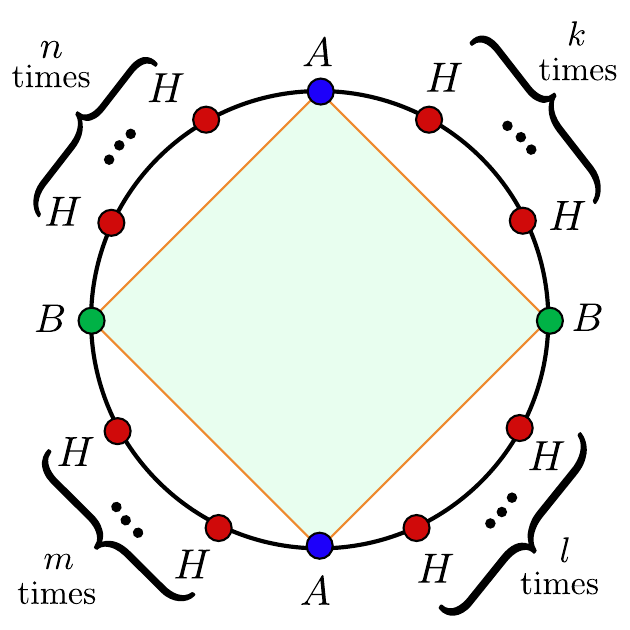}
    \caption{The diagram that contributes to the moment $ m=\brac{A H^kBH^lAH^mBH^n}$ for two traceless operators $A$ and $B$ which have zero initial overlap. Since the large Gaussian matrices $H$ is free from both the deterministic matrices $A$ and $B$, all their mixed free cumulant vanish, and hence the contribution from this diagram is $m=\kappa_4(ABAB) \brac{H^k}\brac{H^l}\brac{H^m}\brac{H^n}$. }
    \label{fig:MomentsOTOC}
\end{figure}

Consider the OTOC $\langle A(0) B(t) A(0) B(t)  \rangle$   between two operators $A$ and $B(t)$, with the time evolution being generated by a random matrix Hamiltonian of rank $\mcD$ drawn from GUE, in the limit $\mcD \rightarrow \infty$. We shall use some results from free probability theory to obtain an expression for this correlator.\footnote{For previous discussions in the literature on the OTOCs for time evolution generated by random matrices, see \cite{Cotler:2017jue, you2018entanglement, Vijay:2018qmx, Bellitti:2019hhn}.} For simplicity, we consider the case that the operators $A$ and $B$ are initially non-overlapping, i.e., $\brac{AB}=\brac{A}\brac{B}$\footnote{E.g., for the case of spin systems, these could be the Pauli spin operators at different sites.}, and they have vanishing first order moment, i.e., $\brac{A}=0=\brac{B}$. Expanding the OTOC, we have 
\begin{equation}
    \text{OTOC}(t)= \sum_{k,l,m,n} \frac{(it)^{k+m}(-it)^{l+n}}{k!~ l!~ m!~n!} \brac{A H^kBH^lAH^mBH^n}~.
\end{equation}
Now, to compute the mixed moment $m=\brac{A H^kBH^lAH^mBH^n}$, we use the fact the large random matrix $H$ of rank $\mcD$ is free from both the deterministic matrices $A$ and $B$ of the same rank, in the limit $\mcD \rightarrow \infty$ \cite{mingo2017free}. Therefore, considering the expansion of the moment in terms of the free cumulants, we see that the only non-zero contribution comes from the diagram in Fig. \ref{fig:MomentsOTOC}, so that it can be written as 
\begin{equation}
    m=\kappa_4(ABAB) \brac{H^k}\brac{H^l}\brac{H^m}\brac{H^n}~.
\end{equation}
Now using the relation $\brac{ABAB}=\kappa_4(ABAB) +2 \kappa_2(AB)$ valid for two operators $A$ and $B$ having vanishing first moments, along with the fact that for a Hamiltonian drawn from the GUE, we have\footnote{Since, $H$ is a Gaussian random matrix, $\brac{H^n}$ is non-zero only when $n$ is even.} $\brac{H^n}=C_{n/2}$, $C_n$ being the Catalan numbers $C_n=\frac{(2n)!}{n! (n+1)!}$, we have 
\begin{equation}
\label{OTOC_free}
    \text{OTOC}(t)= \brac{ABAB} \sum_{k,l,m,n} \frac{(it)^{2(k+m)}(-it)^{2(l+n)}}{(2k)!~ (2l)!~ (2m)!~(2n)!} C_k C_l C_m C_n~= \brac{ABAB} \bigg(\frac{I_1(2 i t)}{it}\bigg)^4~.
\end{equation}
Therefore, the OTOC shows power law decay at early times, and eventually vanishes at later times, consistent with the free probability prediction when the operators $A$ and $B(t)$ are free.  We also emphasize that even though they match with the free probability prediction at early times, the numerical result for finite-dimensional random matrices, as well as the result obtained using the formula for the higher-order averages over the Haar random ensemble, do not go to zero at later times, rather saturate to a small value. Only in the large $\mcD$ limit does the numerical prediction for the OTOC vanish at late times and match with the free probability prediction. 

Before concluding this section, we discuss the following point regarding the difference between the free probability prediction for the OTOC and the one that exists in the literature, computed e.g., using the higher-order moments of the Haar random ensemble. This will highlight the importance of the limit $ \mcD \rightarrow \infty$ as well as  
helps us to understand the role played by the eigenvalue correlation in the random matrix Hamiltonian, which generates the time evolution. First, we compare the following leading order expression for the OTOC obtained in \cite{Cotler:2017jue} 
\begin{equation}
\label{OTOC_haar}
    \text{OTOC}(t)\simeq \brac{ABAB} \frac{\mathcal{R}_4(t)}{\mcD^4}~,~
    \mathcal{R}_4(t)= \brac{|\text{Tr}(e^{- i H t})|^4}=\int \mathcal{D} \lambda \sum_{k,l,m,n}e^{i(E_k+E_l-E_m-E_n)t}~,
\end{equation}
with the OTOC obtained in \eqref{OTOC_free}. Here $\mathcal{R}_4(t)$ is the $4$-point spectral form factor at infinite temperature, its expression obtained using the so-called box-approximation can be written as  \cite{Cotler:2017jue}
\begin{equation}\label{4pt}
\frac{\mathcal{R}_4(t)}{\mcD^4} = r_1^4(t)+ 2 \mcD^{-2} r_2^2(t)-4r_2(t) \big(\mcD^{-2}-\mcD^{-3}\big)-7 \mcD^{-3}r_2(2t)+  4 \mcD^{-3} r_2(3t)+(2 \mcD^{-2}-\mcD^{-3})~,   
\end{equation}
where $r_1(t)$ is the Fourier transform of the density of states of the Hamiltonian, while $r_2(t)$ is the double Fourier transform of the
connected part of the two-point correlation function (which is the sine kernel for GUE \cite{mehta1991random, Cotler:2016fpe}), and their expressions are given by,
\begin{equation}
    r_1^2(t)=|\brac{e^{-it H}} |^2 = \bigg(\frac{J_1(2 t)}{t}\bigg)^2~,~~\text{and}~~ r_2(t)= \begin{cases} 1-\frac{t}{2\mcD}~,~& ~\text{for}~~t<2\mcD~\\
    0~,~& ~\text{for}~~t>2\mcD~.
    \end{cases}
\end{equation}
Thus, $ r^2_1(t)$ is the part of the spectral form factor that comes from the disconnected part of the two-point correlation function.
Now we can observe from the expression in Eq. \eqref{4pt} that in the limit $ \mcD \rightarrow \infty$, the dominant contribution to the OTOC in eq. \eqref{OTOC_haar} comes from the term $r_1^4(t)=|\brac{e^{-it H}} |^4$, i.e. the disconnected part of the spectral form factor, and hence matches with the expression for the OTOC obtained from the free probability theory in eq. \eqref{OTOC_free}. In other words, the correlation between the eigenvalues of a random matrix Hamiltonian does not seem to have an affect on the OTOC in the limit when the rank of the Hamiltonian goes to infinity.\footnote{One can easily check that this is also true for two-point functions.} In the free probability computation in eq. \eqref{OTOC_free} (and the Fig. \ref{fig:MomentsOTOC}), this is a direct consequence of the fact that crossing partitions do not contribute to the moment expansion.

\section{Eigenvalue statistics of operators  as a signature of asymptotic freeness } \label{sec-operatorStatistics}

\begin{figure}[h!]
    \centering
    \includegraphics[width=0.7\linewidth]{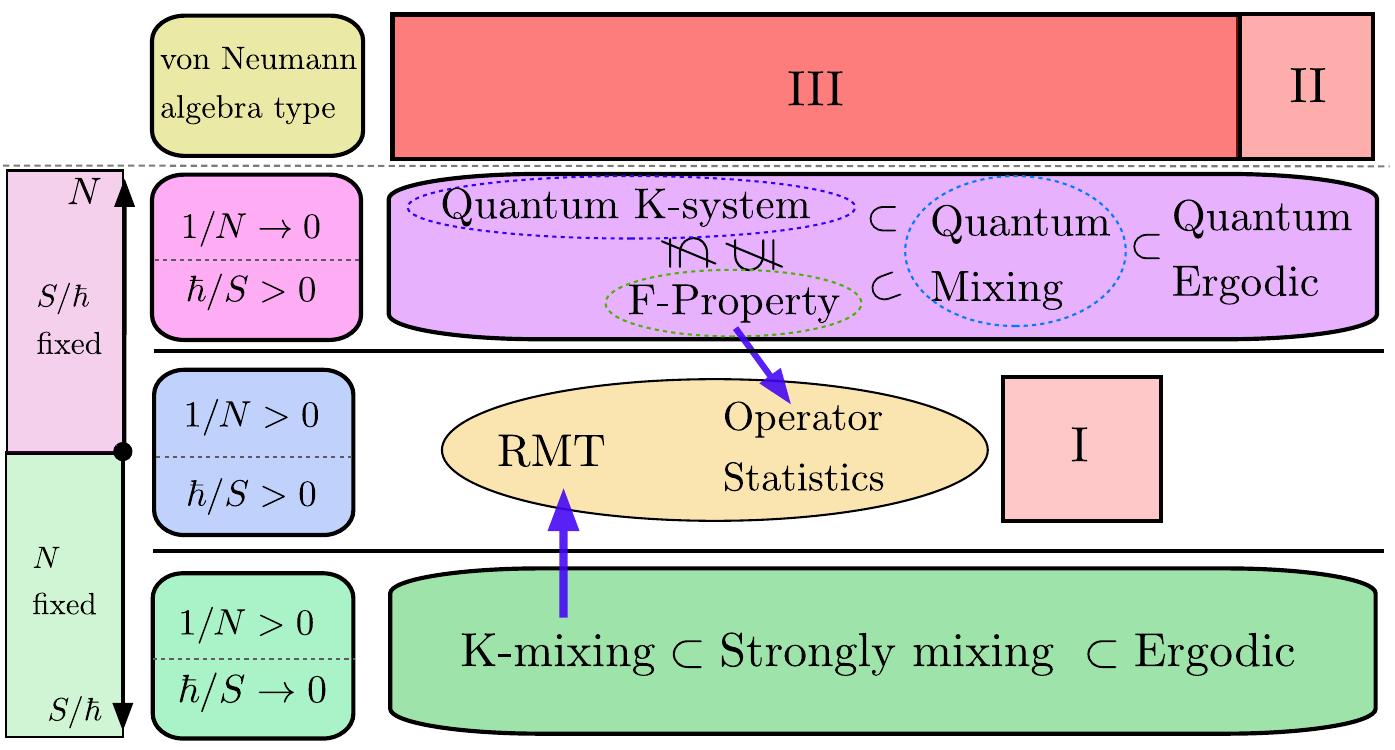}
    \caption{Illustration of the quantum and classical ergodic hierarchies.  $N$ denotes the number of degrees of freedom of the system. Freeness (F-property) is stronger than quantum 2-mixing, but it is neither stronger nor weaker than the quantum K-property.  The corresponding von Neumann algebraic structures associated with different degrees of quantum chaos are indicated. Note that the quantum hierarchy is only meaningful in the $1/N\rightarrow 0$ limit, as for any finite $N$, the algebra remains type I, preventing even ergodicity. Nonetheless, a finite-$N$ system can exhibit signatures of its asymptotic behavior. For example, the emergence of operator spectral statistics governed by free probability theory serves as a robust indicator of asymptotic freeness. }
    \label{fig:FreenessErgodicHierarchy}
\end{figure}

The fact that freeness can only emerge in the large \( \mathcal{D} \)\footnote{We remind the reader that $\mathcal{D}$ denotes the rank of the operators in the algebra.} limit can be understood from an algebraic perspective, particularly in the context of the quantum ergodic hierarchy. First, observe that freeness implies 2-mixing. Since mixing is a property that arises only for type II or III algebras, it follows that freeness can occur only for type II or type III algebras. For finite \( \mathcal{D} \), however, the algebra of observables is always of type I, and therefore, freeness cannot occur. As a result, mixed free cumulants of centered variables (\( A - \langle A \rangle \) and \( B(t) - \langle B(t) \rangle \)) will not generically vanish for finite \( \mathcal{D} \). However, if the dynamics is chaotic, these cumulants are expected to approach a small residual value that scales inversely with the number of degrees of freedom in the system. This expectation was indeed shown to be true for OTOCs in \cite{Huang:2017fng}. Furthermore, if two operators \( A \) and \( B(t) \) are free, the eigenvalue density of their sum \( \rho_{A(0) + B(t)} \) is expected to follow the free convolution prediction, which can emerge even for finite, but sufficiently large \( \mathcal{D} \), and serves as a ``smoking gun" of asymptotic freeness \cite{Chen:2024zfj}. These concepts are illustrated schematically in Figure \ref{fig:FreenessErgodicHierarchy}.

\subsection{Free additive convolution}\label{free_addition}
In this section, we describe how to determine the distribution of the sum of two self-adjoint free random variables. Given two self-adjoint random variables, \( a_1 \) and \( a_2 \), which are free, the distribution of their sum, \( a_1 + a_2 \), is given by the free convolution of their individual distributions, i.e.,  
\begin{equation}
    \rho_{a_1+a_2} = \rho_{a_1} \boxplus \rho_{a_2}.
\end{equation}
This operation, introduced in \cite{voiculescu1986addition}, provides a non-commutative analogue of classical convolution in probability theory. The procedure for computing \( \rho_{a_1 + a_2} \) is outlined below (see \cite{speicher2019lecture, nica2006lectures, Potters,tao2012topics} for comprehensive reviews on the subject, and \cite{Zee:1996qu, zinn1999adding} for discussions related to random matrices). 

For an operator \( a \) with eigenvalue density \( \rho(\lambda) \) compactly supported on \( \mathbb{R} \), the Cauchy transform is defined as 
\begin{equation}\label{Cauchy_tra}
    G(z) = \int_{\mathbb{I}} \textrm{d}\lambda \, \frac{\rho(\lambda)}{z - \lambda}~,~~~z \in \mathbb{C}\,, %\mathbb{I}~,
\end{equation}
where $\mathbb{I}$ is the support of the distribution $\rho(\lambda)$ on the real line. 
For example, for a $\mathcal{D} \times \mcD$ random matrix Hamiltonian ($H$), consider its resolvent defined as, $G_N(z)=\frac{1}{\mcD}\Tr \frac{1}{z-H}=\frac{1}{\mcD}\sum_i \frac{1}{z-\lambda_i}$. Now averaging over the random matrix ensemble, and taking the limit $\mcD \rightarrow \infty$, we have the Cauchy transform of the eigenvalue density: $\mathbb{E}(G_N(z)) \equiv G(z) = \int_{\mathbb{I}} \textrm{d}\lambda \, \frac{\rho(\lambda)}{z - \lambda}$. \footnote{\label{moments_gen}For a self-adjoint operator, say $\mathcal{O}$, the Cauchy transformation of its eigenvalue distribution can also be thought of as the moment generating function 
of this operator, i.e., $G_\mathcal{O}(z)=\sum_{k=0}^{\infty} \frac{\varphi(\mathcal{O}^k)}{z^{k+1}}$, where $\varphi(\mathcal{O}^k)$ denotes the $k$-th order moment of the operator (see Appendix \ref{sec-appA}). E.g., for a random matrix Hamiltonian, we have, $\varphi(H^k)=\mathbb{E}(\Tr (H^k))/\mathcal{D} = \int \rho(\lambda) \lambda^k \textrm{d}\lambda$. The expansion of the $G(z)$ indicates that it is well-behaved at $\infty$, and therefore, having the knowledge of this function near infinity is equivalent to having the knowledge of all the moments of the operator.} 

The eigenvalue density \( \rho(\lambda) \) can be recovered from the Cauchy transform using the Stieltjes inversion formula:  
\begin{equation}
    \rho(\lambda) = -\frac{1}{\pi} \lim_{\epsilon \to 0} \left(\Im G(\lambda + i\epsilon)\right).
    \label{eq: Stieltjes inversion}
\end{equation}
Since \( \rho \) is compactly supported on an interval containing the origin, the inverse of the Cauchy transform, \( G^{-1}(z) \), which we denote as \( \mathcal{B}(z) \), has a pole at zero and can be decomposed as  
\begin{equation}\label{B_and_R}
    G^{-1}(z) \equiv \mathcal{B}(z) = \frac{1}{z} + R(z)~,
\end{equation}
where \( R(z) \) is called the \( R \)-transform of \( \rho \). The \( R \)-transform is essentially the regular part of the inverse Cauchy transform after subtracting its singular component.  Whereas the Cauchy
transform $G(z)$ is analytic in the entire upper half of the complex plane, the $R$-transform is defined for small values of $z$ only and the domain of dependence depends on the eigenvalue density \cite{mingo2017free}. 
Since the inverse function \( \mathcal{B}(z) \) may have multiple branches, the correct physical solution is determined by requiring that the corresponding \( R \)-transform satisfies  
\begin{equation}
    R(0) = 0.
\end{equation}
This condition ensures the proper analytic continuation and consistency of the transformation.

For two free random variables \( a_1 \) and \( a_2 \) with compactly supported eigenvalue densities \( \rho_{a_1} \) and \( \rho_{a_2} \), the \( R \)-transform of their sum satisfies\footnote{The \( R \)-transform of a random variable \( a \) can also be written as \( R_a(z) = \sum_{n=1}^{\infty} \kappa_n(a) z^{n-1} \), where \( \kappa_n(a) \) are the free cumulants of \( a \). For two free self-adjoint random variables \( a \) and \( b \), the cumulants satisfy the property \( \kappa_n(a+b) = \kappa_n(a) + \kappa_n(b) \). This property directly leads to the result that the \( R \)-transform of the sum of \( a \) and \( b \) is the sum of their individual \( R \)-transforms. For a discussion of free cumulants, we refer to Appendix \ref{sec-appA}.}
  
\begin{equation}\label{R_tra_sum}
    R_{\rho_{a_1} \boxplus \rho_{a_2}}(z) = R_{\rho_{a_1}}(z) + R_{\rho_{a_2}}(z),
\end{equation}
for sufficiently small \( |z| \) in \( \mathbb{C} \). This property provides a method for determining the eigenvalue density of the sum of free operators. Given the \( R \)-function, the corresponding Cauchy transform \( G_{\rho_{a_1} \boxplus \rho_{a_2}}(z) \) can be computed using equation (\ref{B_and_R}). By applying the Stieltjes inversion formula (\ref{eq: Stieltjes inversion}), we obtain the eigenvalue density of the sum:  
\begin{equation}\label{conv_density}
    \rho_{a_1 + a_2}(\lambda) = -\frac{1}{\pi} \lim_{\epsilon \to 0} \left(\Im~ G_{\rho_{a_1} \boxplus \rho_{a_2}}(\lambda + i\epsilon)\right).
\end{equation}
Since the Cauchy transform, and consequently the eigenvalue density, may admit multiple solutions, the physically relevant solution must be chosen by ensuring positivity within the appropriate domain. There 
can also be solution for  \( G_{\rho_{a_1} \boxplus \rho_{a_2}}(z) \) which are complex conjugates of each other. In that case, one can take the absolute value of any one of these conjugate solutions to get the correct eigenvalue density (see the example of the convolution of the eigenvalue density
of the sum of two free spin-1 operators that we discuss below).

Next, we illustrate the procedure of free additive convolution with 
two examples.

\paragraph{Example 1. The sum of Spin-1/2 Operators (Bernoulli distribution).}

In this case, the eigenvalue density of either of the two operators follows the Bernoulli distribution
\begin{equation}
    \rho_{\sigma}(\lambda)=\frac{\delta(\lambda-1)+\delta(\lambda+1)}{2}~.
\end{equation}
Since both $\rho_{\sigma_1}$ and $\rho_{\sigma_2}$ \footnote{To simplify the notations, below, we denote $\rho_{\sigma_1}$ and $\rho_{\sigma_2}$ 
as $\rho_1$ and $\rho_2$, respectively. The free additive convolution of them is denoted as $\rho_3$, i.e., $\rho_3=\rho_{1} \boxplus \rho_{2}$. This convention is followed through out the paper.} have compact supports on $\mathbb{R}$, one can obtain the Cauchy transformation \eqref{Cauchy_tra} of this distribution to be 
\begin{equation}
    G_{\rho_1}(z)=\frac{z}{z^2-1}~.
\end{equation}
Inverting $G_{\rho_1}(z)$, and using the definition in eq. \eqref{B_and_R} we find the corresponding $R$-transformation is 
\begin{equation}
    R_{\rho_1}(z)=\frac{-1+\sqrt{4z^2+1}}{2z}~.
\end{equation}
Therefore, for two asymptotic free Pauli operators, we have, according to the relation in eq. \eqref{R_tra_sum},
\begin{equation}
    R_{\rho_3}(z)=\frac{\sqrt{1+4z^2}-1}{z}.
\end{equation}
Next, we obtain $\mathcal{B}_{\rho_3}(z)=\frac{1}{z}+R_{\rho_3}(z)$, and invert it to find the corresponding Cauchy transformation to be given by
\begin{equation}
    G_{\rho_3}(z)=\frac{1}{\sqrt{z^2-4}}~.
\end{equation}
Here, among the two solutions of the quadratic equation for $ G_{\rho_3}(z)$, we have kept only the one that gives rise to a positive eigenvalue density. 
After applying eq. \eqref{conv_density}, we obtain the convolved eigenvalue density of two Pauli operators as
\begin{equation} \label{eq:arcsine}
    \rho_{3}(\lambda)=\rho_{1} (\lambda) \boxplus \rho_{2}(\lambda)=\frac{1}{\pi \sqrt{4-\lambda^2}}, ~~~ \text{for}~~ |\lambda|<2.
\end{equation}
Therefore, the convolved distribution vanishes when $|\lambda|>2$. The distribution in \eqref{eq:arcsine} is known as the \textit{arcsine} distribution in the literature \cite{nica2006lectures, speicher2009free}.

\begin{figure}
\centering
\begin{tikzpicture}
\node at(-5.5,0) {$\rho_{\sigma_1}\Big(=\frac{1}{2}(\delta_{-1}+\delta_{1})\Big)~~~\boxplus~~~ \rho_{\sigma_2}\Big(=\frac{1}{2}(\delta_{-1}+\delta_{1})\Big)=$};
\pgfdeclareimage[width=0.35\textwidth]{img}{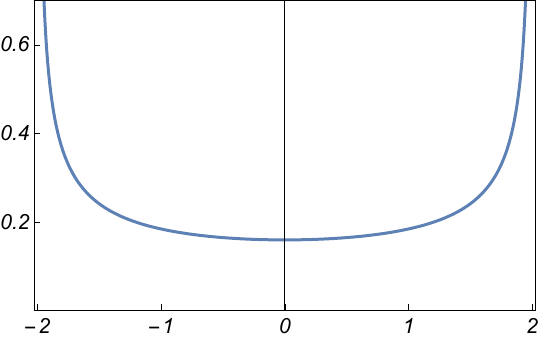}
\node at(1.5,0){\pgfuseimage{img}};
\node at(1.5,-2) {$\rho_{\sigma_1+\sigma_2}=\frac{1}{\pi \sqrt{4-\lambda^2}},~~ \lambda \in (-2,2)$};
\end{tikzpicture}
\caption{Convolved distribution of the sum of two free Pauli operators from free probability theory.}
\label{fig: acsine_plot}
\end{figure}

\begin{figure}[h!]
		\centering
		\includegraphics[width=3.5in,height=2.5in]{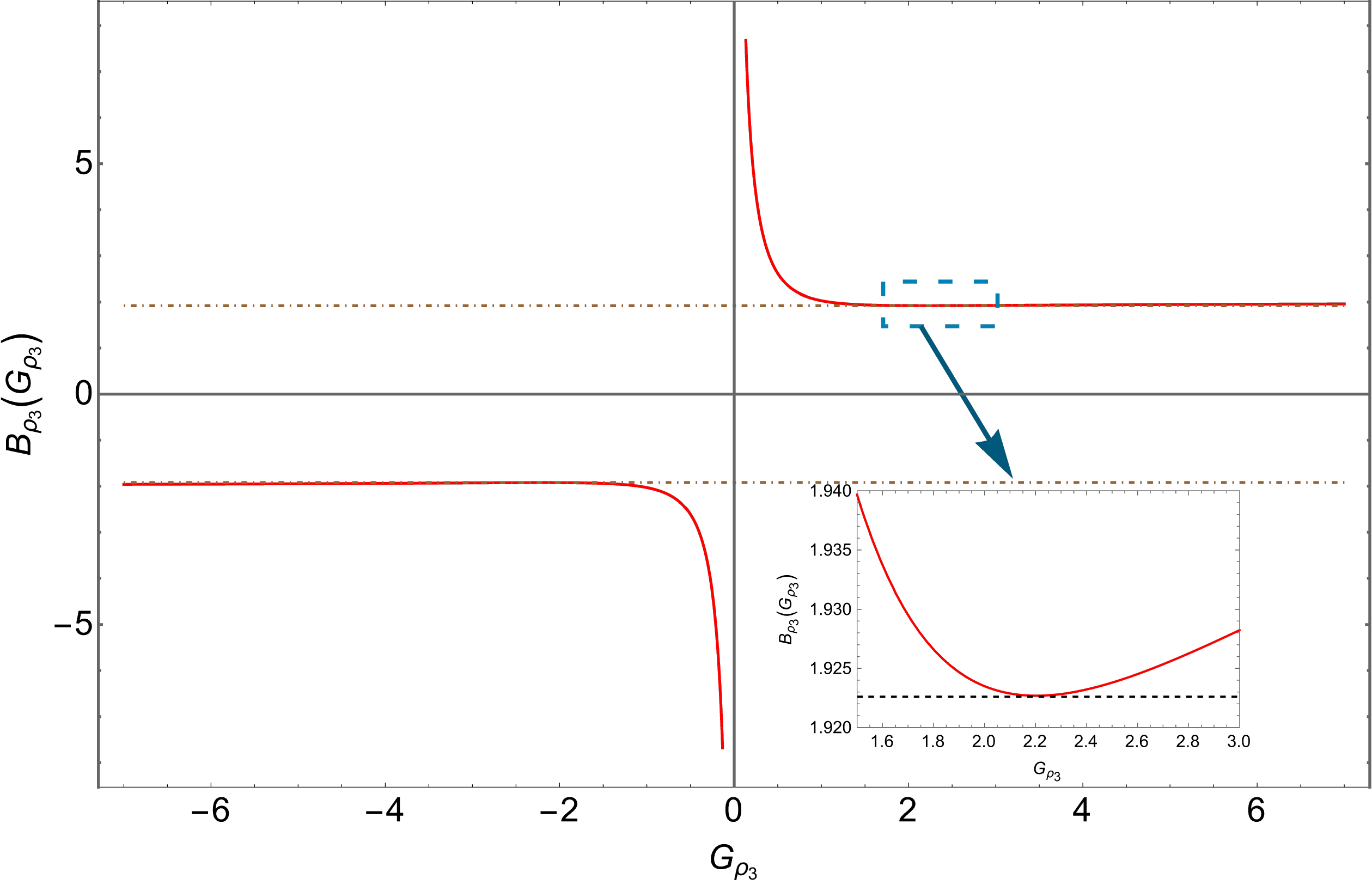}
		\caption{Plot of the function $\mathcal{B}_{\rho_3}(G_{\rho_3}(z))(=z)$. The edges of the spectrum of $\Sigma_3$ are located between a local maximum and minimum of this function.
        We have shown this minimum more clearly in the inset. } 
		\label{fig:B_SPIN1_SUM}
\end{figure}

\paragraph{Example 2. The sum of Spin-1 Operators (generalized Bernoulli distribution).}
As the next example, we consider the sum of two spin-1 operators $\Sigma$, each of which has three eigenvalues so that the spectrum of individual operators follows a generalized Bernoulli distribution of the form 
\begin{equation}
    \rho_{\Sigma}(\lambda)=\frac{\delta(\lambda-1)+\delta(\lambda)+\delta(\lambda+1)}{3}~.
\end{equation}
The Cauchy transform of this distribution can be calculated to 
be given by
\begin{equation}
    G_{\rho_\Sigma}(z)=\frac{3z^2-1}{3z(z^2-1)}~.
\end{equation}
After finding the inverse function, one can get the R-transformation (here, we discard two other $R$-transformations that do not vanish at $z=0$)
\begin{equation}\label{eq: Rfunc spin1}
\begin{split}
    R_{\rho_\Sigma}(z)=\mathcal{B}_{\rho_\Sigma}(z)-\frac{1}{z}=\frac{3z^2+(f(z)-1)^2}{3z f(z)}~,\\
    \text{with}~~ f(z)= \Big(1+3\sqrt{-z^2(1+3z^2+3z^4)}\Big)^{1/3}~.
    \end{split}
\end{equation}
The $R$-transform of the eigenvalue density of the sum of two such operators (say, $\Sigma_1$ and $\Sigma_2$), which are asymptotically free, is the sum of the $R$-transforms of the individual distributions, and hence is given by \footnote{As in the previous example, here $\rho_3$ denotes the eigenvalue density of the summed operator, $\Sigma_3$.}
\begin{equation}
    R_{\rho_3}(z)=\frac{6z^2+2(f(z)-1)^2}{3z f(z)}~.
\end{equation}
From this, we first obtain $\mathcal{B}_{\rho_3}(z)=R_{\rho_3}(z)+\frac{1}{z}$ and calculate the corresponding Cauchy transformation by using the inversion formula $\mathcal{B}_{\rho_3}(G_{\rho_3}(z))=z$. 
There are multiple solutions of this equation for the Cauchy transform $G_{\rho_3}(z)$. To obtain the correct eigenvalue density, we first plot
the function $\mathcal{B}_{\rho_3}(G_{\rho_3}(z))(=z)$, shown in Fig. \ref{fig:B_SPIN1_SUM}. There is a region on the vertical axis (from around $-1.92$ to $1.92$) where the function  $\mathcal{B}_{\rho_3}$
does not attain real values, and hence, this corresponds to the region where 
the eigenvalues of the sum operator lie. Furthermore, as we have shown in the inset of Fig. \ref{fig:B_SPIN1_SUM}, this exists between the
local minimum and maximum of the function $\mathcal{B}_{\rho_3}(G_{\rho_3})$. More precisely, we determine the approximate edges of the spectrum of $\Sigma_3$ to be between $-1.9227$ to $1.9227$. 

Having determined the range of the spectrum, we now determine the density 
of eigenvalues of the convolved operator. 
As we are eventually looking for a convolved probability distribution, it should be positive definite in an interval in $\mathbb{R}$, which, in this case, is approximately between $-1.9227$ to $1.9227$.

\begin{figure}
\centering
\begin{tikzpicture}
\node at(-5.5,0) {$\rho_{1}\Big(=\frac{1}{3}(\delta_{-1}+\delta_0+\delta_{1})\Big)~~\boxplus~~\rho_{2}\Big(=\frac{1}{3}(\delta_{-1}+\delta_0+\delta_{1})\Big)=$};
\pgfdeclareimage[width=0.35\textwidth]{img}{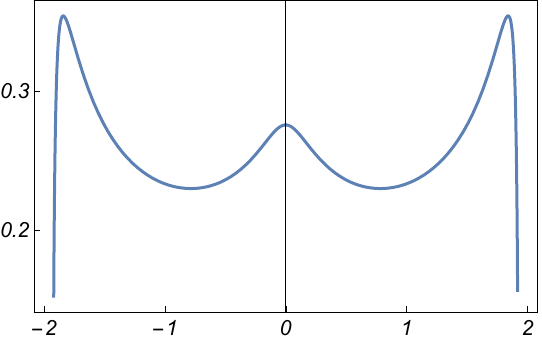}
\node at(2.5,0){\pgfuseimage{img}};
\node at(2.5,-2) {$\rho_{3}$,  $\lambda \in (-1.9227,1.9227)$};
\end{tikzpicture}
\caption{Convolved distribution for the eigenvalue density of the sum of two free spin-1 operators according to free probability theory. The approximate edges of the spectrum, as determined from the plot of the function $\mathcal{B}_{\rho_3}(G_{\rho_3}(z))(=z)$ in Fig. \ref{fig:B_SPIN1_SUM} are, respectively, -1.9227 and 1.9227.  }
\label{fig: spin1_plot}
\end{figure}

Since the complex conjugated pair of solutions has the same absolute
value of their imaginary parts, from the Stieltjes inverse formula, we 
can obtain the eigenvalue density of the sum of two spin-1 operators to be 
\begin{equation} 
     \rho_{3}(\lambda)= \lim_{\epsilon \rightarrow 0} \Big|\frac{1}{\pi}\Im~ G_{\rho_3}(\lambda+i\epsilon)\Big|~.
\end{equation}
This is plotted in Fig. \ref{fig: spin1_plot}. Furthermore, as one can check, here taking the limit $\epsilon \rightarrow 0$ gives the 
same distribution as directly substituting $\epsilon=0$, it is possible to obtain the following analytical expression for the distribution shown in Fig. \ref{fig: spin1_plot} as well, 
\begin{equation} 
    \rho_{3}(\lambda)
    =\bigg|\frac{ f_1^{2}(\lambda)-4 - 9 \lambda^4 + 33 \lambda^2 }{3 \sqrt{3} \,\pi \lambda \left(\lambda^2-4\right) f_1(\lambda)}\bigg|~,~~~
    -1.9227 \leq \lambda \leq 1.9227~,
    \label{eq: sumdis_spin1}
\end{equation}
where we have defined the function
\begin{equation}
    f_1(\lambda)=\left( 8+9 \left(3 \lambda^4-30 \lambda^2+70\right) \lambda^2+27 \sqrt{-\lambda^2 \left(\lambda^2-4\right)^2 \left(9 \lambda^4-33 \lambda^2-1\right)}\right)^{1/3}~.
\end{equation}
To the best of our knowledge, the distribution in Equation \eqref{eq: sumdis_spin1} for the sum of two free spin-1 operators is first derived in this work.

\subsubsection{Free additive convolution for arbitrary probability measures } \label{sec-numFreeConvolution}

In this section, we briefly review the method described in Section 5 of~\cite{speicher2019lecture} to compute the free additive convolution of arbitrary probability measures. See also Appendix B of \cite{Pollock:2025acf}. This approach is useful for studying more general algebras, beyond those involving spin-1/2 and spin-1 operators that appear in the models considered in this work. 

Consider two free operators, $a$ and $b$, with distributions $\rho_a$ and $\rho_b$, whose Cauchy transforms are denoted by $G_a(z)$ and $G_b(z)$, respectively. Their corresponding $R$-transforms are denoted as $R_a(z)$ and $R_b(z)$.  We first define the {\it subordination functions} $\omega_a(z)$ and $\omega_b(z)$  as follows:
\begin{equation}
    \omega_a(z) = z - R_b(G_{\rho_a \boxplus \rho_b }(z))\,,~~~~ \omega_b(z) = z - R_a(G_{\rho_a \boxplus \rho_b }(z))\,,
\end{equation}
where $G_{\rho_a \boxplus \rho_b }(z)$ denotes the Cauchy transform of $\rho_a \boxplus \rho_b$. Then one can show that,
\begin{equation} \label{eq-numCauchySum}
    G_{\rho_a \boxplus \rho_b}(z) = G_a(\omega_a(z)) = G_b(\omega_b(z))\,.
\end{equation}
From this, the eigenvalue distribution of the sum of these two operators can be obtained via the inverse Stieltjes transformation formula:
\begin{equation} \label{eq-inverseNumSum}
\rho_{a+b}(\lambda) = -\lim_{\epsilon \rightarrow 0} \frac{1}{\pi} \Im G_{\rho_a \boxplus \rho_b}(\lambda + i \epsilon)\,.
\end{equation}

Therefore, the problem of determining the distribution $\rho_{a+b}$ can be solved by finding either of the two subordination functions, \( \omega_a \) or \( \omega_b \), and then using Eqs.~(\ref{eq-numCauchySum}) and (\ref{eq-inverseNumSum}). This can be done numerically as follows. First, one introduces the auxiliary functions:
\begin{equation}\label{auxi_fn}
    H_a(z) = \frac{1}{G_a(z)} - z\,,~~~~ H_b(z) = \frac{1}{G_b(z)} - z\,.
\end{equation}
These functions satisfy the following fixed-point equation for $\omega_a(z) \in \mathbb{C}^+$ (the upper half-plane)~\cite{speicher2019lecture}:
\begin{equation} \label{eq-iteration}
    \omega_a(z) = z + H_b\left[ z + H_a(\omega_a(z)) \right].
\end{equation}
For eigenvalue distributions corresponding to the sum of spin-1/2 and spin-1 operators, these equations can be solved analytically. More generally, they can be solved numerically in an iterative manner. For a fixed value of $z$, one starts with an initial guess for $\omega_a(z)$ and updates it iteratively using Eq.~(\ref{eq-iteration}) until convergence. Once $\omega_a(z)$ is determined, Eqs.~(\ref{eq-numCauchySum}) and (\ref{eq-inverseNumSum}) allow for the numerical computation of $\rho_{a+b}$.  

This method only requires knowledge of the Cauchy transform of \( a \) and \( b \), which is not too difficult to obtain. Below, we present a couple of detailed examples illustrating how to apply the subordination function method to numerically compute the free convolution prediction.

\paragraph {Example 3. Sum of Spin-2 operators}
To exemplify the use of subordination function methods, we now consider the sum of two free spin-2 operators, which one with eigenvalue distribution given by 
\begin{equation}
    \rho_{s=2}(\lambda)=\frac{ \delta(\lambda-2)+\delta(\lambda-1)+\delta(\lambda)+\delta(\lambda+1)+\delta(\lambda+2)}{5}\,.
\end{equation}
The corresponding Cauchy distribution is given by
\begin{equation}
    G_{s=2}(z)=\int d\lambda\frac{\rho_{s=2}(\lambda)}{z-\lambda}=\frac{5 \left(z^2-3\right) z^2+4}{5 \left(z^5-5 z^3+4 z\right)}\,.
\end{equation}
The auxiliary function can then be obtained as
\begin{equation} \label{eq:Hauxiliary}
    H_{s=2}(z)=\frac{1}{G_{s=2}(z)}-z=\frac{2 z \left(8-5 z^2\right)}{5 \left(z^2-3\right) z^2+4}\,.
\end{equation}
Now, we can numerically solve \eqref{eq-iteration} for $H_a(z)$ given by \eqref{eq:Hauxiliary} to find $\omega_a(z)$. Then we can employ Eqs.~(\ref{eq-numCauchySum}) and (\ref{eq-inverseNumSum}) to numerically find the distribution for the sum of two free spin-2 operators. 

\paragraph {Example 4. Sum of Spin-$s$ operators in the large $s$ limit}
We now consider the sum of two free spin-$s$ operators, denoted as $X^{(s)}$, in the large $s$ limit. The density of eigenvalues of each operator is given by:
\begin{equation}
    \rho_{X^{(s)}}(\lambda)=\frac{1}{2s+1}\sum_{j=-s}^{s} \delta(\lambda-j)\,.
\end{equation}
To simplify the analysis, it is convenient to consider the normalized operator $x^{(s)}=X^{(s)}/s$, whose density of eigenvalues in the large-$s$ limit can be well-approximated by a uniform distribution with support between -1 and 1, namely:
\begin{equation}
    \rho_{x^{(s)}}(\lambda)=\begin{cases}
        \frac{1}{2}, ~~~\lambda \in [-1,1]\,,\\
        0, ~~~\text{otherwise}\,.
    \end{cases}
\end{equation}
The  Cauchy transformation corresponding to this distribution  is simply given by
\begin{equation}\label{Cauchy_larges}
    G_{\rho^{(s)}}(z)=\frac{1}{2} \log \left(\frac{z+1}{z-1}\right)\,.
\end{equation}
From this, we get the expression for the $R$-transformation to be 
\begin{equation}
    R_{\rho^{(s)}}(z)=\coth z-\frac{1}{z}~.
\end{equation}

\begin{figure}[h!]
		\centering
		\includegraphics[width=3in,height=2in]{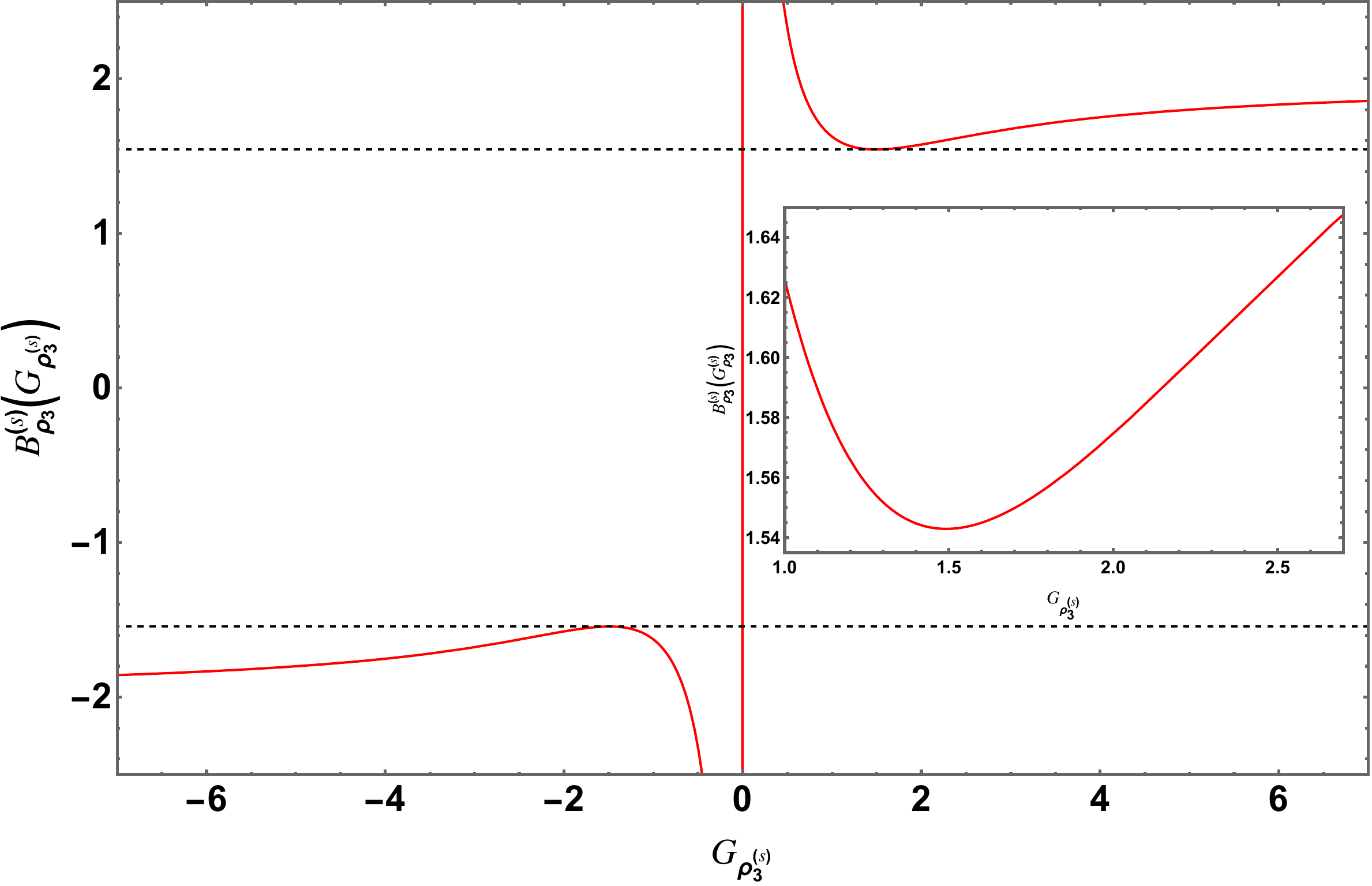}
		\caption{Plot of the function $\mathcal{B}_{\rho_3^{(s)}}(G_{\rho_3^{(s)}}(z))(=z)$. The edges of the spectrum of $x_3^{(s)}$ are located between a local maximum and minimum of this function, shown by the dashed black lines.
        We have shown the minimum of the function more clearly in the inset. } 
		\label{fig:B_large_spin}
\end{figure}

Now considering the free additive convolution of two spin-$s$ operators having the same large value of $s$, $x_3^{(s)}=x_1^{(s)}+x_2^{(s)}$ the $R$-transformation of the sum is given by $R_{\rho_3^{(s)}}(z)=2R_{\rho_1^{(s)}}(z)$.
Therefore, the Cauchy transformation ($G_{\rho_3^{(s)}}(z)$) of the convolved distribution 
$\rho_3(\lambda)=\rho_{x_1^{(s)}}(\lambda)\boxplus\rho_{x_2^{(s)}}(\lambda)$ satisfies the following transcendental equation \cite{Potters, speicher2009free},
-p0s\begin{equation}
  \mathcal{B}_{\rho_3^{(s)}}(G_{\rho_3^{(s)}}(z))=z ~~\rightarrow~~  2 \coth (G_{\rho_3^{(s)}}(z))-\frac{1}{G_{\rho_3^{(s)}}(z)}-z=0~.
\end{equation}
This equation can not be solved exactly for the Cauchy transformation. However, one can determine the range of the spectrum of the operator $x_3^{(s)}$ by plotting the inverse of $G_{\rho_3}(z)$, which has a local maximum and a local minimum. The eigenvalues of $x_3^{(s)}$ lie within this region where there is no real solution of the equation  $\mathcal{B}_{\rho_3^{(s)}}(G_{\rho_3^{(s)}}(z))=z$. As can be seen from the plot in Fig. \ref{fig:B_large_spin}, the edges of the spectrum are approximately at $-1.543$ and $1.543$. 

%\KP{I have added the above plot to determine the range of the spectrum for large spin addition. Also, changed the notations a bit to be consistent with the ones used previously. } \VJ{Thx!}

To obtain the density of eigenvalues, we use the subordination method described in section \ref{sec-numFreeConvolution}. The auxiliary function \eqref{auxi_fn} corresponding to the Cauchy transformation in \eqref{Cauchy_larges} reads
\begin{equation}
    H_{x^{(s)}}(z)=\frac{2-z \log \left(\frac{z+1}{z-1}\right)}{\log \left(\frac{z+1}{z-1}\right)}\,.
\end{equation}
The same approximation in terms of a uniform distribution works well for the non-normalized operator $X^{(s)}$, for which the corresponding auxiliary function is given by
\begin{equation}
H_{X^{(s)}}(z) = \frac{2s + 1-z \log \left(\frac{z+s}{z-s}\right) }{\log \left(\frac{z+s}{z-s}\right)}\,.
\end{equation}
Again, we can employ Eqs.~(\ref{eq-numCauchySum}) and (\ref{eq-inverseNumSum}) to numerically find the distribution for the sum of two free spin-$s$ operators in the large $s$ limit.

\begin{figure}[h!]
    \centering
    \begin{tabular}{cc}
        \includegraphics[width=0.45\textwidth]{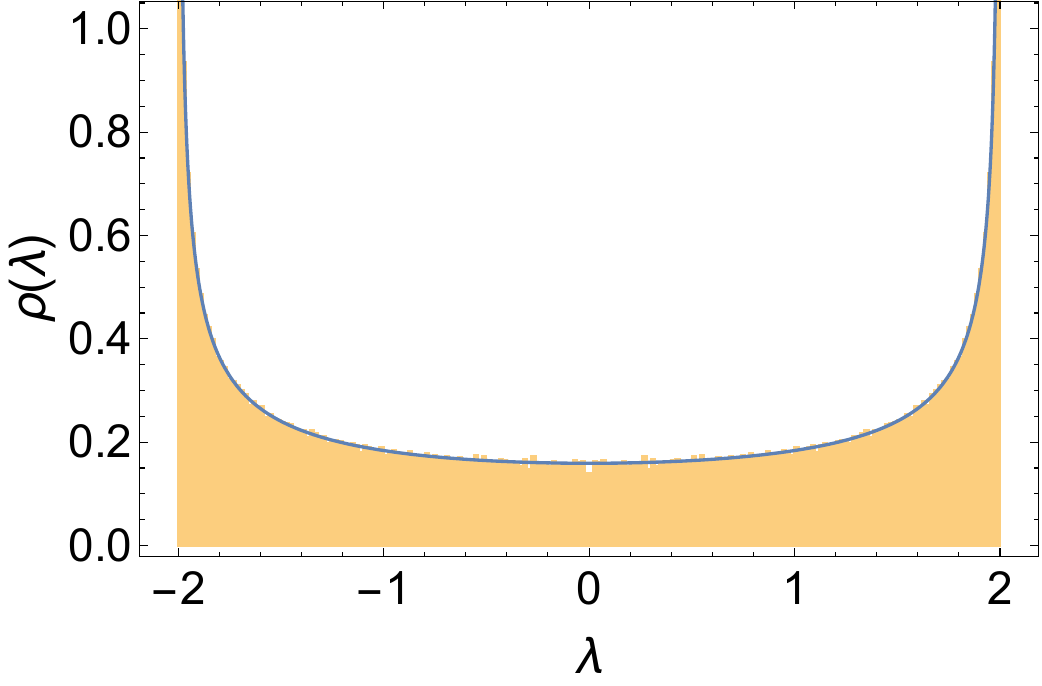} &
        \includegraphics[width=0.45\textwidth]{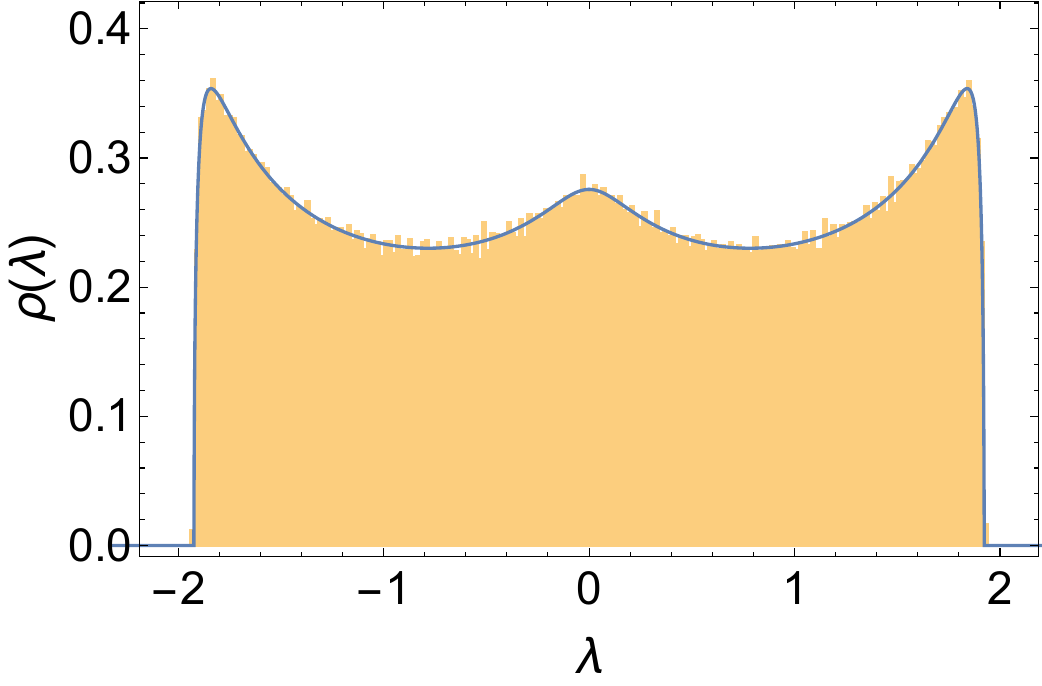} 
    \end{tabular}
     \caption{Density of eigenvalues for the operator sum $X^{(s)}_i + U^{\dagger} X^{(s)}_i U$, where $X^{(s)}$  spin-$s$ operators and $U = e^{-i t M}$ represents the time evolution operator. Here we set $t=50$. The matrix $M$ is drawn from a GOE with zero mean and variance $2/d_s$, where $d_s = (2s+1)^L$ is the Hilbert space dimension for a spin-$s$ chain with $L$ sites. The histograms are based on 100 independent realizations of $M$. The left panel corresponds to spin-$1/2$ operators with $(s,L)=(1/2,10)$, while the right panel shows results for spin-1 operators with $(s,L)=(1,6)$. The blue curves represent the free probability prediction.
}
\label{fig:FreeConvolutionRM}
\end{figure}

\begin{figure}[h!]
    \centering
    \begin{tabular}{cc}
        \includegraphics[width=0.45\textwidth]{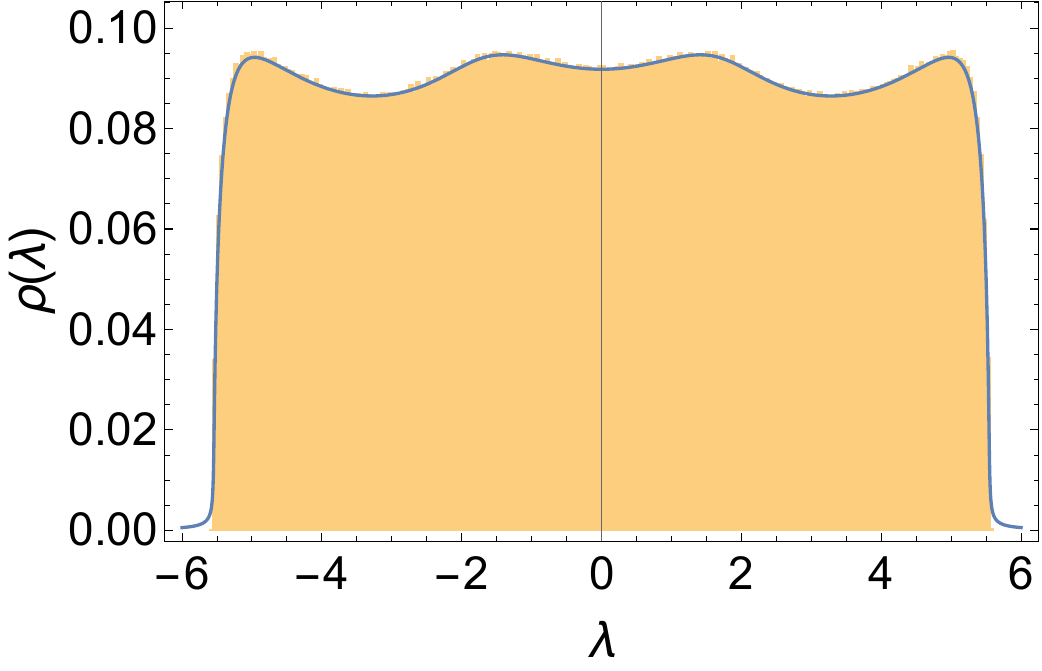} & 
        \includegraphics[width=0.45\textwidth]{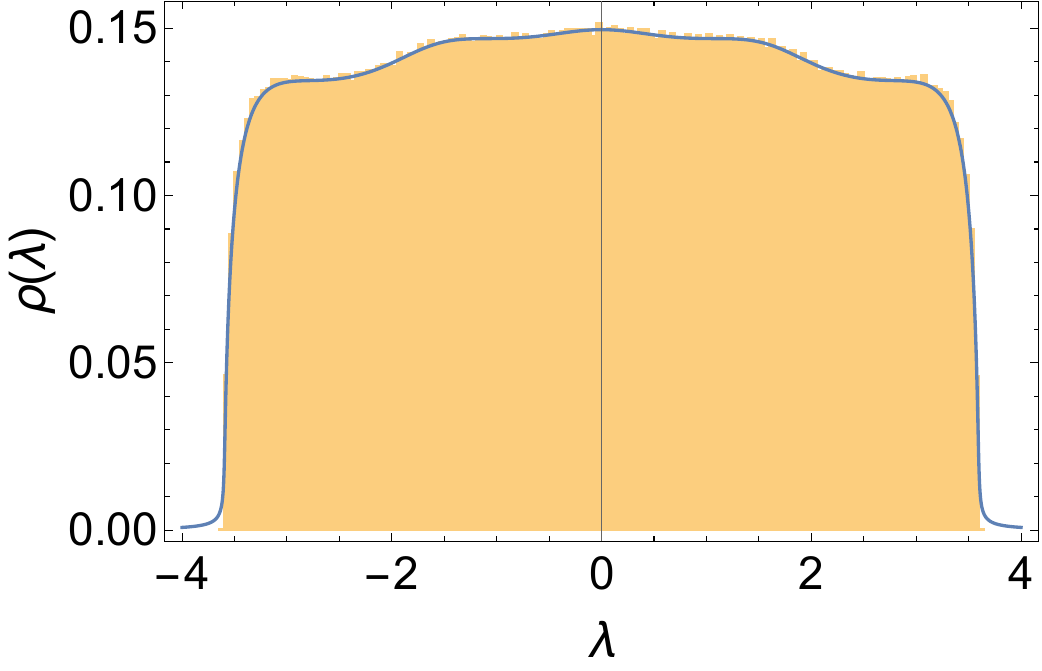} \\
        (a) spin-3/2 & (b) spin-2 \\
        \includegraphics[width=0.45\textwidth]{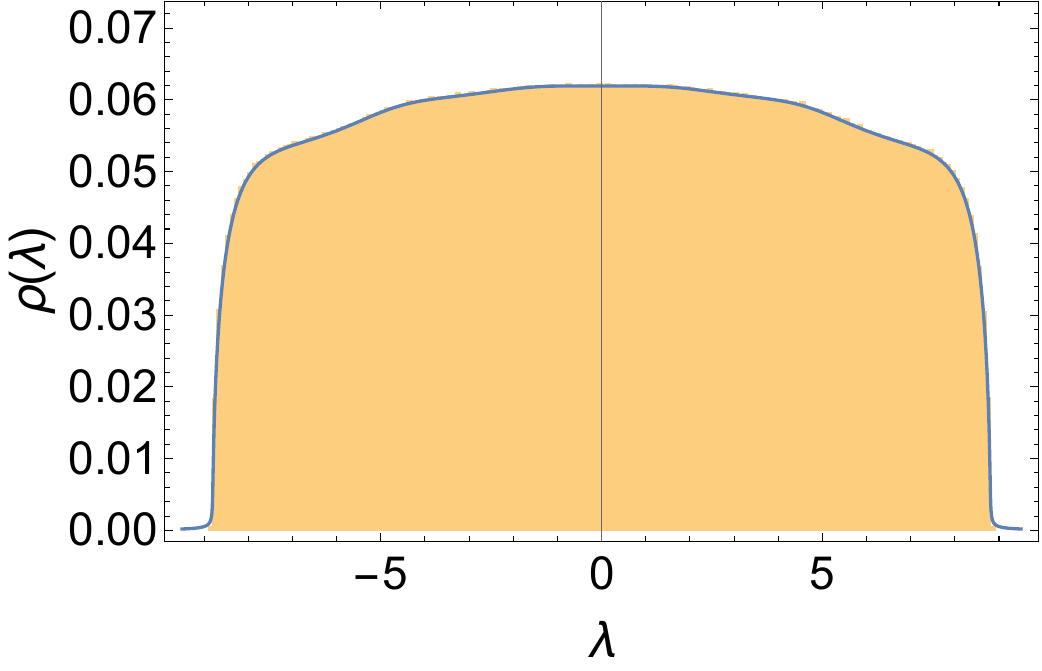} &  
        \includegraphics[width=0.45\textwidth]{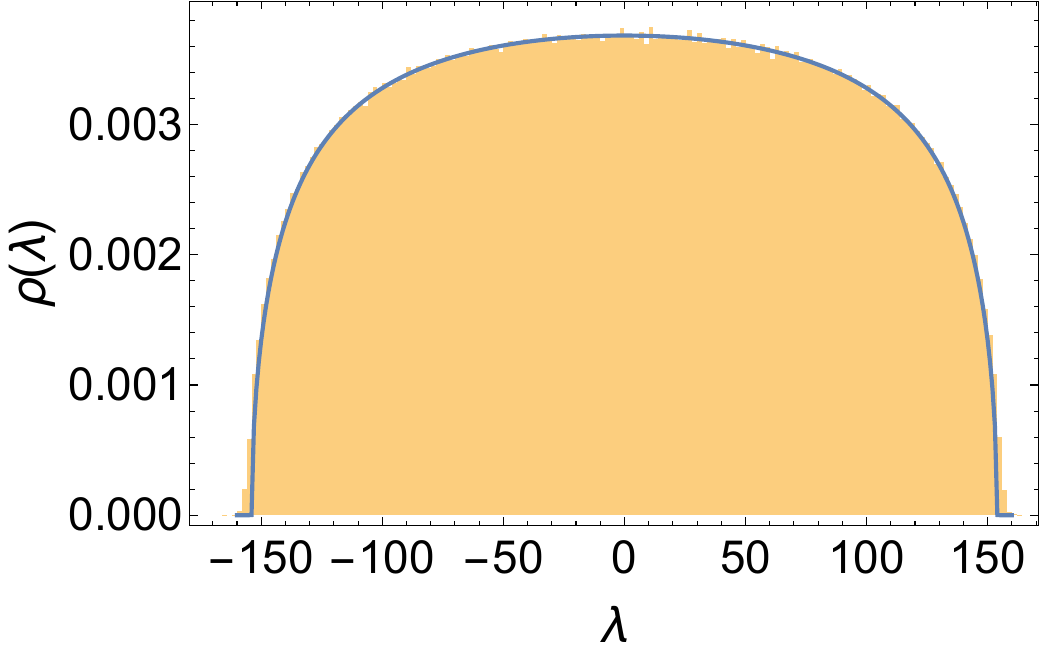} \\
        (c) spin-5/2 & (d) spin-100 \\
    \end{tabular}
   \caption{
    Density of eigenvalues for the operator sum $X^{(s)}_i + U^{\dagger} X^{(s)}_i U$, where $X^{(s)}$ are spin-$s$ operators and $U = e^{-i t M}$ represents the time evolution operator. Here we set $t = 50$. The matrix $M$ is drawn from a GOE with zero mean and variance $2/d_s$, where $d_s = (2s+1)^L$ is the Hilbert space dimension for a spin-$s$ chain with $L$ sites. Panels correspond to (a) spin-3/2 operators with $L = 5$, (b) spin-2 operators with $L = 4$, (c) spin-5/2 operators with $L = 4$, and (d) spin-100 operators with $L = 1$. The histograms in (a), (b), and (c) are based on 1000 independent realizations of $M$, while (d) is based on 5000 realizations. The operators are normalized such that their distributions read: $\rho_{s=3/2} = \frac{1}{4}(\delta_{-3} + \delta_{-1} + \delta_{1} + \delta_{3})$, $\rho_{s=2} = \frac{1}{5}(\delta_{-2} + \delta_{-1} + \delta_0 + \delta_{1} + \delta_{2})$, $\rho_{s=5/2} = \frac{1}{6}(\delta_{-5} + \delta_{-3} + \delta_{-1} + \delta_{1} + \delta_{3} + \delta_{5})$, and $\rho_{X^{(s)}}(\lambda) = \frac{1}{2s+1}$ for $\lambda \in [-s,s]$, and zero otherwise. The blue curves represent the free probability prediction obtained using the subordination function method.
}

    \label{fig:freeConvolutionRM2}
\end{figure}

\paragraph{Numerics versus Free probability Predictions}
Now, we apply the subordination function methods explained above to compute the free additive convolution for the sums of spin-$s$ operators for several values of $s$. We compare the free probability prediction with numerical results obtained by direct diagonalization of two operators of the form
\begin{equation} \label{eq-sumXs}
    X^{(s)}_i + U^\dagger X^{(s)}_i U,
\end{equation}
where the time-evolution operator is \( U = e^{-i t\,M} \), with \( M \) drawn from a Gaussian orthogonal ensemble (GOE), and \( X^{(s)}_i \) are generalized spin-\( s \) operators (see Section~\ref{sec-higherSpinModel} for a precise definition). We compare the numerically obtained spectra with the free additive convolution prediction discussed in Section~\ref{sec-numFreeConvolution}. For a chain with \( L \) sites, these operators have dimension \( d_s = (2s+1)^L \). The variance of \( M \) was fixed at \( 2/d_s \). Figure \ref{fig:FreeConvolutionRM} shows the spectrum of \( X^{(s)}_i + U^{\dagger} X^{(s)}_i U \) for spin-\( 1/2 \) operators (left panel) and spin-1 operators (right panel), along with the free convolution prediction for the corresponding operators (see eqs. \eqref{eq:arcsine} and \eqref{eq: sumdis_spin1}). Figure \ref{fig:freeConvolutionRM2} shows the spectrum of \( X^{(s)}_i + U^{\dagger} X^{(s)}_i U \) for spin-\( 3/2 \), spin-2, spin-\( 5/2 \) and spin-100 operators, along with the numerical free convolution prediction for the corresponding operators, obtained using the method described in Section \ref{sec-numFreeConvolution}. These results demonstrate that when the dynamics are driven by a random matrix, the spectrum of \( X^{(s)}_i + U^{\dagger} X^{(s)}_i U \) aligns with the predictions of free probability. This is expected in light of \eqref{eq-fundamentalResult} and the discussion presented in section \ref{sec:free_time_evol}, but has not yet been formally proved.

\subsection{Decorrelated ensembles and the role of eigenvectors}
\label{sec:decorrelatedEnsembles}

When considering any of the Hermitian Gaussian random matrix ensembles, there are two key aspects that these share: correlations between their eigenvalues, which result in the Wigner--Dyson level-spacing distribution and sine kernel universality arising from energy-density pair correlations, and delocalization of their eigenvectors. Eigenvectors of Gaussian RMTs are uniformly distributed over the corresponding orthogonal/unitary/ symplectic group according to the Haar measure and become isotropically delocalized with respect to the standard basis in $\mathbb{R}^{\mcD}$~\cite{Erdos_2009,Orourke:2016eig}. In contrast, the eigenvectors of the Poissonian ensemble are completely localized in the standard basis in $\mathbb{R}^{\mcD}$. These two aspects are intimately tied to each other and form part of the same underlying RMT behaviour \cite{Denton:2019pka}. It is thus difficult to imagine a situation in which we could have one of these properties without the other. However, one can still ask whether both aspects are necessary for triggering the asymptotic freeness between operators when one of them evolves in time via a Hamiltonian drawn from such Gaussian RMT ensembles. One way to test this, for example, is to artificially remove correlations from the eigenvalues while retaining the delocalized properties of the eigenvectors as well as the same average density of states, which in this case in the large-$\mcD$ limit is given by the Wigner semicircle.

To be precise, suppose that we draw one Hamiltonian $H$ from a Hermitian Gaussian RM ensemble (with fixed variance $\sigma^{2}_{\textrm{GUE}}=1/\mcD$, $\sigma^{2}_{\textrm{GOE}}=2/\mcD$, $\sigma^{2}_{\textrm{GSE}}=1/(2\mcD)$ and zero mean) and find its eigenvectors $\vert n\rangle$ and eigenvalues $E_{n}$. For simplicity, we can imagine that these eigenvalues are arranged in an ascending order with respect to $n$ together with their corresponding eigenvectors. The eigenvalues of $H$ will give rise to a discrete density of states $\rho (E)$ whose average $\overline{\rho(E)}$ will tend to the Wigner semicircle law $\rho_{W}(E)=\sqrt{4-E^{2}}/(2\pi)$ (for $\vert E\vert \leq 2$) in the thermodynamic limit $\mcD\rightarrow \infty$. The true eigenvalues $E_{n}$ of $H$ will show correlations in the way described above due to fluctuations around the average distribution $\overline{\rho(E)}$. However, if we now independently draw $\mcD$ variables $\lbrace E_{n}'\rbrace$ from $\overline{\rho(E)}$ (or equivalently from the Wigner semicircle distribution $\rho_{W}(E)$) then these will no longer show the above correlations, while still displaying the same average density distribution as the original correlated eigenvalues. Because of this we can call them ``decorrelated'' eigenvalues. In particular, the level-spacing distribution will be Poissonian. We can now construct a new Hamiltonian $H'$ from the original eigenvectors $\vert n \rangle$ and the new set of decorrelated eigenvalues $E'_{n}$ arranged ascendingly. Moreover, we can perform this procedure for as many Hamiltonians as required, since all Hamiltonians drawn from Gaussian RMTs are diagonalizable and have real eigenvalues. In the context of this work, we will refer to such a class of random matrix ensembles, ``decorrelated'' Gaussian ensembles.\footnote{This is same spirit as the ``Poissonian Hamiltonian ensembles'' described in~\cite{Balasubramanian:2022tpr}.} We now ask whether time evolution under such a class of Hamiltonians produces asymptotic freeness or not.

In Appendix~\ref{app-decorrensembles} we numerically show that at least in the case of decorrelated GOE matrices, and for operators of spins $1/2$ and $1$, the asymptotic freeness prediction arises for operator sums of the form $X_{i}(0)+X_{j}(t)$ for $i=j$ and $i\neq j$. Moreover, the operator statistics resulting from the time evolution generated by the decorrelated GOE matrices matches the statistics obtained from the true time evolution by the GOE matrices. This seems to suggest that, at least in the context of GOE matrices, time evolution under Hamiltonians with delocalized eigenvectors is sufficient to give rise to asymptotic freeness.

This analysis, along with the emergence of free convolution near integrability for certain operators, suggests that freeness can arise even without random matrix behavior, depending on the choice of operators. Since freeness reflects strong operator growth and scrambling -- in the sense of vanishing of OTOCs and their higher order counterparts -- this highlights a subtle distinction between operator growth and chaos defined via random matrix universality. While random matrix behavior ensures freeness for any operators (the F-property), a weaker, operator-dependent form of freeness may still emerge in its absence. This aligns with the findings of \cite{Dowling:2023hqc}, which show that scrambling is necessary but not sufficient for random matrix chaos. From the perspective of freeness, random matrix chaos implies universal freeness, but freeness itself can appear in a more limited, operator-specific way when conserved quantities obstruct full random matrix universality.

\section{Asymptotic Freeness in the Mixed-Field Ising Model}\label{Mixed_Ising}

As the first example of studying the emergence of asymptotic freeness in realistic many-body quantum systems, we consider the one-dimensional mixed-field Ising model with open boundary conditions:
\begin{equation} \label{eq:Hspin1/2}
    H_0=-\sum_{i=1}^{L-1} Z_i Z_{i+1}-\sum_{i=1}^{L}\left( h_x X_i+h_z Z_i \right)\,,
\end{equation}
where $X_i, Y_i,$ and $Z_i$ are Pauli matrices acting on the $i$th site. For example:
\begin{align}
    \label{spinops}
    X_{i}=\left(\mathds{1}_{2}\right)^{\otimes(i-1)}\otimes \sigma_{x}\otimes\left(\mathds{1}_{2}\right)^{\otimes(L-i)}~,
\end{align}
with similar formulas for $Y_i$ and $Z_i$. For a fixed number of sites $L$, the dimension of the Hilbert space is $d=2^L$. 

The model (\ref{eq:Hspin1/2}) is integrable if either $h_x=0$ or $h_z=0$, and it is non-integrable if both parameters are finite and nonzero. In particular, for $(h_x,h_z)=(-1.05,0.5)$, this model displays nearest-neighbor level spacing statistics well-described by random matrix theory associated with chaotic dynamics. See for instance~\cite{Ba_uls_2011, Craps:2019rbj}.

For any values of $(h_x,h_z)$, this model has parity symmetry -- the Hamiltonian (\ref{eq:Hspin1/2}) commutes with the parity operator
\begin{equation} \label{eq:parity}
    \hat{\Pi}=
    \begin{cases}
       \hat{P}_{1,L} \, \hat{P}_{2,L-1} \, \cdots \, \hat{P}_{\frac{L}{2},\frac{L+2}{2}} \,,\,\,\quad \text{for}\,\, L \,\, \text{even}\\
       \hat{P}_{1,L} \, \hat{P}_{2,L-1} \, \cdots \, \hat{P}_{\frac{L-1}{2},\frac{L+3}{2}} \,,\,\,\, \text{for}\,\, L \,\, \text{odd}
    \end{cases}
\end{equation}
where the permutation operator
\begin{equation}
    \hat{P}_{i,j}=\frac{1}{2}\left(\mathds{1}_{d}+X_iX_j+Y_i Y_j+Z_i Z_j \right)
\end{equation}
permutes the spin configuration of the sites $i^\text{th}$ and $j^\text{th}$. The action of the parity operator on a given state can be understood by imagining a mirror at one edge of the chain. The action of $\hat{\Pi}$ on a given spin configuration (state) returns the mirror image of that spin configuration.

\paragraph{Model with a random magnetic field.}
We also study a variant of the Ising model where parity symmetry is broken by introducing a random magnetic field:
\begin{equation} \label{eq-HamilDisorder}
    H_d=\sum_{i=1}^{L} \epsilon_i X_i\,,
\end{equation}
where $\epsilon_i$ is drawn from a Gaussian distribution with average zero and unit variance. In this case, the full Hamiltonian takes the form\footnote{A similar model featuring a random magnetic field was investigated in \cite{Znidaric:2008vkt}.}

\begin{equation} \label{eq-fullHising}
    H=H_0+g H_d\,.
\end{equation}
This modified model is particularly advantageous for spectral statistics analysis for two reasons. First, the absence of parity symmetry simplifies the level-spacing statistics analysis, as there is no need to decompose the Hamiltonian into block diagonal form. Second, by analyzing multiple realizations of the random magnetic field in Eq.~(\ref{eq-HamilDisorder}), it is possible to gather a large number of eigenvalues (on the order of half a million) without significantly increasing the system size. This approach allows for efficient computations while maintaining statistical reliability.

\subsection{Spectral statistics of the Hamiltonian}
\begin{figure}[h!]
    \centering
    \begin{tabular}{cc}
        \includegraphics[width=0.45\textwidth]{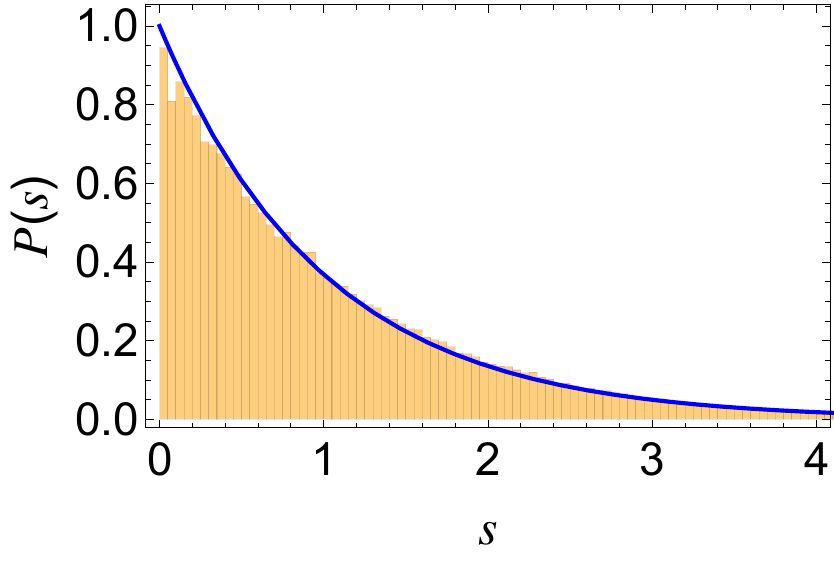} & 
        \includegraphics[width=0.45\textwidth]{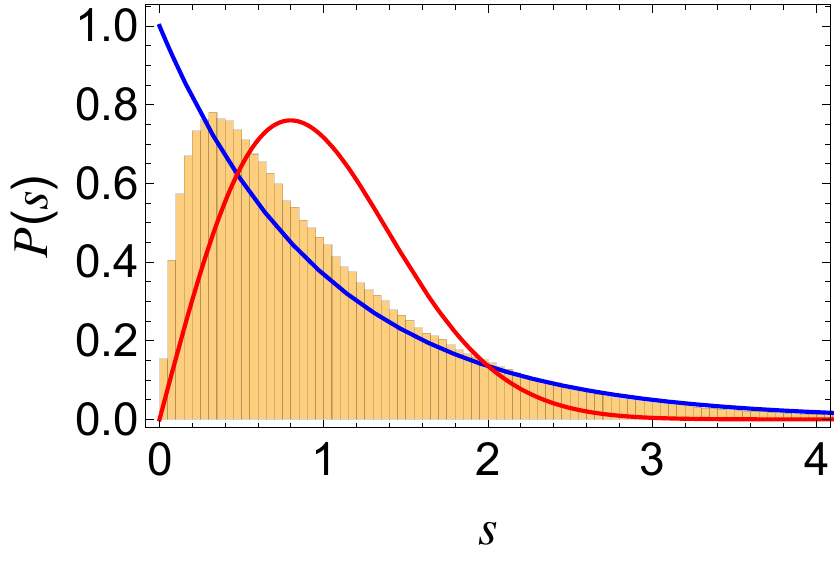} \\
        (a) & (b)  \\
        \multicolumn{2}{c}{\centering\includegraphics[width=0.45\textwidth]{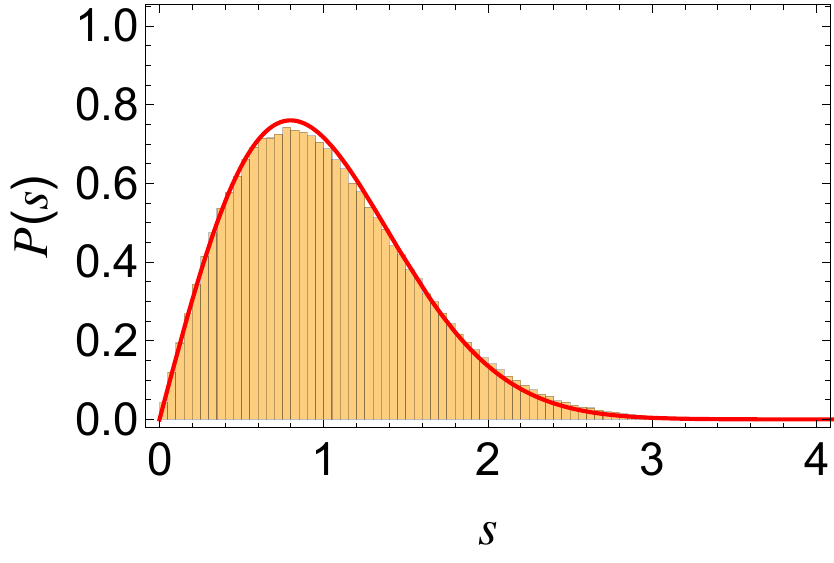}}  \\
       \multicolumn{2}{c}{(c)} \\
    \end{tabular}
     \caption{
Spectral statistics for 1000 realizations of the Hamiltonian (\ref{eq-fullHising}) with \(L = 10\) and \(g = 0.2\). Panels correspond to: (a) the near-integrable regime at \((h_x, h_z) = (-1, 0)\),  (b) the intermediate regime at \((h_x, h_z) = (-1.05, 0.03)\), and (c) the chaotic regime at \((h_x, h_z) = (-1.05, 0.5)\). The values of the corresponding average $r$-parameter are $0.38$, $0.45$, and $0.53$, respectively. The blue line represents the Poisson distribution, while the red line corresponds to the Wigner-Dyson distribution for \(\nu = 1\) (GOE). The analysis focuses on the central 50\% of the spectrum, excluding the outermost 25\% on each edge. }
\label{fig:LevelSpacingIsingH}
\end{figure}

The nearest-neighbor level spacing statistics of the Hamiltonian (\ref{eq-fullHising}) is shown in Fig.~\ref{fig:LevelSpacingIsingH}. After fixing $g=0.2$, the level spacing statistics is well described by a Poisson distribution for $(h_x,h_z)=(-1,0)$, which transitions to a GOE-type Wigner-Dyson distribution for $(h_x,h_z)=(-1.05,0.5)$.

\begin{figure}[h!]
    \centering
    \begin{tabular}{ccc}
        \includegraphics[width=0.45\textwidth]{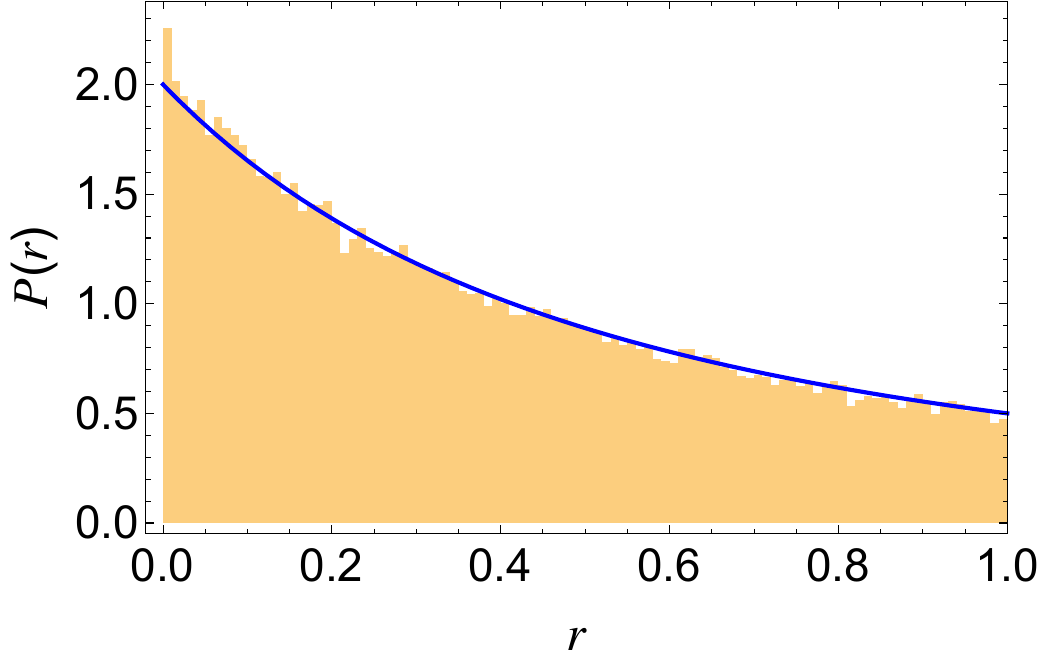} &
        \includegraphics[width=0.45\textwidth]{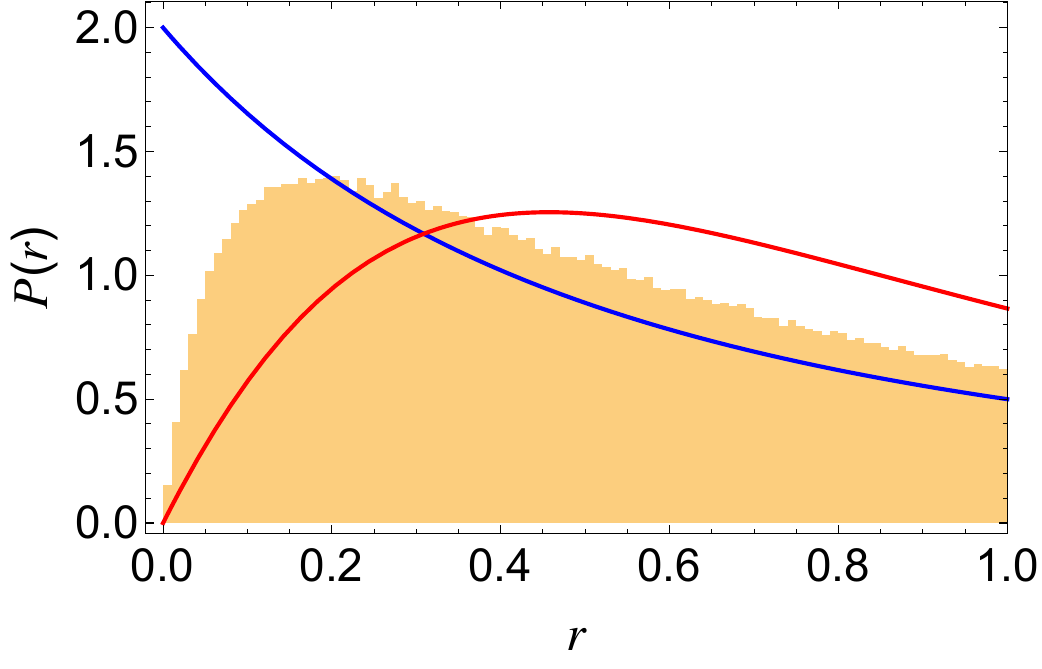} \\
        (a) & (b) \\
        \multicolumn{2}{c}{\centering\includegraphics[width=0.45\textwidth]{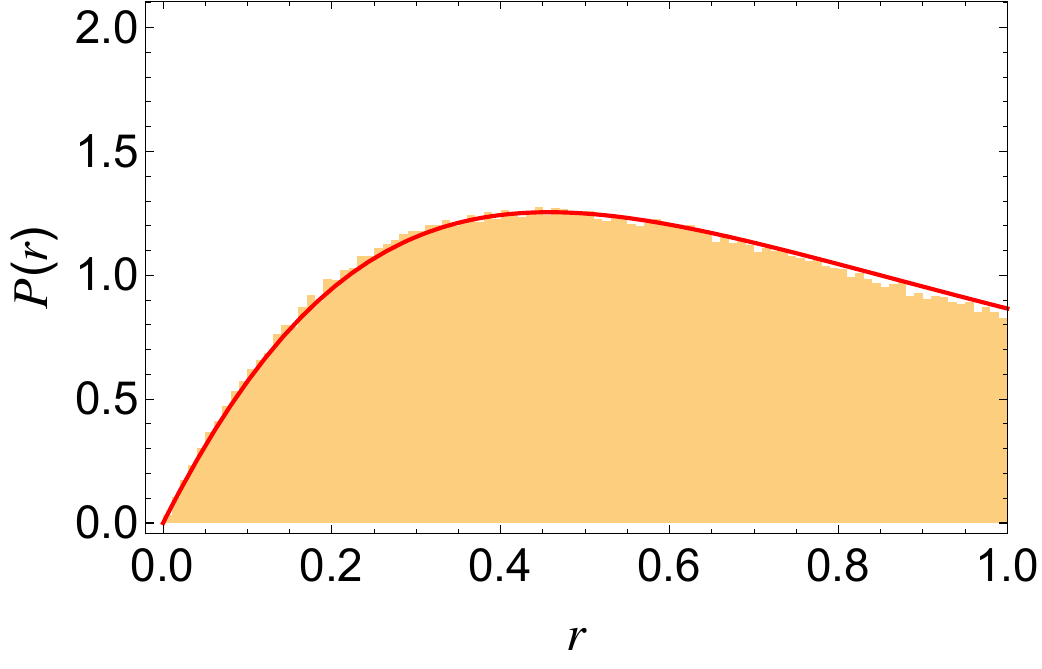}} \\
         \multicolumn{2}{c}{(c)} \\
    \end{tabular}
     \caption{
$r$-parameter statistics for 1000 realizations of the Hamiltonian (\ref{eq-fullHising}) with \(L = 10\) and \(g = 0.2\). Panels correspond to: (a) the near-integrable regime at \((h_x, h_z) = (-1, 0)\), (b) the intermediate regime at \((h_x, h_z) = (-1.05, 0.03)\), and (c) the chaotic regime at \((h_x, h_z) = (-1.05, 0.5)\). The values of the corresponding average $r$-parameter are $0.38$, $0.45$, and $0.53$, respectively. The blue line represents the Poisson $r$-parameter distribution, while the red line corresponds to the Wigner-Dyson $r$-parameter distribution for \(\nu = 1\) (GOE). The analysis focuses on the central 50\% of the spectrum, excluding the outermost 25\% on each edge.}
\label{fig:rparamaterStatisticsIsingH}
\end{figure}

\begin{figure}
    \centering
    \includegraphics[width=0.5\linewidth]{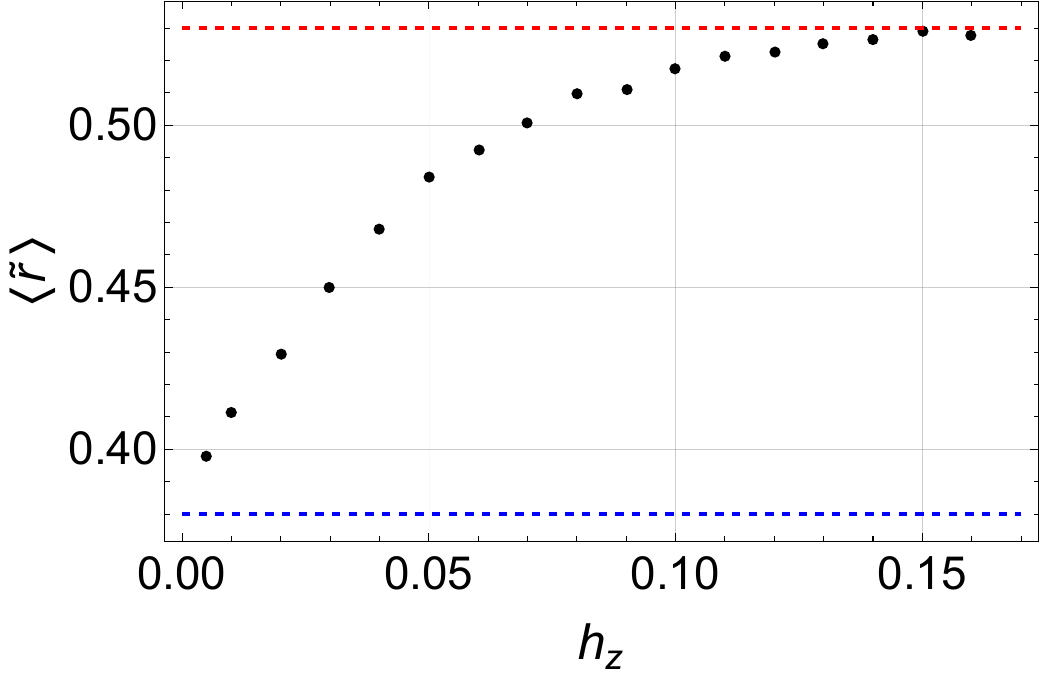}
    \caption{Average $r$-parameter versus $h_z$ for 1000 realizations of the Hamiltonian (\ref{eq-fullHising}) with $h_x = -1$, $g = 0.2$, and $L = 10$. The dashed blue and red lines indicate the average $r$-parameter for spectra whose level spacing distributions follow Poisson and GOE-type Wigner-Dyson statistics, respectively, with values of approximately 0.38 and 0.53. The analysis focuses on the central 50\% of the spectrum, excluding the outermost 25\% on each edge.}
    \label{fig:RPARAMETERvsHZ}
\end{figure}

In spin chains, such as the Hamiltonian (\ref{eq-fullHising}), the density of states typically follows a Gaussian distribution, which simplifies the unfolding of the spectrum. However, for more general Hamiltonians, like the mass-deformed SYK model, the density of states transitions from a Wigner semicircle distribution to a deformed Gaussian as one moves from the chaotic to the integrable regime. This transition complicates both the determination of the average density of states and the unfolding procedure. To bypass these complications, it is convenient to instead focus on the 
$r$-parameter statistics, which do not require any unfolding. A brief review of the 
$r$-parameter statistics is provided in Appendix \ref{app:r-parameter}. Figure \ref{fig:rparamaterStatisticsIsingH} displays the $r$-parameter statistics for the Hamiltonian (\ref{eq-fullHising}) at the chaotic, intermediate, and near-integrable points. The degree of chaoticity, in terms of random matrix behavior, can be evaluated using the average $r$-parameter. This parameter takes a value of 0.38 for uncorrelated spectra, typical of integrable systems, and 0.53 for spectra that exhibit GOE-type random matrix statistics. In Figure \ref{fig:RPARAMETERvsHZ}, we show how the average $r$-parameter evolves from 0.38 to 0.53 as a function of the coupling $h_z$.

\begin{figure}[ht]
    \centering
    \begin{tabular}{cc}
        \includegraphics[width=0.45\textwidth]{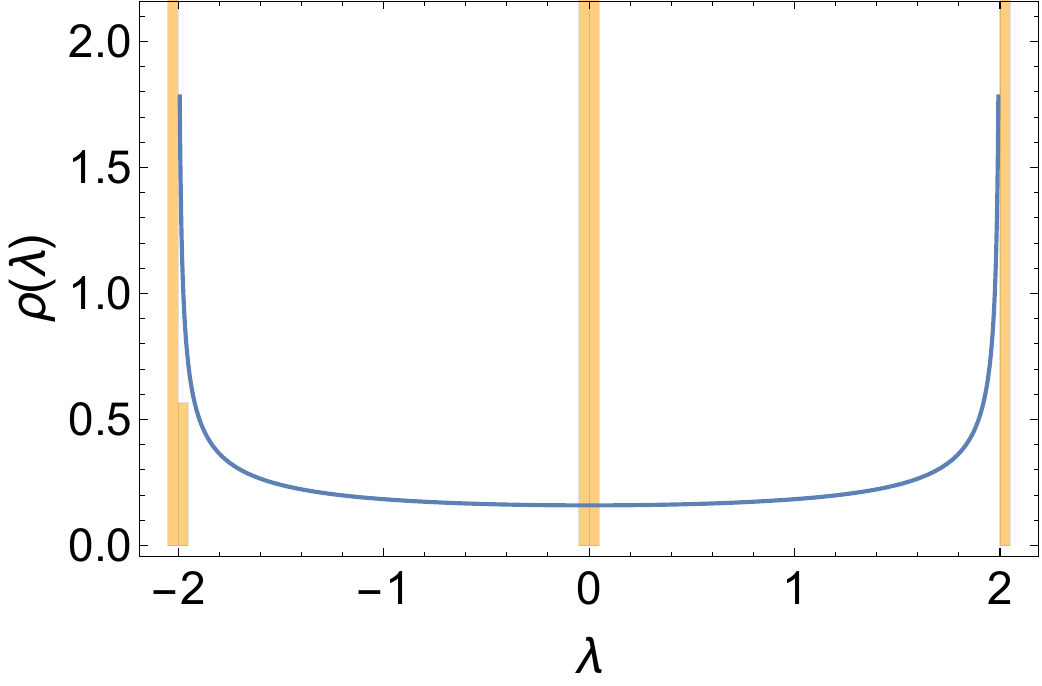} \put(-120,100){$t=1$} & \includegraphics[width=0.45\textwidth]{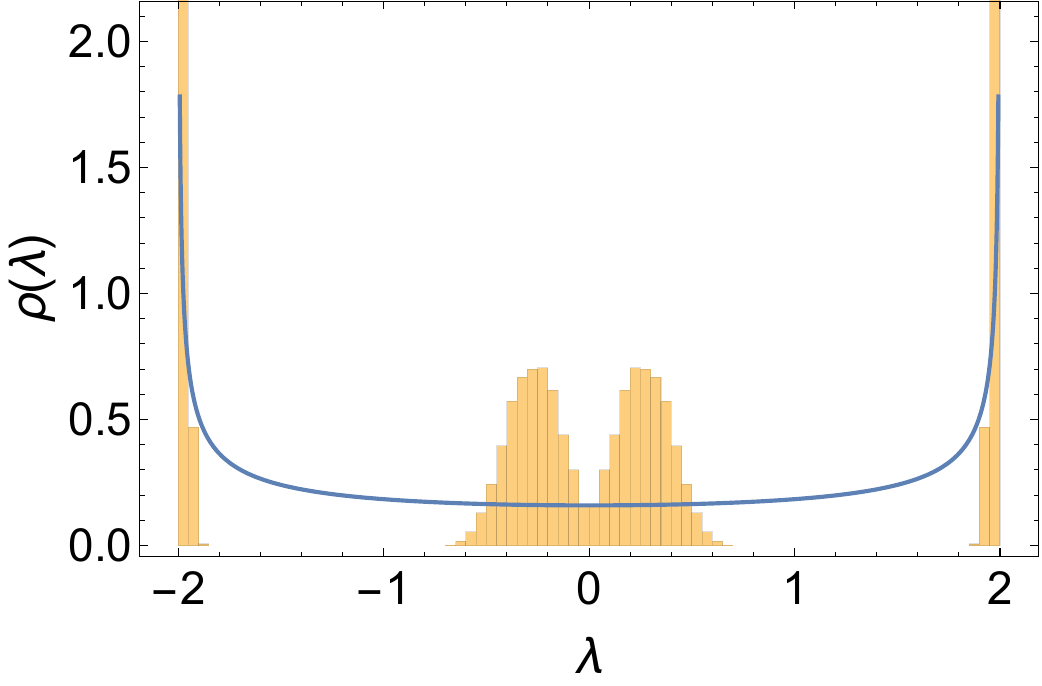}\put(-120,100){$t=5$} \\
        %(a) Caption for figure 1 & (b) Caption for figure 2 \\
        \includegraphics[width=0.45\textwidth]{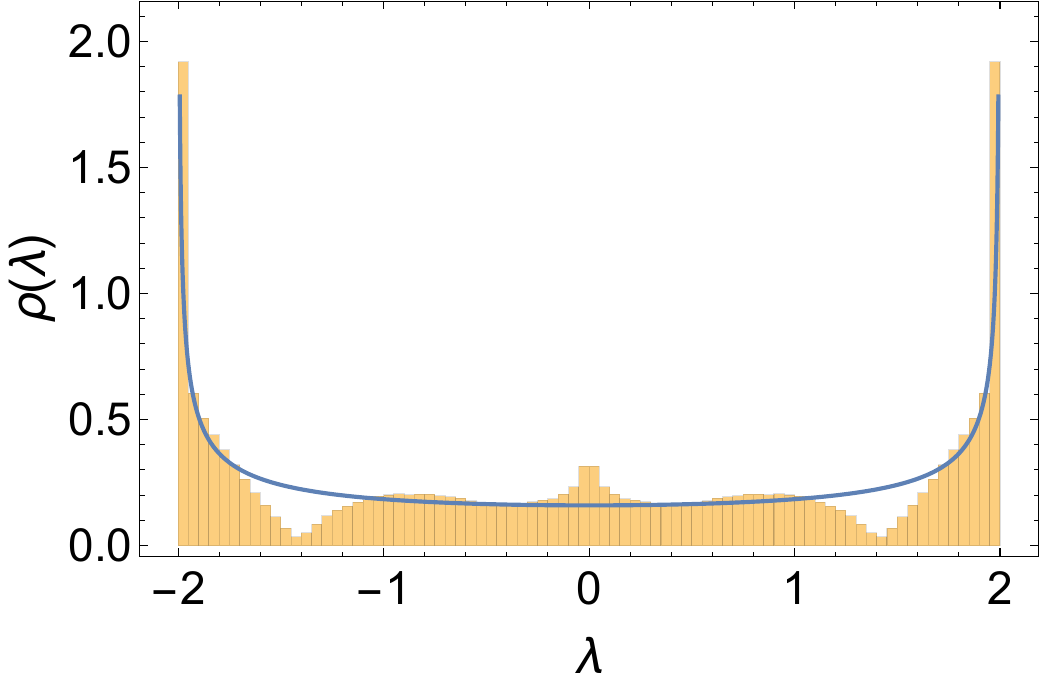}\put(-120,100){$t=7$} & \includegraphics[width=0.45\textwidth]{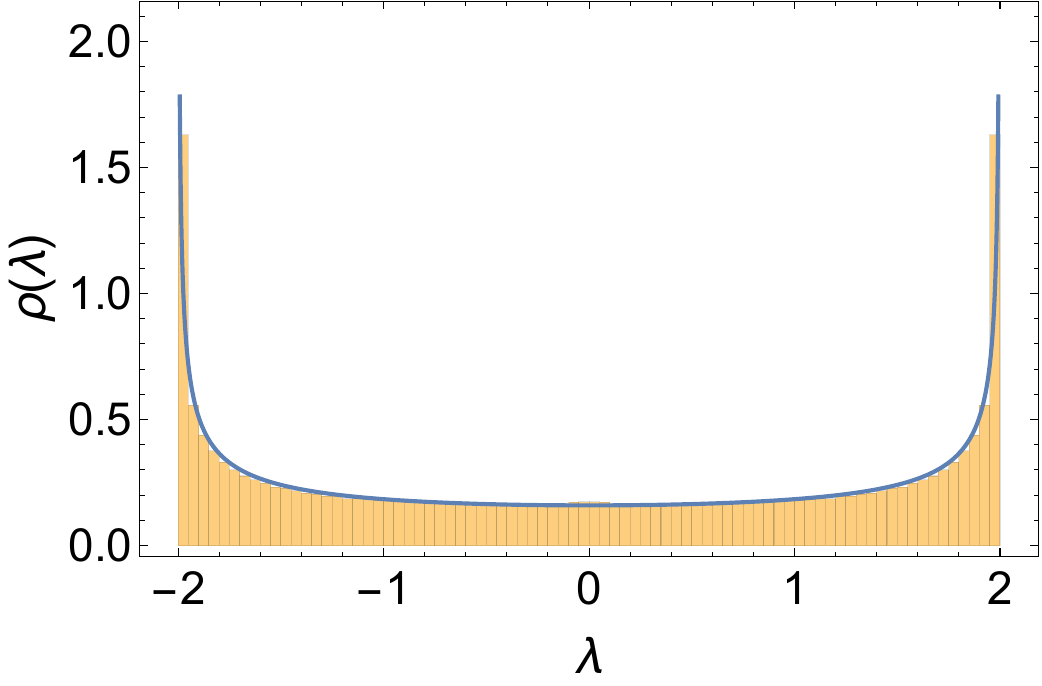}\put(-120,100){$t=20$} \\
        %(c) Caption for figure 3 & (d) Caption for figure 4 \\
    \end{tabular}
    \caption{Spectrum of eigenvalues the operator $Z_1(0)+Z_{10}(t)$ for increasing values of time at the chaotic point. Here we fix $(h_x,h_z,g)=(-1.05, 0.5, 0.2)$, and $L=10$ and consider 100 realizations of the Hamiltonian~(\ref{eq-fullHising}). The number of bins in each histogram is $100$. }
    \label{fig:emergenceArcSineLaw} 
\end{figure}

\subsection{Approximate emergence of asymptotic freeness}
\begin{figure}[ht]
    \centering
    \begin{tabular}{cc}
        \includegraphics[width=0.45\textwidth]{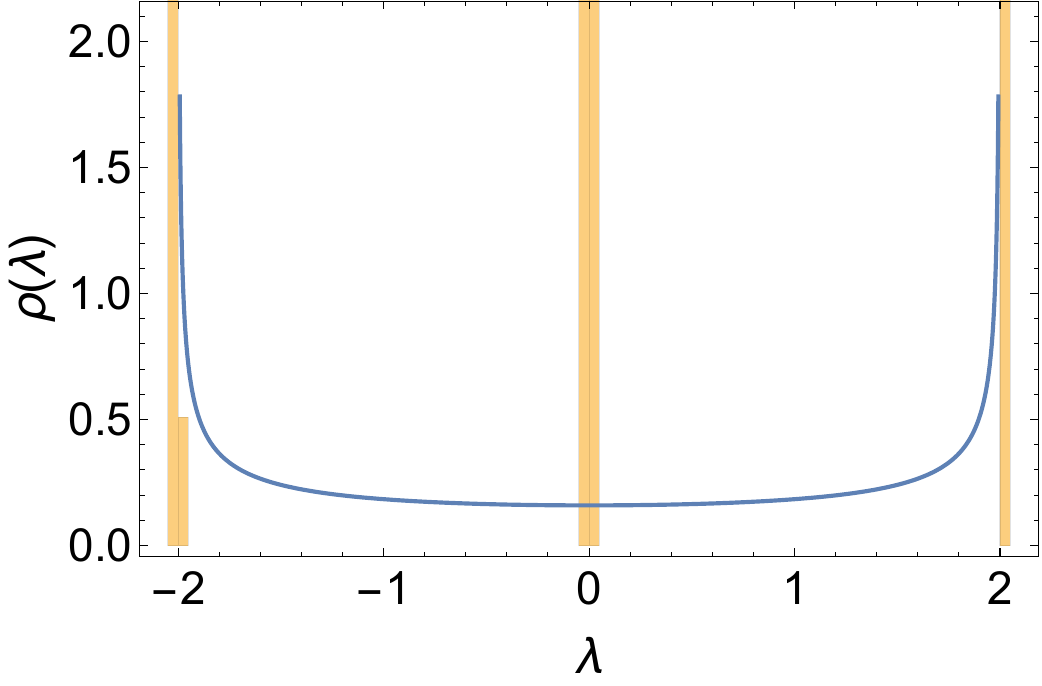} \put(-120,100){$t=1$} & \includegraphics[width=0.45\textwidth]{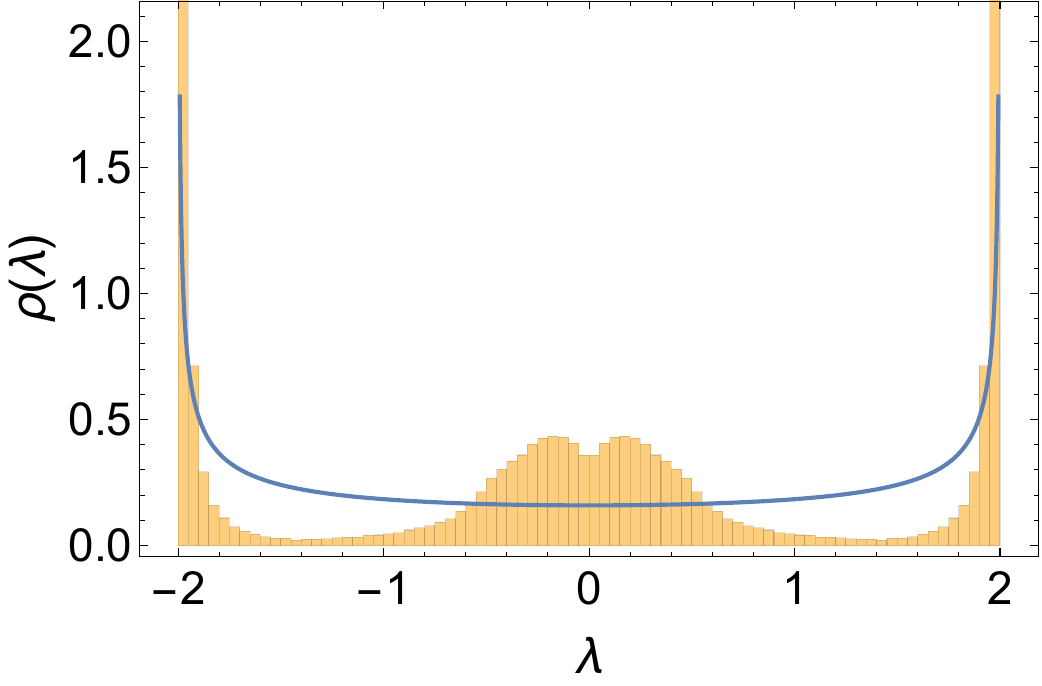}\put(-120,100){$t=1 \times 10^3$} \\
        %(a) Caption for figure 1 & (b) Caption for figure 2 \\
        \includegraphics[width=0.45\textwidth]{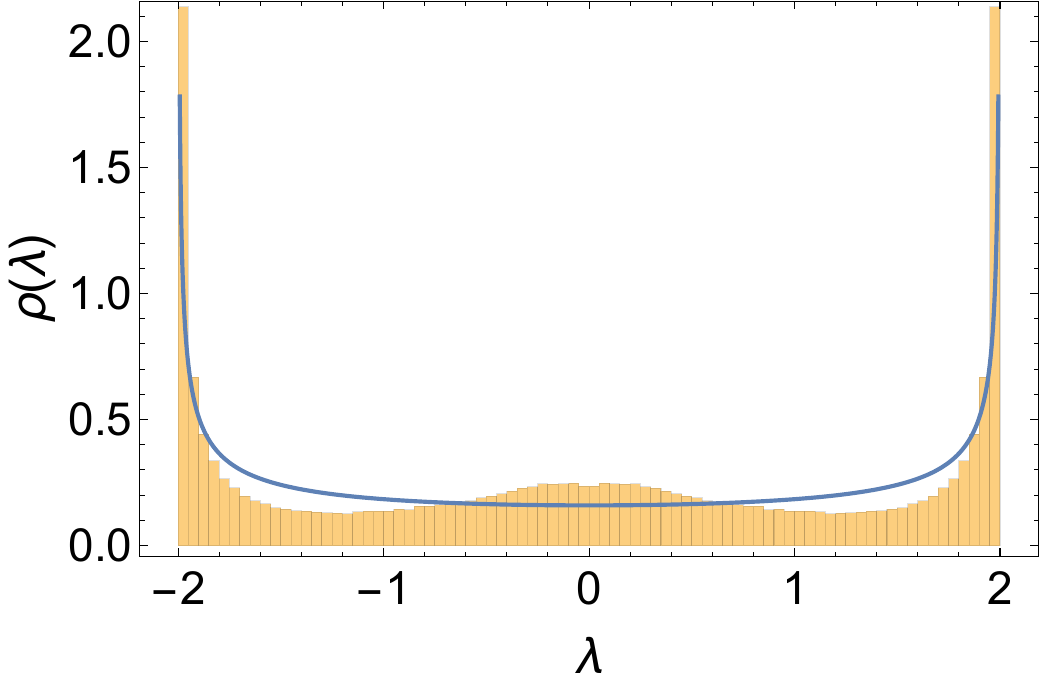}\put(-120,100){$t=5 \times 10^3$} & \includegraphics[width=0.45\textwidth]{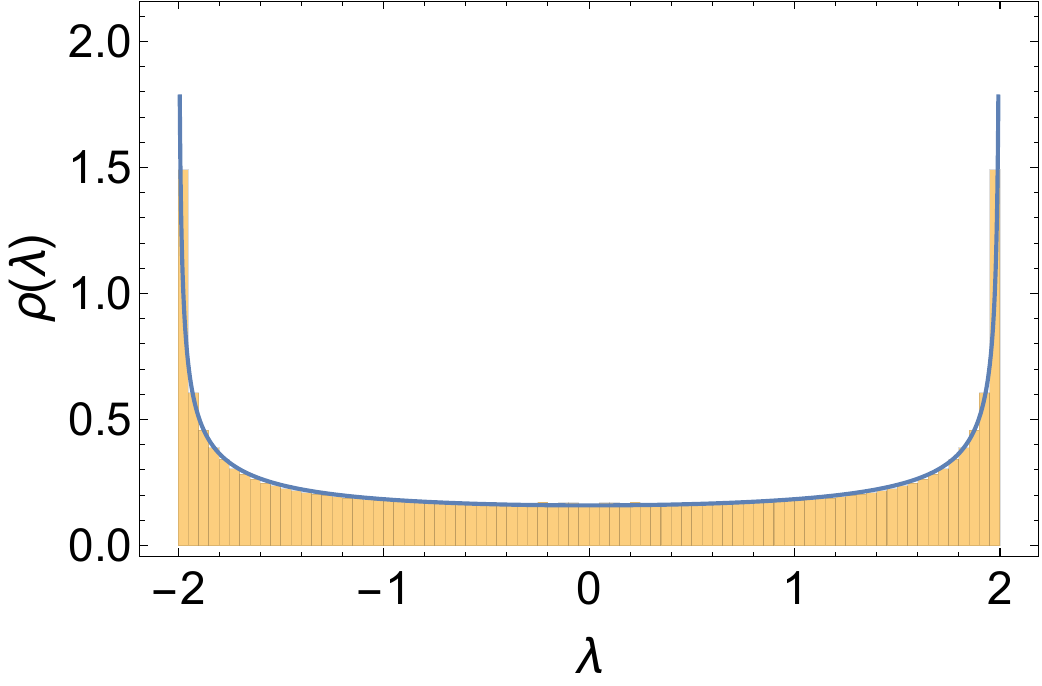}\put(-120,100){$t=24 \times 10^3$} \\
        %(c) Caption for figure 3 & (d) Caption for figure 4 \\
    \end{tabular}
    \caption{ Spectrum of eigenvalues of the operator \( Z_1(0) + Z_{10}(t) \) for increasing time at the near-integrable point. The parameters are set to \( (h_x, h_z, g) = (-1, 0.005, 0.2) \) and \( L = 10 \), with 100 realizations of the Hamiltonian~(\ref{eq-fullHising}). The number of bins in each histogram is $100$.}
\label{fig:emergenceArcSine2} 
\end{figure}

In this section, we examine the spectral statistics of operators of the form \( Z_i(0) + Z_j(t) \), where the generalized Pauli operator is time-evolved under the Hamiltonian~(\ref{eq-fullHising}). Both \( Z_i(0) \) and \( Z_j(t) \) have eigenvalues \( \pm 1 \), characterized by a Bernoulli distribution. When these operators are asymptotically free, the spectrum of their sum is expected to follow an arcsine distribution (see Example 1 in section \ref{free_addition}): 
\begin{equation} \label{eq:arcisinelaw}
f(\lambda) =
\begin{cases} 
\frac{1}{\pi  \sqrt{4-\lambda^2}}, & \text{if } -2 < \lambda < 2, \\
0, & \text{otherwise.}
\end{cases}
\end{equation}
Asymptotic freeness is anticipated in chaotic dynamics but not in integrable dynamics. Figure~\ref{fig:emergenceArcSineLaw} illustrates the spectrum of \( Z_1(0) + Z_{10}(t) \) at the chaotic point of the Hamiltonian~(\ref{eq-fullHising}), while Figure~\ref{fig:emergenceArcSine2} shows the same for a near-integrable point. The arcsine law manifests as long as the dynamics are non-integrable; however, the timescale at which it emerges grows rapidly as the system approaches the integrable limit. In fully integrable dynamics, the arcsine law does not emerge. 
In the following, we introduce a methodology to determine the time scale at which the arcsine law emerges.

\vspace{1cm}

\paragraph{Methodology to determine the arcsine time.}
To identify the time at which the arcsine law emerges, we employ a least-squares procedure. Specifically, we compute the following quantity, which quantifies the deviation of the eigenvalue distribution from the arcsine law: 
\begin{equation}
    \chi^2 = \sum_i \left( \frac{\rho(\lambda_i) - f(\lambda_i)}{f(\lambda_i)} \right)^2\,,
\end{equation}
where $\rho(\lambda_i; t)$ represents the density of eigenvalues obtained from the histogram of eigenvalues of $A(0) + B(t)$ at time $t$, and $f(\lambda_i)$ is defined in Eq.~(\ref{eq:arcisinelaw}). 

\begin{figure}[h!]
    \centering
    \begin{tabular}{cc}
        \includegraphics[width=0.45\textwidth]{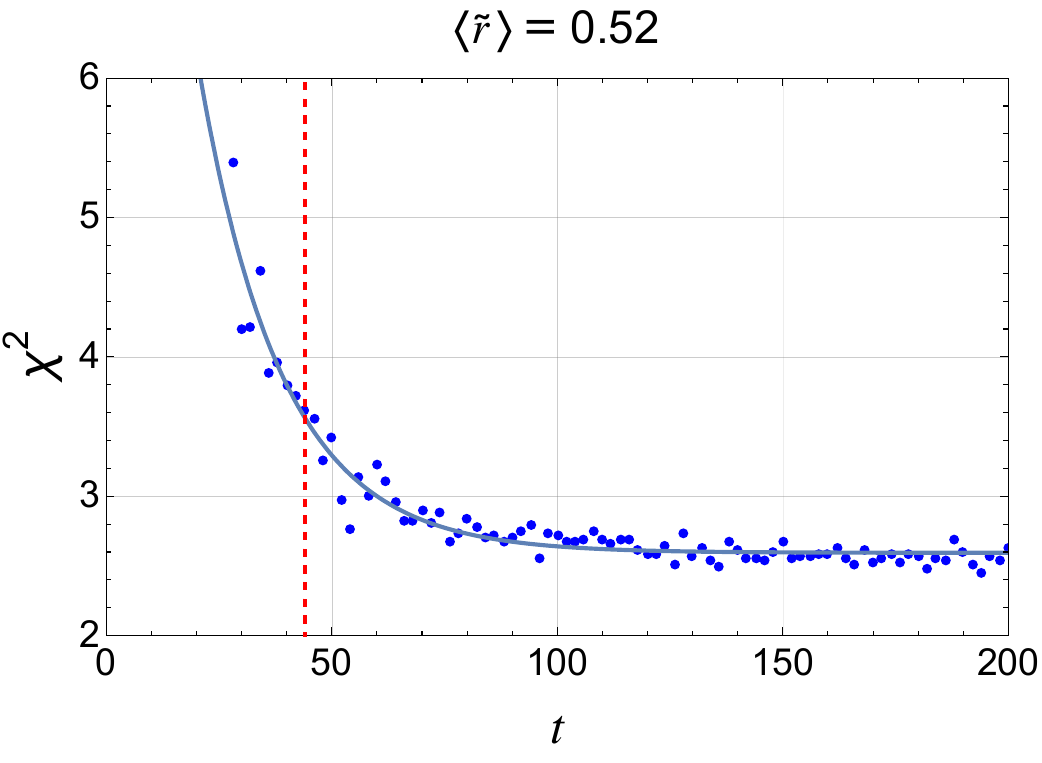} &
        \includegraphics[width=0.45\textwidth]{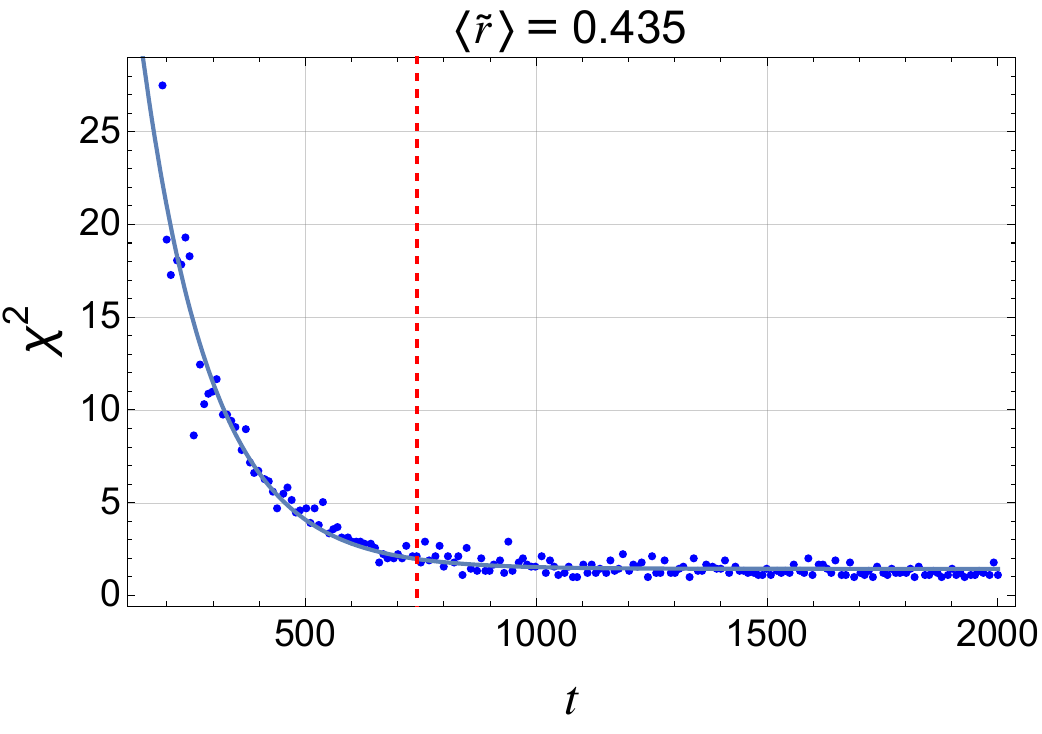} 
    \end{tabular}
     \caption{Time evolution of $\chi^2$ for the operator $Z_1(0)+Z_{10}(t)$ for two distinct average $r$-parameter values: $\langle \tilde{r} \rangle = 0.52$ (left panel) and $\langle \tilde{r} \rangle = 0.435$ (right panel). Both panels share fixed parameters $g = 2$, $L = 10$, and $h_x = -1$, while $h_z$ differs between the panels: $h_z = 0.1$ (left) and $h_z = 0.023$ (right).  The fitted function, $F(t) = A e^{-(t-t_0)/t_d} + A$, is shown as a continuous line, where $A$, $t_0$, and $t_d$ are fitting parameters. The arcsine times, estimated as $t_\text{arcsine} = t_0 + t_d$, are $t_\text{arcsine} = 44 \pm 1$ (left) and $t_\text{arcsine} = 742 \pm 17$ (right) and are marked by vertical dashed red lines. The analysis is based on a single realization of the Hamiltonian~(\ref{eq-fullHising}). 
}
\label{fig:chi2versustime}
\end{figure}

\begin{figure}[h!]
    \centering
    \begin{tabular}{cc}
    \includegraphics[width=0.45\linewidth]{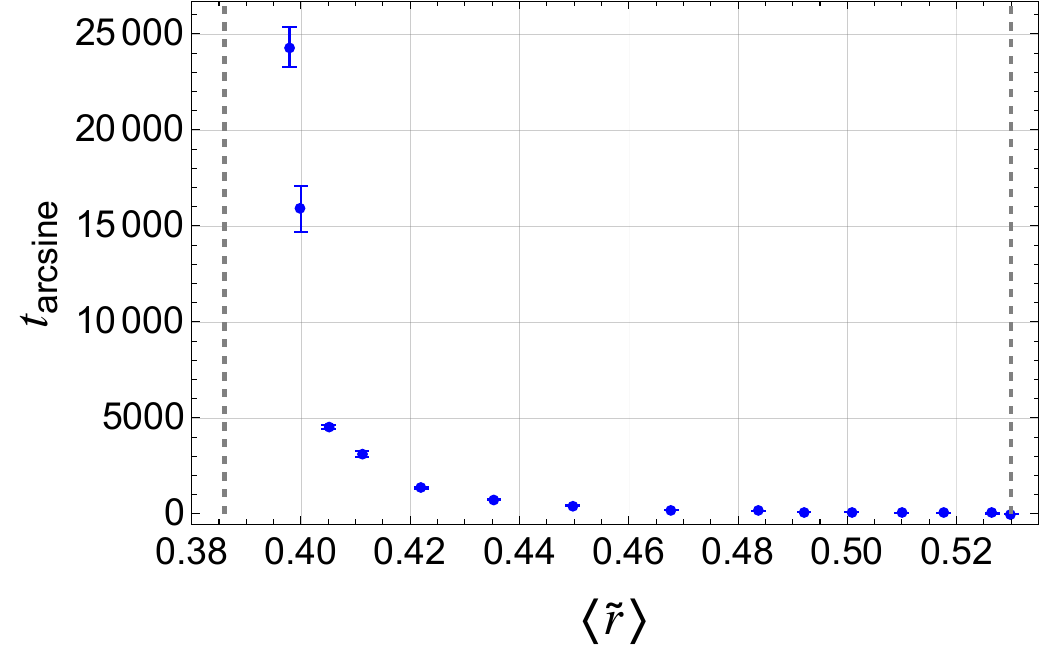}&
    \includegraphics[width=0.45\linewidth]{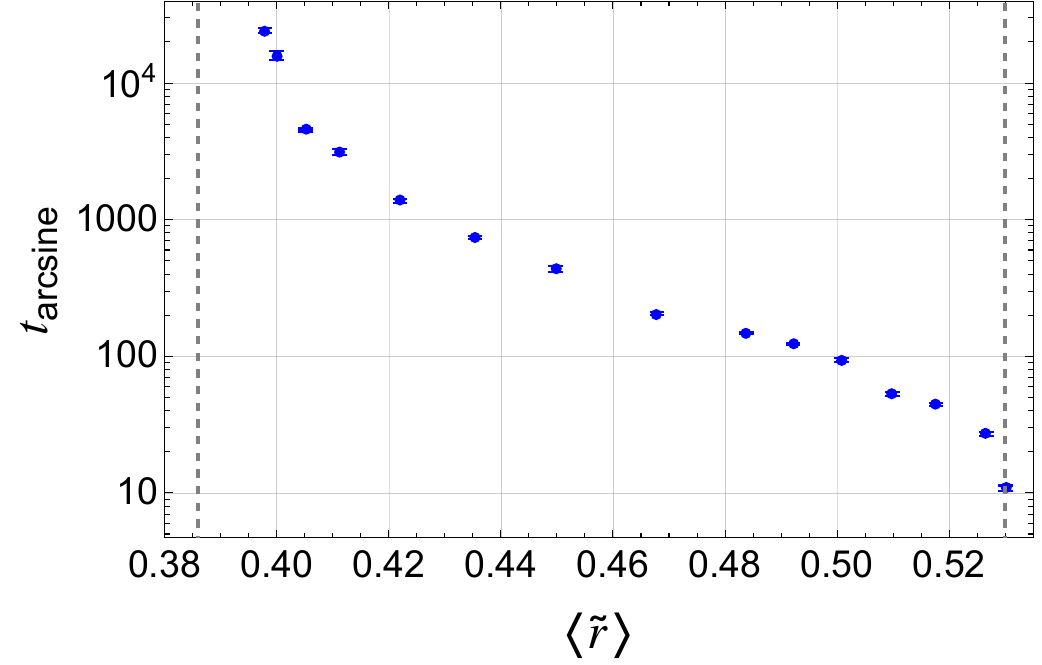}
    \end{tabular}
    \caption{Time scale for the emergence of the arcsine law for the operator $Z_1(0)+Z_{10}(t)$ as a function of the average $r$-parameter, $\langle \tilde{r} \rangle$.  The left/right panel shows the results in a linear/logarithmic scale. The parameters are fixed at $g = 0.2$ and $L = 10$, while $h_x$ and $h_z$ are varied to span $\langle \tilde{r} \rangle$ from 0.53 (indicating GOE statistics) to approximately 0.38 (indicating Poisson statistics). Vertical gray dashed lines mark these characteristic values of $\langle \tilde{r} \rangle$. The analysis is based on a single realization of the Hamiltonian~(\ref{eq-fullHising}).}
    \label{fig:arcsinetimeversusrparameter}
\end{figure}

The arcsine time scale is estimated as follows: at early times, $\rho(\lambda_i; t)$ deviates significantly from the arcsine distribution, resulting in large values of $\chi^2$. As the time parameter $t$ increases, $\chi^2$ decreases until it reaches an approximately constant value, around which it oscillates. To determine the time scale at which $\chi^2$ stabilizes, we fit a function of the form 
\begin{equation}
 F(t) = A \, e^{-(t - t_0) / t_d} + A\,,
\end{equation}
to the $\chi^2$ data points, starting from the time scale near saturation up to later times where the data oscillates around a constant value. From this fit, we extract the parameters $A$, $t_0$, and $t_d$, and define the arcsine time as $t_\text{arcsine} = t_0 + t_d$. It is important to note that this timescale does not exactly correspond to the point where $\chi^2$ becomes constant, since $F(t_\text{arcsine}) = A/e + A$. Instead, it marks the time scale at which the arcsine law emerges. See Figure~\ref{fig:chi2versustime}.
Alternatively, one could define the time scale $t^{(n)}_\text{arcsine} = t_0 + n \,t_d$, with $n = 2, 3, \dots$, at which $F(t_\text{arcsine}) = A/e^n + A$. This alternative definition yields qualitatively similar results. Figure \ref{fig:arcsinetimeversusrparameter} illustrates how the arcsine time depends on the average $r$-parameter. 

One important limitation of the above analysis is that it is based on a single realization of the Hamiltonian (\ref{eq-fullHising}). This choice was made to enable the study of the spectral properties of operators as a function of time and to perform a least-squares analysis within a reasonable computational time. Specifically, we consider a single realization of the Hamiltonian (\ref{eq-fullHising}) with \(L=10\), resulting in an analysis based on only 1024 eigenvalues, which introduces substantial statistical uncertainties. Nevertheless, the analysis effectively illustrates how the arcsine time depends on the degree of chaoticity of the system and diverges as one approaches integrability. In a later section, a more refined analysis, based on 1000 realizations of the Hamiltonian (\ref{eq-fullHising}), reveals small deviations from the arcsine law, particularly for the \(X\)-Pauli operators.

\subsection{Fluctuations on top of the arcsine law} \label{sec-fluctuationsSpinHalf}

In random matrix theory, the average density of states follows the Wigner semicircle law, with fluctuations exhibiting universal properties described by the Wigner-Dyson distribution of the appropriate universality class. Since the arcsine law governs the average density of states for operators of the form $A(0) + B(t)$ in non-integrable systems, it is natural to investigate whether the fluctuations on top of it also exhibit universal properties. A precise spectral statistics analysis requires a sufficiently large number of eigenvalues\footnote{We thank Antonio M. Garcia-Garcia for discussions on this point.}. To achieve this, the analysis below was performed using 1000 realizations of the Hamiltonian \eqref{eq-fullHising}, yielding a total of 1,024,000 eigenvalues. The level spacing and $r$-parameter statistics were computed using the central 50\% of the spectrum, resulting in approximately half a million eigenvalues.

\begin{figure}[h!]
    \centering
    \begin{tabular}{ccc}
        \includegraphics[width=0.45\textwidth]{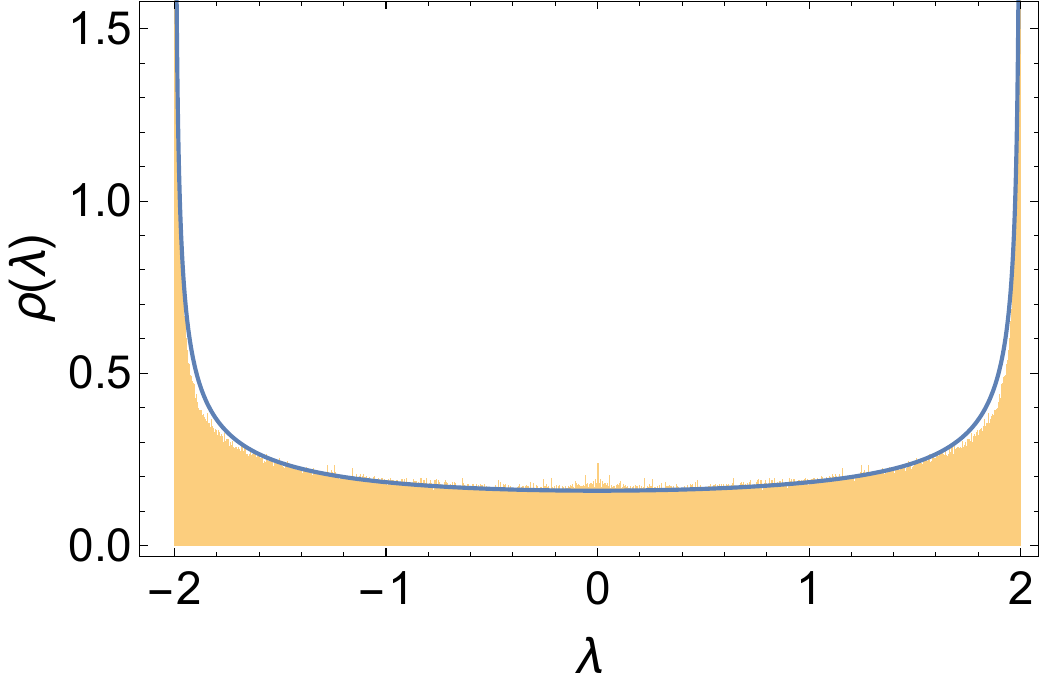} &
        \includegraphics[width=0.45\textwidth]{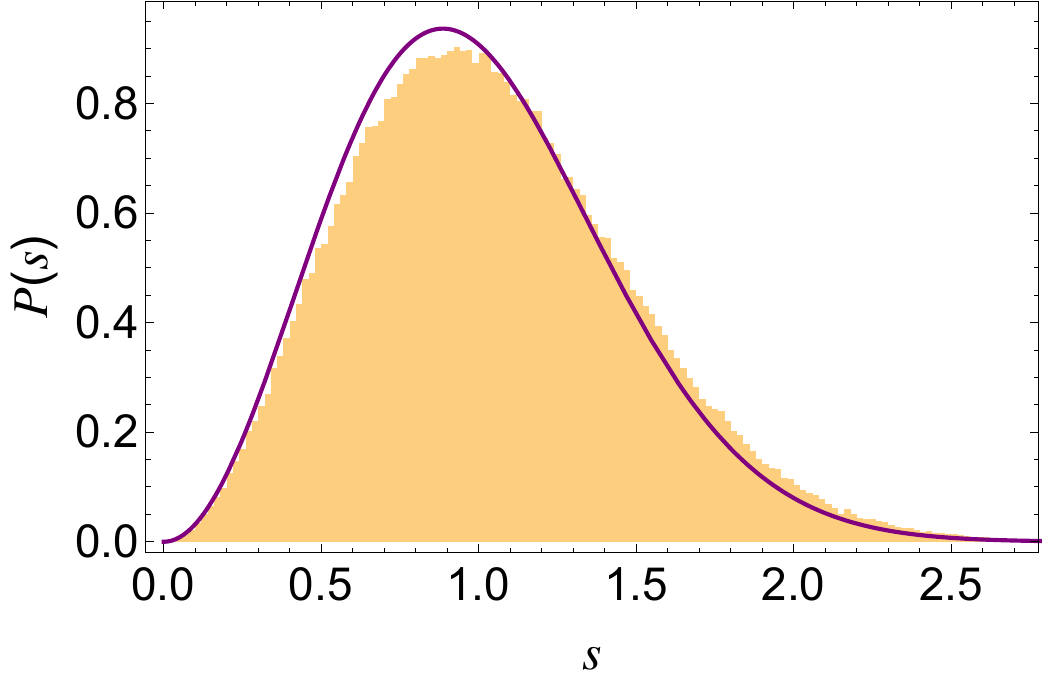} \\
         (a) Eigenvalue density & (b) Level spacing distribution  \\
        \multicolumn{2}{c}{\centering\includegraphics[width=0.45\textwidth]{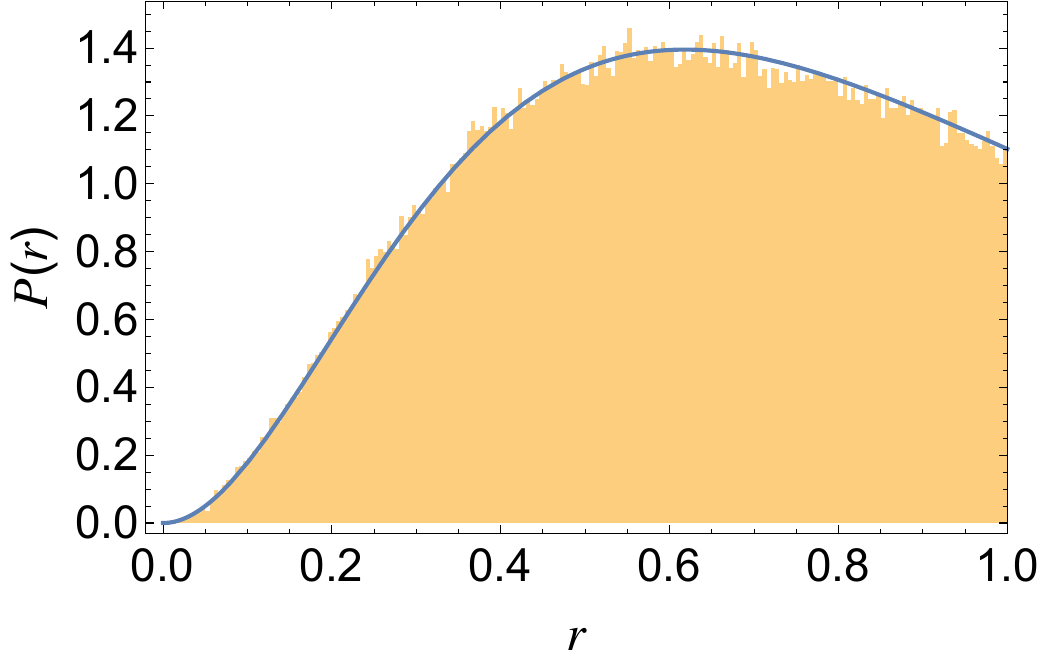}} \\
        \multicolumn{2}{c}{(c) $r$-parameter distribution}\\
    \end{tabular}
    \caption{
        Eigenvalue density, level spacing distribution, and $r$-parameter distribution for the operator $X_1 + X_{10}(t)$ based on 1000 realizations of the Hamiltonian \eqref{eq-fullHising} at the chaotic point $(h_x, h_z, g) = (-1.05, 0.5, 0.2)$, with $L=10$ and $t=50$. The $r$-parameter distribution was fitted using the form \eqref{eq: r-dis-N=3}, yielding the Dyson index $\nu=2.000\pm0.002$. The solid lines represent (a) the arcsine distribution, (b) the Wigner-Dyson distribution with $\nu=2$, and (c) the Wigner $r$-parameter distribution with $\nu=2$. The corresponding average $r$-parameter was found to be $\langle \tilde{r} \rangle = 0.60$. The analysis focuses on the central 50\% of the spectrum, excluding the outermost 25\% at each edge.
    }
    \label{fig:fluctuationsIsingChaotic}
\end{figure}

\begin{figure}[h!]
    \centering
    \begin{tabular}{ccc}
        \includegraphics[width=0.45\textwidth]{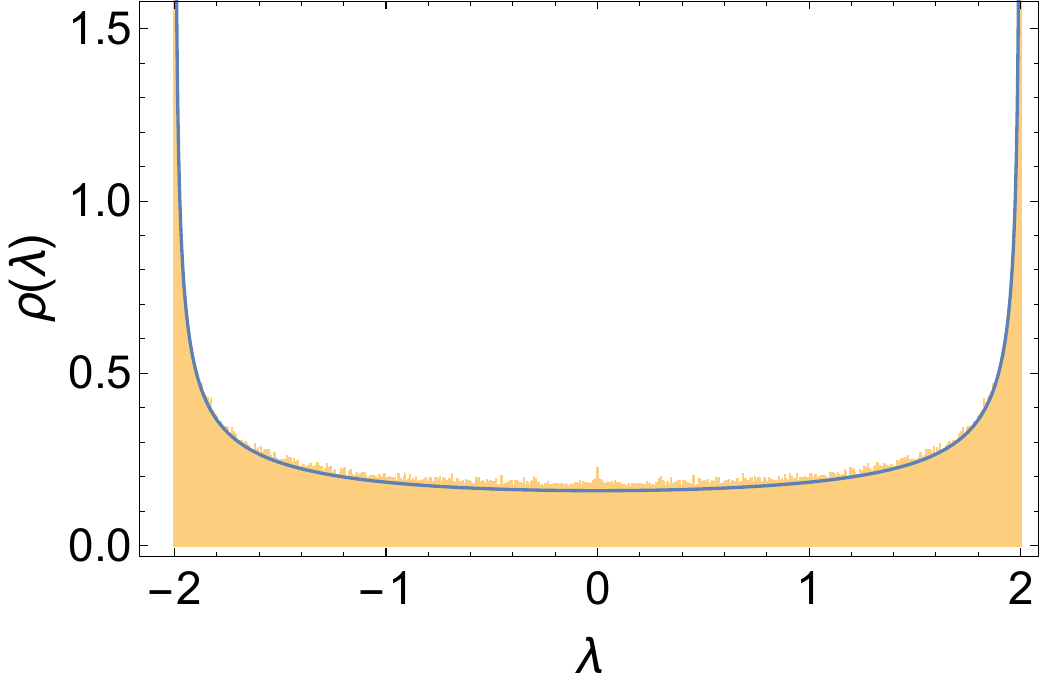} &
        \includegraphics[width=0.45\textwidth]{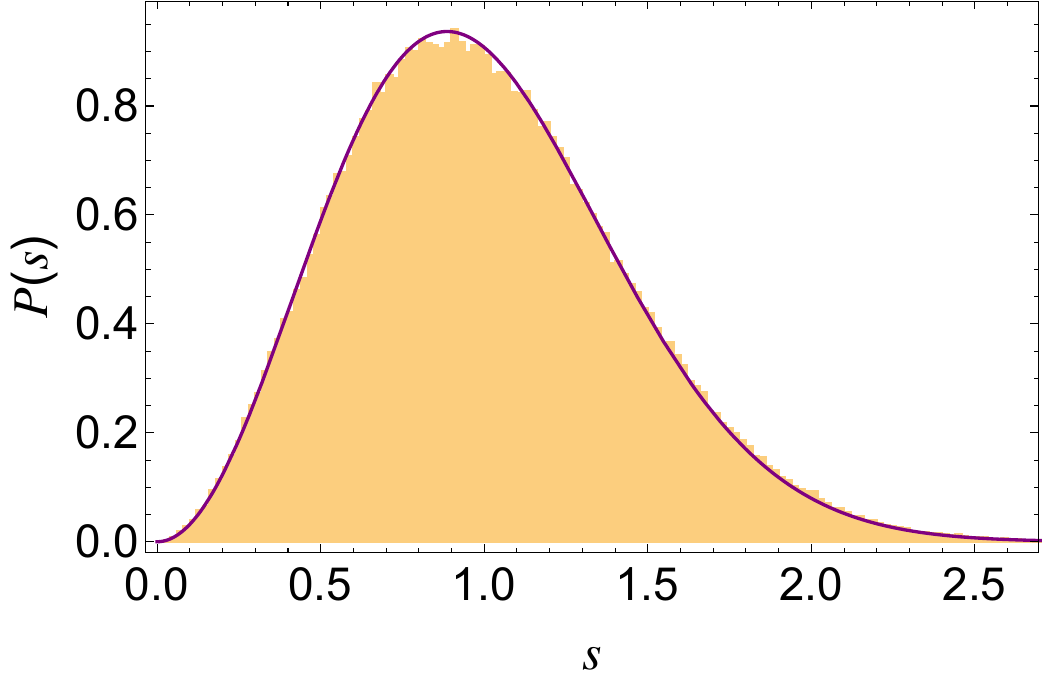} \\
         (a) Eigenvalue density & (b) Level spacing distribution \\
        \multicolumn{2}{c}{\centering \includegraphics[width=0.45\textwidth]{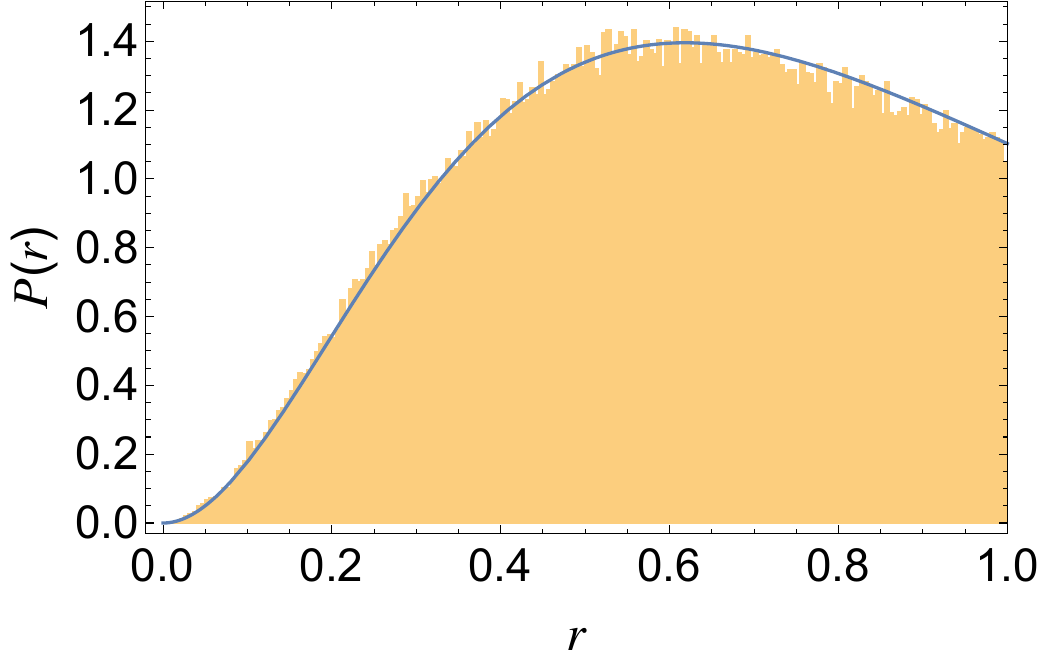}} \\
        \multicolumn{2}{c}{(c) $r$-parameter distribution} \\
    \end{tabular}
    \caption{
        Eigenvalue density, level spacing distribution, and $r$-parameter distribution for the operator $Z_1 + Z_{10}(t)$ based on 1000 realizations of the Hamiltonian \eqref{eq-fullHising} at the chaotic point $(h_x, h_z, g) = (-1.05, 0.5, 0.2)$, with $L=10$ and $t=50$. The $r$-parameter distribution was fitted using the form \eqref{eq: r-dis-N=3}, yielding the Dyson index $\nu=2.000\pm0.002$. The solid lines represent (a) the arcsine distribution, (b) the Wigner-Dyson distribution with $\nu=2$, and (c) the Wigner $r$-parameter distribution with $\nu=2$. The corresponding average $r$-parameter was found to be $\langle \tilde{r} \rangle = 0.60$. The analysis focuses on the central 50\% of the spectrum, excluding the outermost 25\% at each edge.
    }
    \label{fig:fluctuationsIsingChaoticC}
\end{figure}

\paragraph{Fluctuations at the chaotic point.}
Figure \ref{fig:fluctuationsIsingChaotic} presents the eigenvalue density, the level spacing distribution, and the $r$-parameter distribution for the operator $X_1 + X_{10}(t)$, based on 1000 realizations of the Hamiltonian \eqref{eq-fullHising} at the chaotic point $(h_x, h_z, g) = (-1.05, 0.5, 0.2)$, with $L=10$ and $t=50$. In panel (a), the improved statistics reveal deviations from the arcsine law, particularly near the edges of the spectrum, indicating that the arcsine law emerges only in an approximate sense. In panel (b), the level spacing statistics of the operator closely resemble the Wigner-Dyson distribution with Dyson index $\nu=2$, corresponding to the GUE of random matrices. However, some deviations are observed, possibly due to imperfect unfolding of the spectrum, which was performed using the arcsine law as the average density of eigenvalues. Panel (c) shows that the $r$-parameter distribution is well described by the Wigner $r$-parameter distribution with $\nu=2$, yielding an average value of $\langle \tilde{r} \rangle = 0.60$.
Figure \ref{fig:fluctuationsIsingChaoticC} presents similar results for the operator \(Z_1 + Z_{10}(t)\), with the key difference that the arcsine law and the Wigner-Dyson level spacing distribution fit the data very well in this case.

\begin{figure}[h!]
    \centering
    \begin{tabular}{ccc}
        \includegraphics[width=0.45\textwidth]{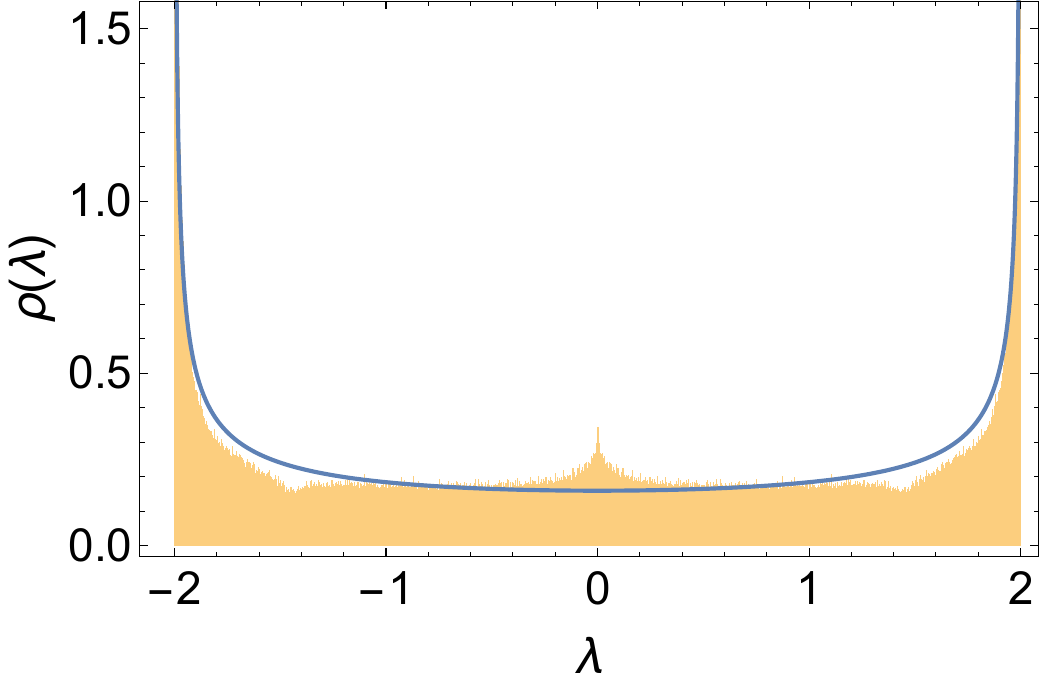} &
        \includegraphics[width=0.45\textwidth]{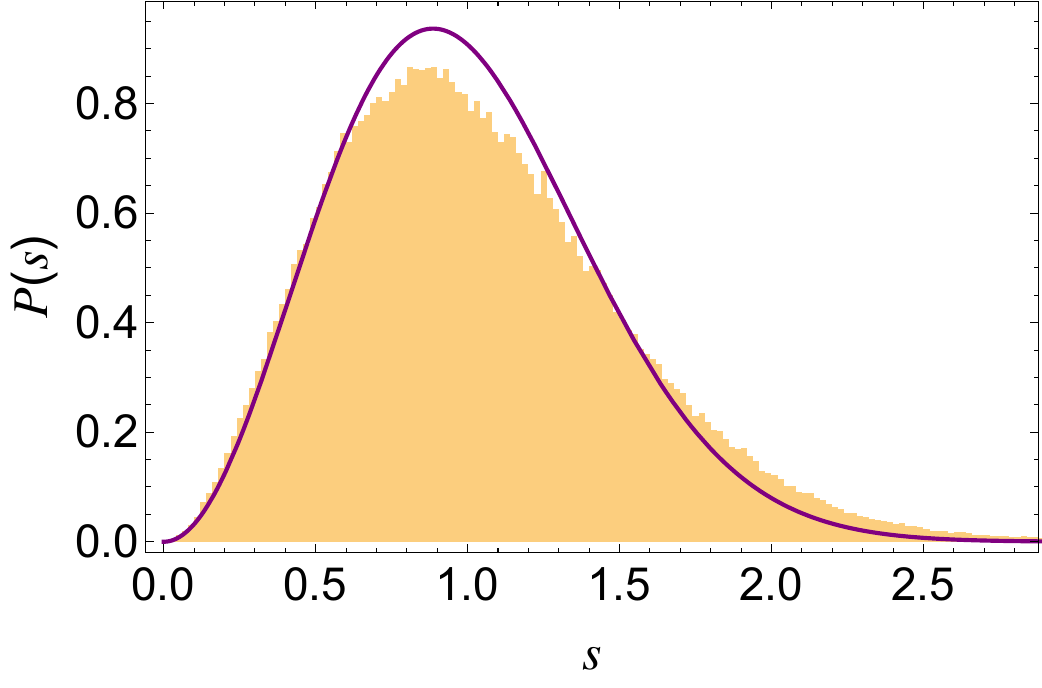} \\
         (a) Eigenvalue density & (b) Level spacing distribution  \\
       \multicolumn{2}{c}{\centering \includegraphics[width=0.45\textwidth]{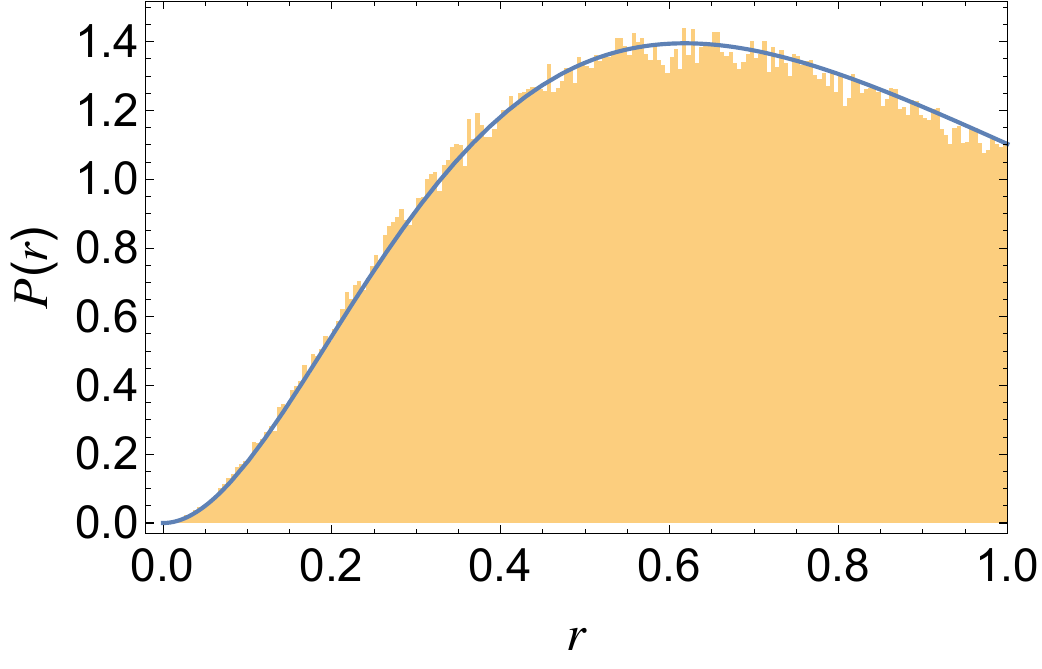}} \\
         \multicolumn{2}{c}{(c) $r$-parameter distribution} \\
    \end{tabular}
    \caption{
        Eigenvalue density, level spacing distribution, and $r$-parameter distribution for the operator $X_1 + X_{10}(t)$ based on 1000 realizations of the Hamiltonian \eqref{eq-fullHising} at the near-integrable point $(h_x, h_z, g) = (-1, 0.005, 0.2)$, with $L=10$ and $t=5 \times 10^4$. The $r$-parameter distribution was fitted using the form \eqref{eq: r-dis-N=3}, yielding the Dyson index $\nu=2.001\pm0.002$. The solid lines represent (a) the arcsine distribution, (b) the Wigner-Dyson distribution with $\nu=2$, and (c) the Wigner $r$-parameter distribution with $\nu=2$. The corresponding average $r$-parameter was found to be $\langle \tilde{r} \rangle = 0.60$. The analysis focuses on the central 50\% of the spectrum, excluding the outermost 25\% at each edge.
    }
    \label{fig:fluctuationsIsingNearIntegrable}
\end{figure}

\begin{figure}[h!]
    \centering
    \begin{tabular}{ccc}
        \includegraphics[width=0.45\textwidth]{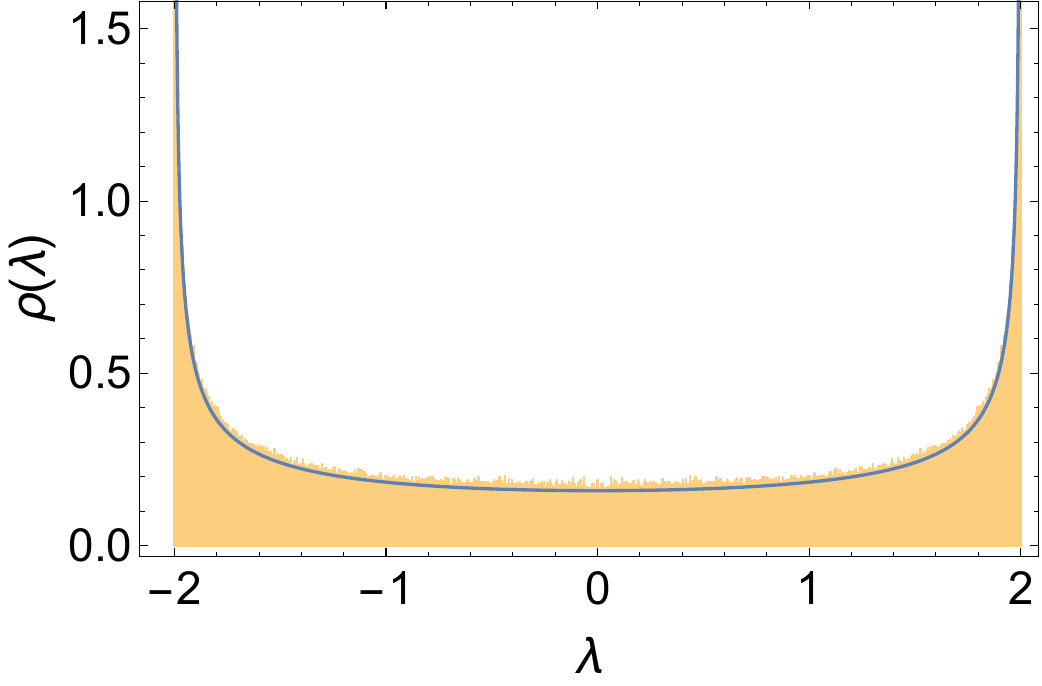} &
        \includegraphics[width=0.45\textwidth]{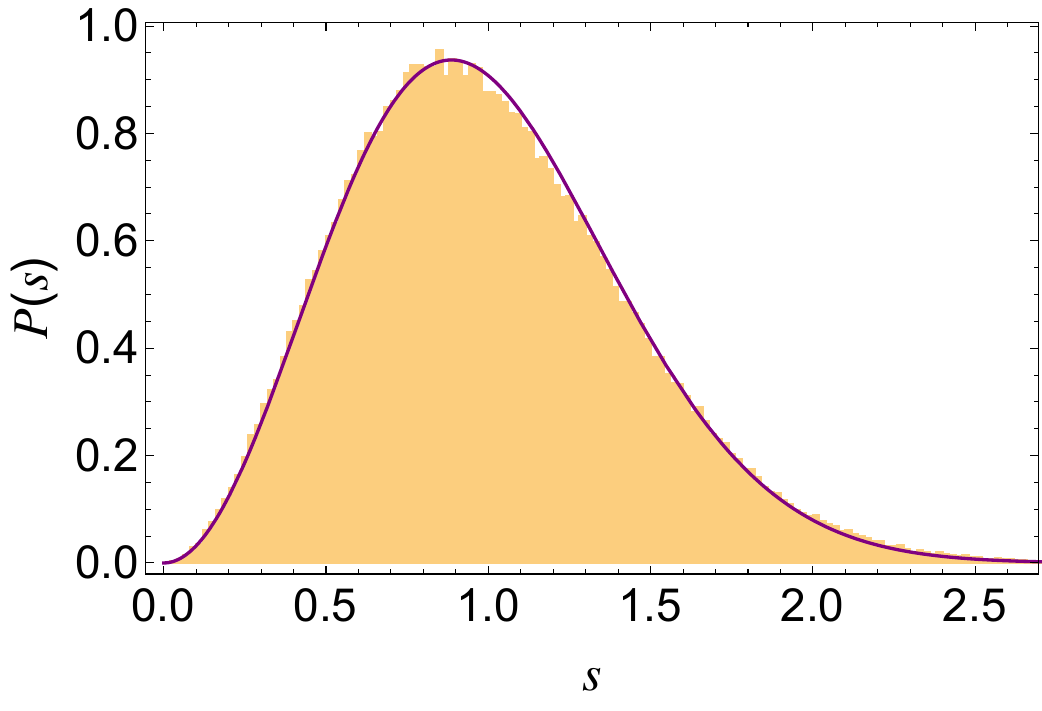} \\
        (a) Eigenvalue density & (b) Level spacing distribution  \\
       \multicolumn{2}{c}{\centering \includegraphics[width=0.45\textwidth]{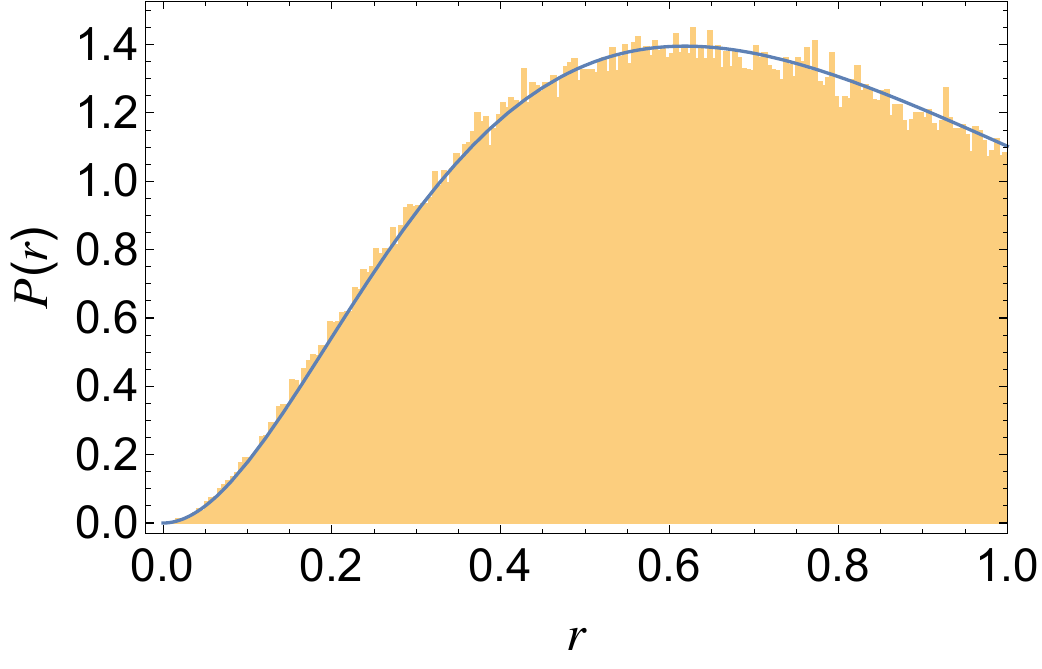}} \\
         \multicolumn{2}{c}{(c) $r$-parameter distribution} \\
    \end{tabular}
    \caption{
        Eigenvalue density, level spacing distribution, and $r$-parameter distribution for the operator $Z_1 + Z_{10}(t)$ based on 1000 realizations of the Hamiltonian \eqref{eq-fullHising} at the near-integrable point $(h_x, h_z, g) = (-1, 0.005, 0.2)$, with $L=10$ and $t=5 \times 10^4$. The $r$-parameter distribution was fitted using the form \eqref{eq: r-dis-N=3}, yielding the Dyson index $\nu=2.001\pm0.002$. The solid lines represent (a) the arcsine distribution, (b) the Wigner-Dyson distribution with $\nu=2$, and (c) the Wigner $r$-parameter distribution with $\nu=2$. The corresponding average $r$-parameter was found to be $\langle \tilde{r} \rangle = 0.60$. The analysis focuses on the central 50\% of the spectrum, excluding the outermost 25\% at each edge.
    }
    \label{fig:fluctuationsIsingNearIntegrableZ}
\end{figure}

\paragraph{Fluctuations at the near-integrable point.}
Figure \ref{fig:fluctuationsIsingNearIntegrable} presents the eigenvalue density, the level spacing distribution, and the $r$-parameter distribution for the operator $X_1 + X_{10}(t)$, based on 1000 realizations of the Hamiltonian \eqref{eq-fullHising} at the near-integrable point $(h_x, h_z, g) = (-1.0, 0.005, 0.2)$, with $L=10$ and $t=5 \times 10^4$. In panel (a), the improved statistics reveal deviations from the arcsine law even at very large times, which are substantially larger compared to the chaotic case, indicating that the arcsine law has not emerged at this time scale, highlighting limitations in analyzing the emergence of the arcsine law based on a single realization of the Hamiltonian \eqref{eq-fullHising}. As can be seen from panel (b), the level spacing statistics of the operator exhibit level spacing repulsion and resemble the Wigner-Dyson distribution with $\nu=2$, though with larger deviations compared to the chaotic case, possibly due to imperfect unfolding of the spectrum, which was performed using the arcsine law as the average density of eigenvalues. Panel (c) shows that the $r$-parameter distribution is well described by the Wigner $r$-parameter distribution with $\nu=2$, yielding an average value of $\langle \tilde{r} \rangle = 0.60$.
Figure \ref{fig:fluctuationsIsingNearIntegrableZ} presents similar results for the operator \(Z_1 + Z_{10}(t)\), with the key difference that the arcsine law and the Wigner-Dyson level spacing distribution fit the data very well in this case.

The primary takeaway from the above numerical analysis is that the free probability prediction for the spectrum of the operator $X_1+X_{10}(t)$ -- specifically, the arcsine law -- emerges only approximately within the time scales considered for a single realization of the Hamiltonian \eqref{eq-fullHising}. Despite this approximate emergence, the arcsine law appears even near the integrable point, with fluctuations exhibiting random matrix behavior. This is particularly evident in the $r$-parameter statistics. The observed deviations in the level spacing statistics are likely due to imperfections in the unfolding process of the spectrum. Both the chaotic and near-integrable points exhibit universal level spacing statistics for the eigenvalues  $X_1 + X_{10}(t)$  with the key distinction being that, in the near-integrable case, this behavior emerges only at much later times. By contrast, for the operator $Z_1+Z_{10}(t)$, the arcsine law fits the data well even in the near-integrable point, and the fluctuations are well described by a Wigner-Dyson distribution.

In finite numerical simulations, free cumulants in the model \eqref{eq-fullHising} do not vanish due to finite-size effects.
The above results show that while the dynamics drives $Z_1$ and $Z_{10}(t)$ towards asymptotic freeness -- producing a spectrum for their sum that respects the free probability prediction -- the same is not entirely true for $X$-Pauli operators, in which one observes small deviations from the free probability prediction even at the chaotic point. This suggests that the finite size effects for mixed cumulants involving $X$-Pauli operators are larger than the ones involving $Z$-Pauli operator, which is indeed what is observed in numerical simulations. The underlying reason why some operators exhibit larger finite-size effects than others lies in the fact that the Hamiltonian does not \emph{treat} all operators equally. In the version of the Ising model we consider, for instance, the Hamiltonian takes the schematic form
$ H = \sum_i Z_i Z_{i+1} + Z_i + X_i.$
If we exchange $X \leftrightarrow Z$, we find that cumulants involving $X$ tend to thermalize more effectively (i.e., exhibit smaller residual values) than those involving $Z$.

\subsection{Spatial Dependence of $t_\text{arcsine}$}
\begin{figure}[h!]
    \centering
    \begin{tabular}{cc}
        \includegraphics[width=0.45\linewidth]{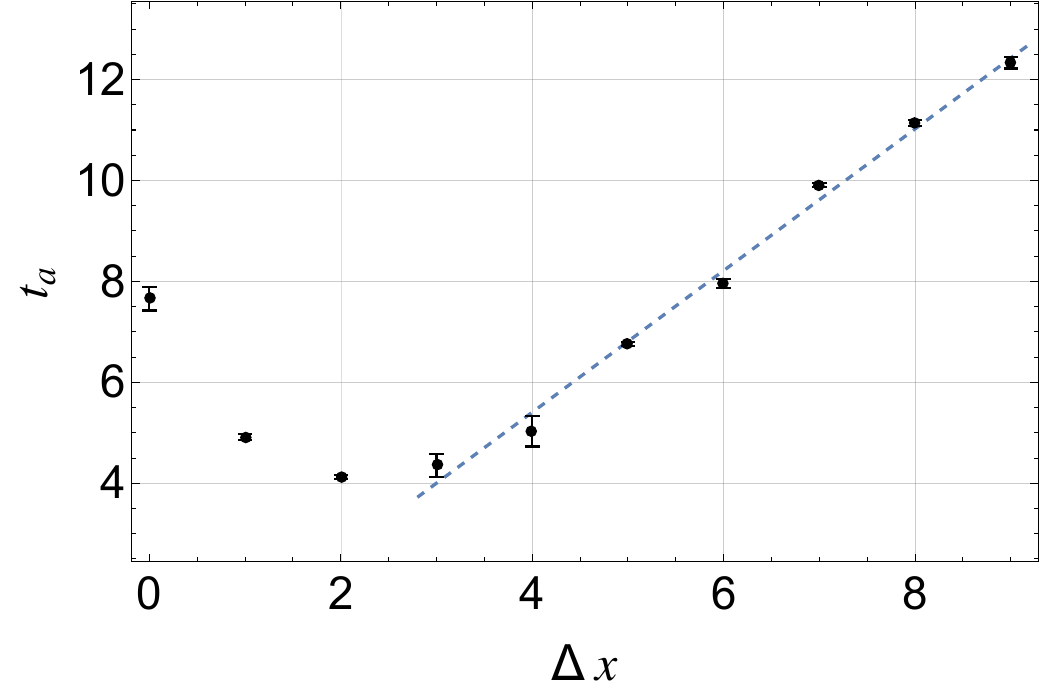} &
        \includegraphics[width=0.45\linewidth]{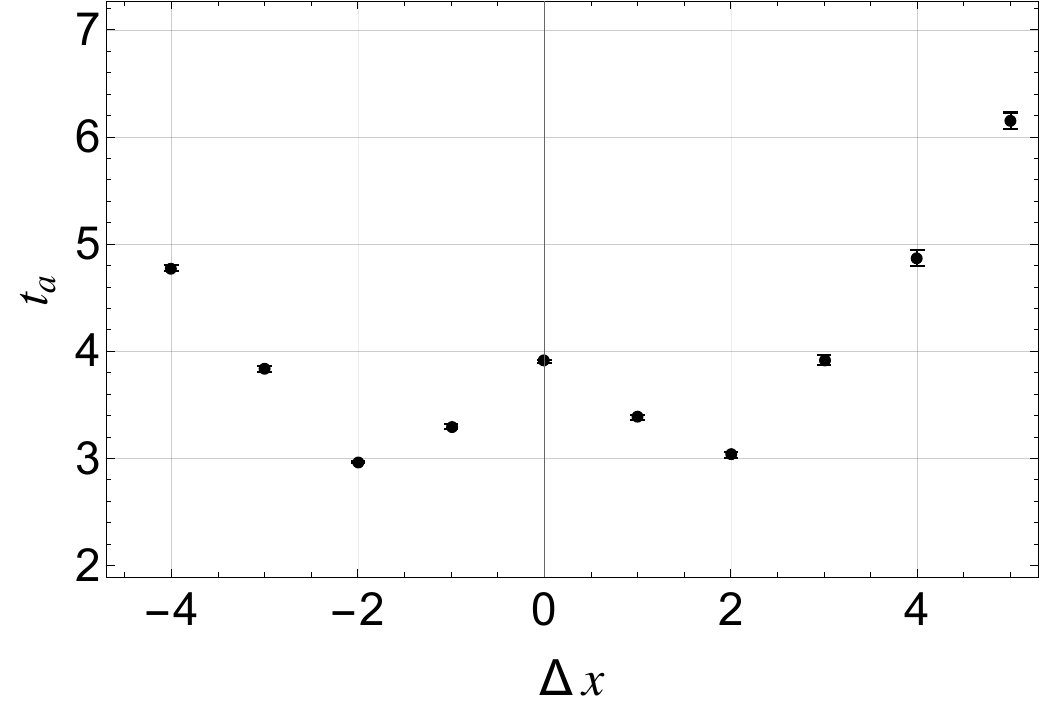} \\
        (a) Open boundary conditions &
        (b) Closed boundary conditions
    \end{tabular}
    \caption{Arcsine time $t_{\text{arcsine}}$ as a function of spatial separation $\Delta x$, illustrating the emergence of free probability behavior for operators of the form $Z_i(0) + Z_{i+\Delta x}(t)$. Results correspond to the chaotic regime with parameters $(h_x, h_z, g) = (-1.05, 0.5, 0.2)$ and system size $L = 10$. The arcsine time was extracted via a $\chi^2$ analysis over 50 independent realizations of the Hamiltonian~\eqref{eq-fullHising}. A linear fit for $\Delta x \geq 3$ yields $t_{\text{arcsine}} = a + b\, \Delta x$, with $a = 3 \pm 4$ and $b = 1.42 \pm 0.07$. Panel (a) shows results with open boundary conditions and $i = 1$, while panel (b) uses closed boundary conditions and $i = 5$.
    }
    \label{fig:tfvsdeltax}
\end{figure}

In this subsection, we investigate the spatial dependence of $t_\text{arcsine}$, the characteristic time scale at which the free probability prediction becomes accurate for operators of the form $Z_i(0) + Z_{i+\Delta x}(t)$. Specifically, we analyze how $t_\text{arcsine}$ varies with the spatial separation between the operators, defined as $\Delta x$. To account for possible edge effects, we consider both open and closed boundary conditions. The results are shown in Fig.~\ref{fig:tfvsdeltax}.

Since freeness between $Z_i(0)$ and $Z_{i+\Delta x}(t)$ indicates that both operators do not commute, having a large commutator, we expect that the arcsine time for distant operators should be larger as compared to operators which are close to each other. Moreover, assuming that the support of the operator $Z_{i+\Delta x}(t)$ grows linearly with time, we expect a linear behavior of $t_\text{arcsine}$ with $\Delta x$. In fact, after a non-monotonic behavior for small values of $\Delta x$, we observe that $t_\text{arcsine}$ indeed grows linearly $\Delta x$. We expect the delay on the onset of the free probability prediction to be controlled by $\Delta x/v_F$, where $v_F$ is some velocity scale. From the linear fit we estimate $v_F = 0.70\pm 0.03$ for $(h_x, h_z, g) = (-1.05, 0.5, 0.2)$ and system size $L = 10$. It would be interesting to compare $v_F$ with the butterfly velocity extracted from OTOCs. However, a precise comparison requires much larger system sizes and is beyond the scope of this paper.

For both open and closed boundary conditions, the arcsine time exhibits a non-monotonic dependence on the spatial separation for small values of $\Delta x$. Specifically, $t_a$ initially decreases with increasing $\Delta x$, reaching a minimum at $|\Delta x| = 2$, and subsequently grows linearly, as expected. One possible explanation for this behavior is that the presence of the Ising term $Z_i Z_{i+1}$ facilitates the emergence of freeness between a time-evolved operator and its neighboring operators -- causing it to fail strongly to commute with them -- more readily than with itself at time zero. However, as more distant operators are considered, the expected delay in the onset of the free probability regime emerges, since the local Ising interaction cannot influence distant points instantaneously.

\section{Asymptotic Freeness in the High-Spin Mixed-Field Ising Model} \label{sec-higherSpinModel}
In this section, we consider a spin-$s$ generalization of the mixed-field Ising model (\ref{eq-HamilDisorder}), where the spin-\(\frac{1}{2}\) particles at each site are replaced by spin-$s$ representations of \(SU(2)\), and study the emergence of asymptotic freeness of operators under time evolution generated by this Hamiltonian. We start with the model proposed in~\cite{Craps:2019rbj}:
\begin{equation} \label{eq-HSIsing}
    H^{(s)} = -2 \frac{\sqrt{s(s+1)}}{\sqrt{3}} \left[ \sum_{i=1}^{L-1} Z^{(s)}_i Z^{(s)}_{i+1} + \sum_{i=1}^{L} \left( h_x X^{(s)}_i + h_z Z^{(s)}_i \right) + g\sum_{i=1}^{L} \epsilon_i X^{(s)}_i \right],
\end{equation}
where \(X^{(s)}_i\), \(Y^{(s)}_i\), and \(Z^{(s)}_i\) are the matrices corresponding to the spin-\(s\) representation of \(SU(2)\) and $\epsilon_i$ is drawn from a Gaussian distribution with average zero and unit variance. The dimension of the Hilbert space is $d_s=(2s+1)^L$. The overall normalization is chosen such that for \(s = \frac{1}{2}\), we recover the model (\ref{eq-HamilDisorder}). In order for this model to have well-defined classical limit $s \rightarrow \infty$, one needs to normalize the basic operators as follows:
\begin{equation}
    X^{(s)}_i X^{(s)}_i + Y^{(s)}_i Y^{(s)}_i + Z^{(s)}_i Z^{(s)}_i = 3\, \mathds{1}_{d_s},
\end{equation}
and
\begin{equation} \label{eq-commutations}
    [X^{(s)}_j, Y^{(s)}_k] = i \frac{\sqrt{3}}{\sqrt{s(s+1)}} \delta_{jk} Z^{(s)}_j,
\end{equation}
with similar relations for cyclic permutations of the above equation. Note that the right-hand side of \eqref{eq-commutations} becomes zero in the classical limit ($s \rightarrow \infty$).

\begin{figure}[h!]
    \centering
    \begin{tabular}{cc}
    \includegraphics[width=0.45\linewidth]{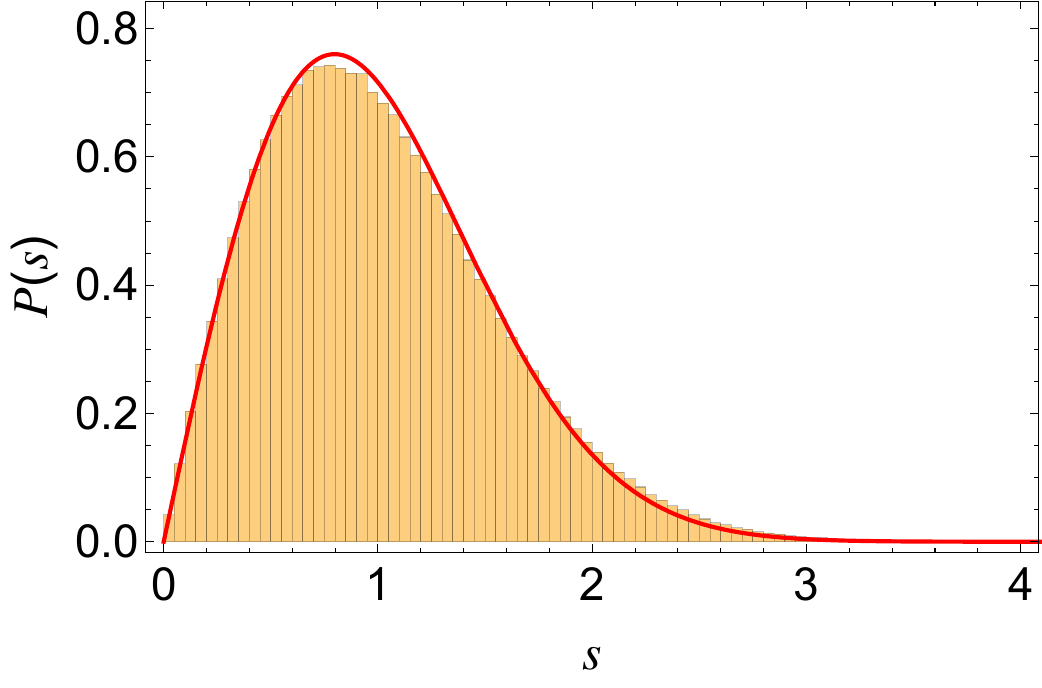}&
    \includegraphics[width=0.45\linewidth]{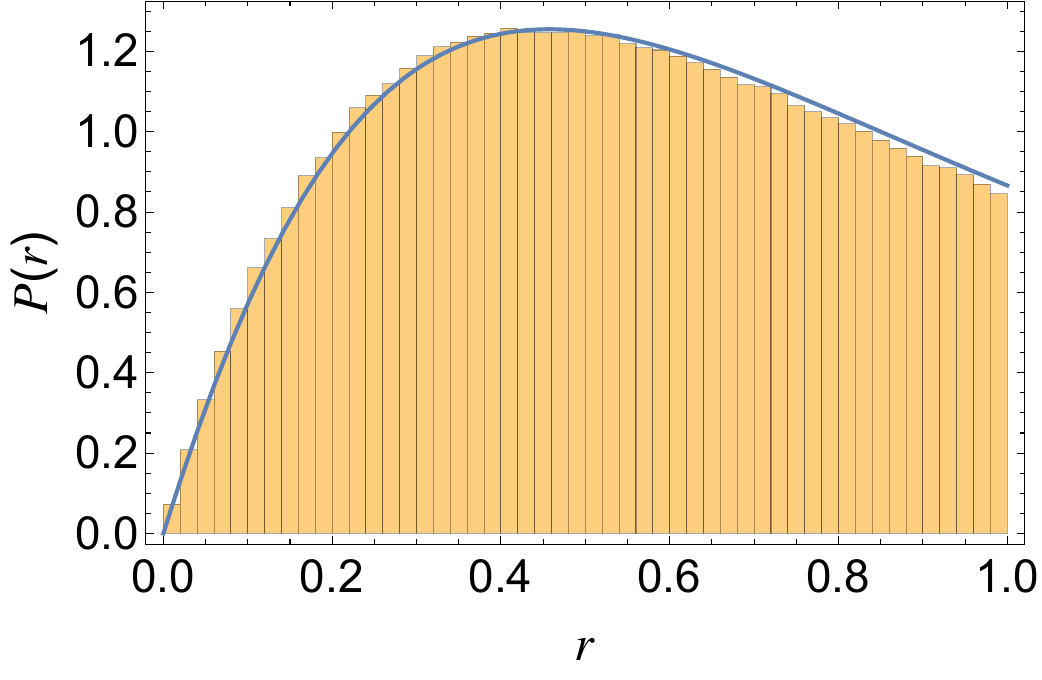}
    \end{tabular}
    \caption{Level spacing statistics (left panel) and $r$-parameter statistics (right panel) for 1000 realizations of the Hamiltonian~(\ref{eq-HSIsing}) with parameters \( (h_x, h_z, g) = (-1.05, 0.5, 0.2) \), \( s = 1 \), and \( L = 7 \). The average $r$-parameter obtained is \( \langle \tilde{r} \rangle = 0.53 \). The red and blue curves represent the Wigner-Dyson distributions for level spacing and $r$-parameter, respectively, corresponding to the $\nu=1$ (GOE). The analysis focuses on the central 50\% of the spectrum, excluding the outermost 25\% on each edge.}
    \label{fig:SpectralStatisticsHSIsing}
\end{figure}

\paragraph{Spectral statistics.} 
The Hamiltonian (\ref{eq-HSIsing}) does not appear to exhibit any integrable point for \( s \geq 1 \), in contrast to the case \( s = \frac{1}{2} \), which is integrable when \( g = 0 \) and either \( h_x \) or \( h_z \) vanish. However, for the specific parameter choice \( (h_x, h_z, g) = (-1.05, 0.5, 0.2) \), the system still displays random matrix statistics for \( s \geq 1 \). Figure \ref{fig:SpectralStatisticsHSIsing} presents the level spacing statistics and the $r$-parameter statistics of the Hamiltonian for 1000 realizations with parameters \( (h_x, h_z, g) = (-1.05, 0.5, 0.2) \), \( s = 1 \), and \( L = 7 \). 

It is important to note that the presence of the random magnetic field term, \( g\sum_i \epsilon_i X^{(s)}_i \), breaks the parity symmetry of the model. This symmetry breaking significantly simplifies the spectral statistics analysis, as it eliminates the need to analyze separate symmetry sectors individually.

\subsection{Approximate emergence of asymptotic freeness}
\begin{figure}[ht]
    \centering
    \begin{tabular}{cc}
        \includegraphics[width=0.45\textwidth]{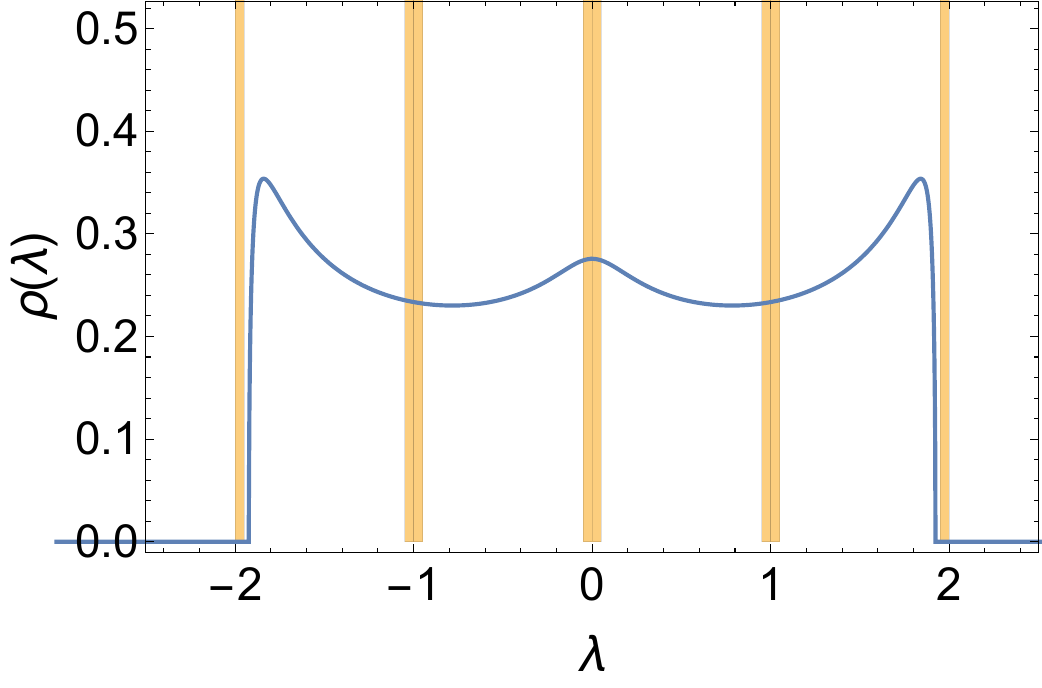} \put(-125,110){$t=0.01$} & \includegraphics[width=0.45\textwidth]{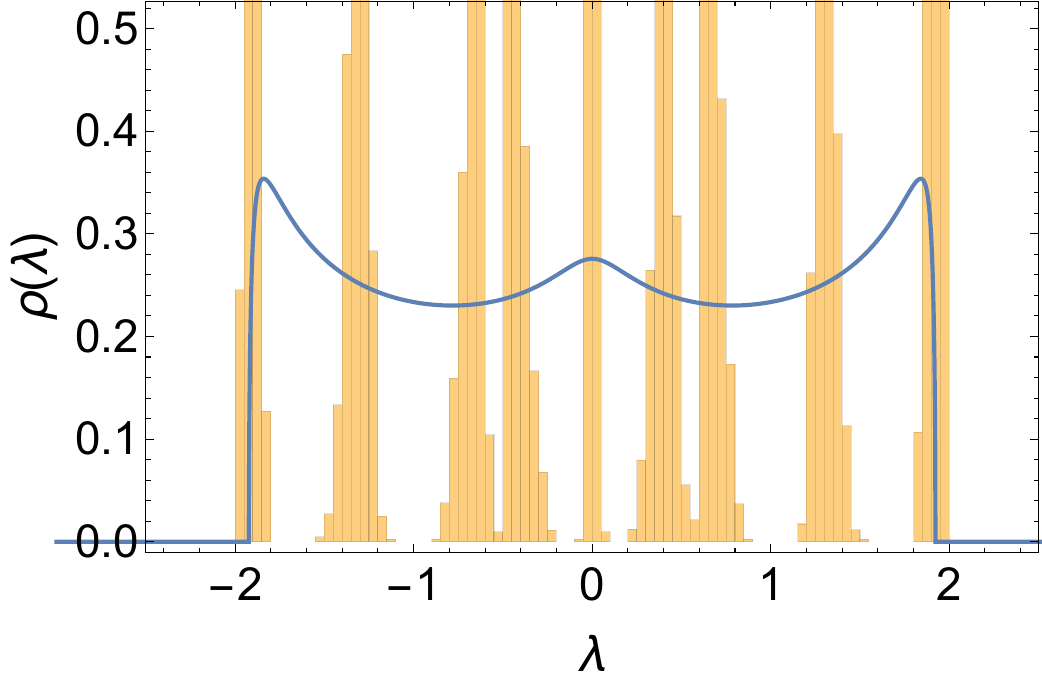}\put(-112,110){$t=1$} \\
        %(a) Caption for figure 1 & (b) Caption for figure 2 \\
        \includegraphics[width=0.45\textwidth]{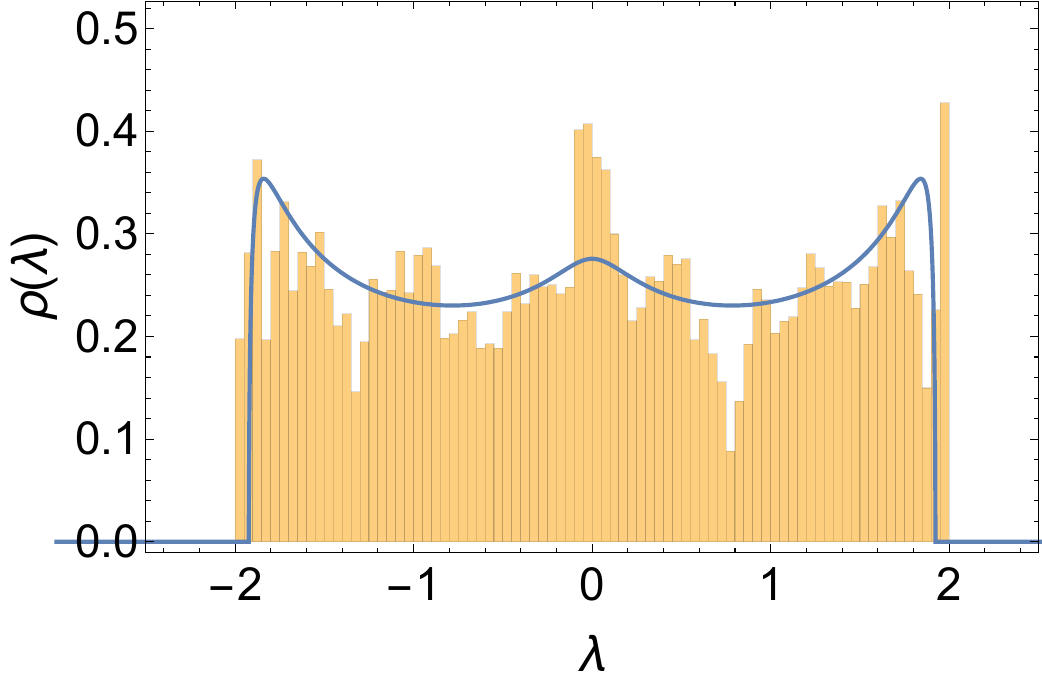}\put(-120,110){$t=2$} & \includegraphics[width=0.45\textwidth]{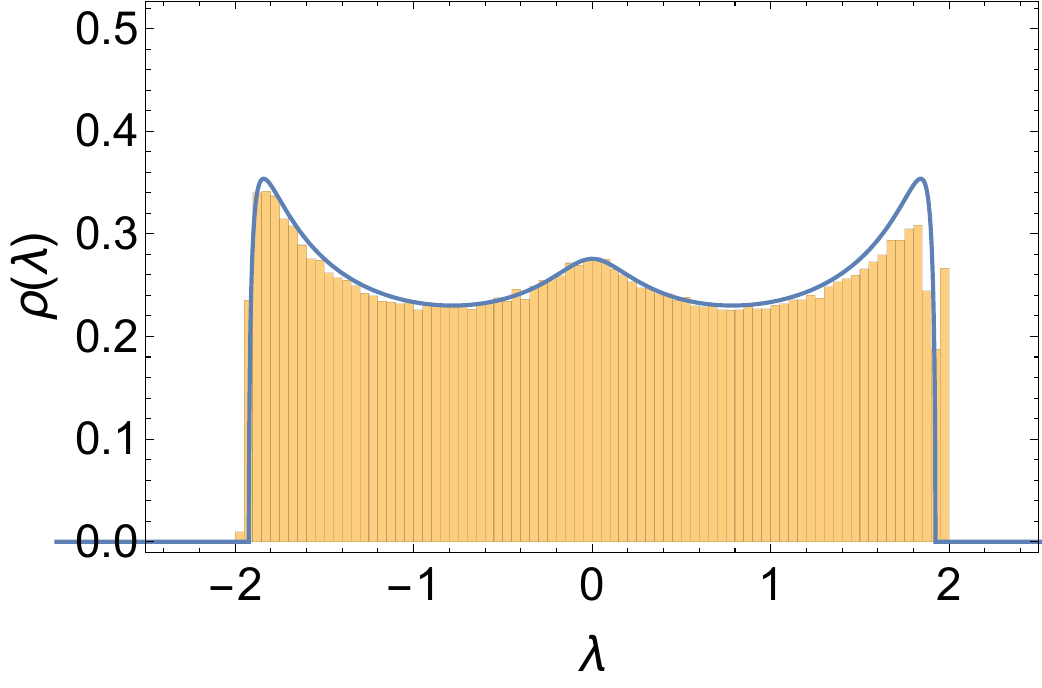}\put(-120,110){$t=5$}\\
        %(c) Caption for figure 3 & (d) Caption for figure 4 \\
    \end{tabular}
    \caption{Evolution of the eigenvalue spectrum of the operator \( \sqrt{2/3}(Z^{(s)}_1(0) + Z^{(s)}_1(t)) \) with increasing time at the chaotic point. The parameters are fixed as \( (h_x, h_z, g) = (-1.05, 0.5, 0.2) \), \( s = 1 \), and \( L = 6 \), with 100 realizations of the Hamiltonian~(\ref{eq-HSIsing}). The distribution predicted by free probability theory (\ref{eq: sumdis_spin1}) (depicted by the blue continuous line) becomes evident around \( t = 5 \). The number of bins in each histogram is $100$.  
 }
    \label{fig:emergenceFPPs1} 
\end{figure}

In the spin-1 case, the fundamental operators of the theory, such as \( X^{(s)}_i \), and their time-evolved counterparts, \( X^{(s)}_i(t) \), have spectra characterized by a generalized Bernoulli distribution with eigenvalues \( \pm \sqrt{3/2}, 0 \). For convenience, we rescale these operators by a factor of \( \sqrt{2/3} \), ensuring their spectra become \( \pm1,0 \). As shown in Section \ref{free_addition} (Example 2), the free probability prediction for the spectrum of the sum of two free variables with generalized Bernoulli distribution with eigenvalues $\pm1,0$ is given by Eq.~(\ref{eq: sumdis_spin1}). Figure \ref{fig:emergenceFPPs1} shows the spectrum of eigenvalues of the operator \( \sqrt{2/3}(Z_1(0) + Z_1(t)) \) with increasing time for 100 realizations of the Hamiltonian~(\ref{eq-HSIsing}) at the chaotic point. The distribution predicted by free probability theory (\ref{eq: sumdis_spin1}) becomes evident around \( t = 5 \).

\subsection{Fluctuations}
\begin{figure}[h!]
    \centering
    \begin{tabular}{cc}
    \includegraphics[width=0.45\linewidth]{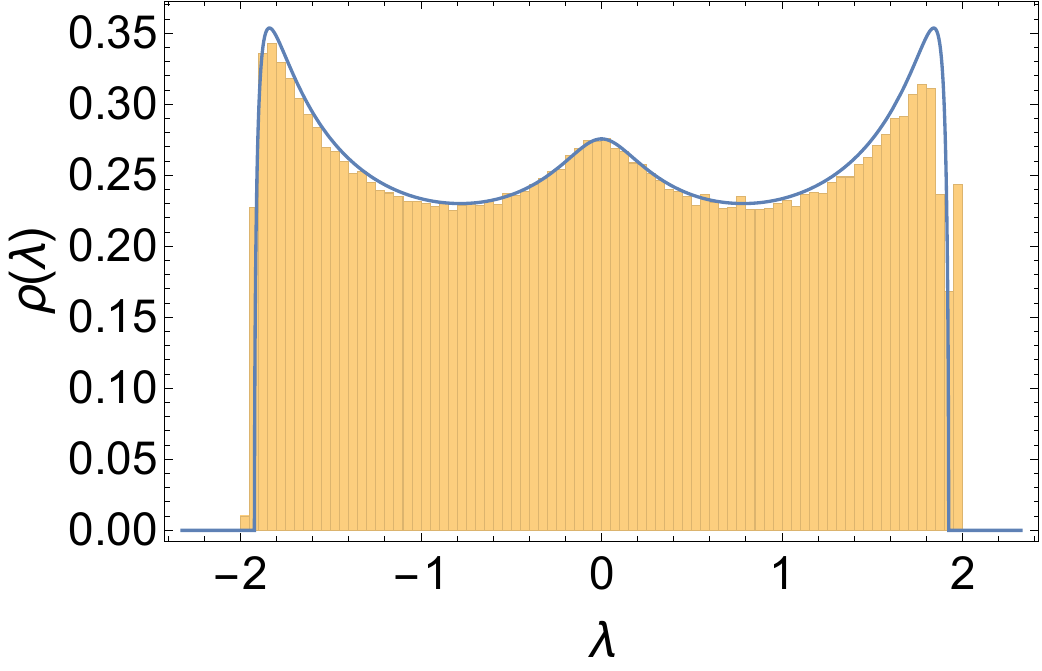}&
    \includegraphics[width=0.45\linewidth]{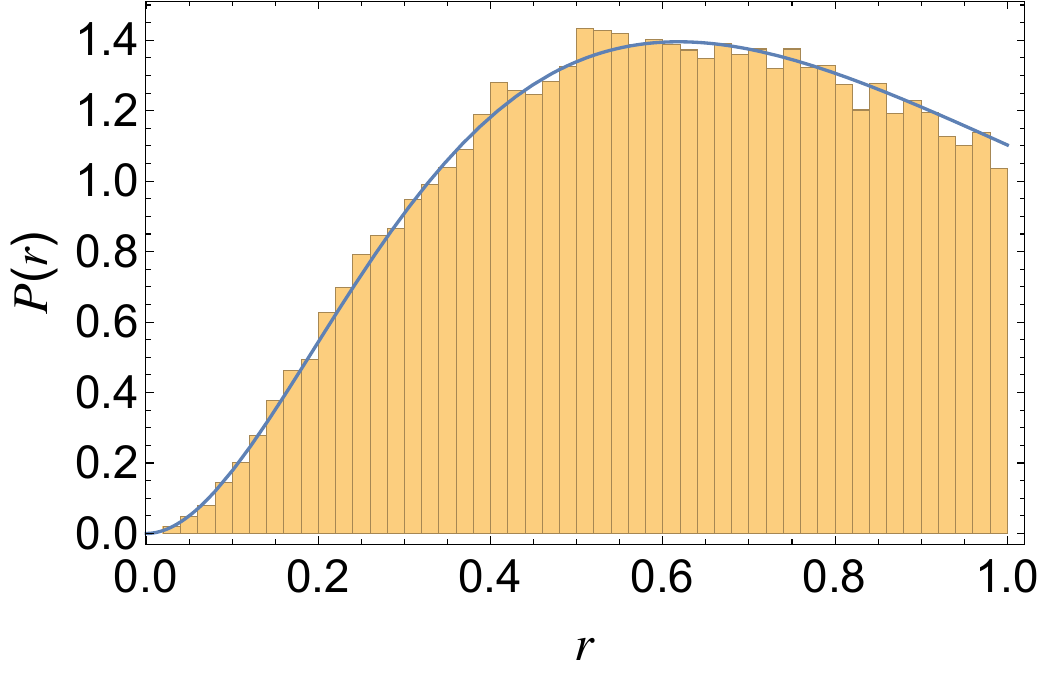}\\
    (a) Eigenvalue density & (b) $r$-parameter distribution
    \end{tabular}
    \caption{Spectral statistics of the operator $\sqrt{2/3}(Z_1^{(s)} + Z_{6}^{(s)}(t))$ based on 100 realizations of the Hamiltonian \eqref{eq-HSIsing} at the chaotic point \( (h_x, h_z, g) = (-1.05, 0.5, 0.2) \), with \( s = 1 \), \( L = 6 \) and $t=50$. The solid lines represent (a) the free probability prediction and (b) the Wigner $r$-parameter distribution for $\nu=2$. The $r$-parameter distribution was fitted using the form \eqref{eq: r-dis-N=3}, yielding the Dyson index \( \nu = 1.999 \pm 0.004 \). The average $r$-parameter was found to be $\langle \tilde{r} \rangle = 0.60$.  The analysis focuses on the central 50\% of the spectrum, excluding the outermost 25\% on each edge.}

    \label{fig:fluctuationsHS}
\end{figure}

\begin{figure}[h!]
    \centering
    \begin{tabular}{cc}
    \includegraphics[width=0.45\linewidth]{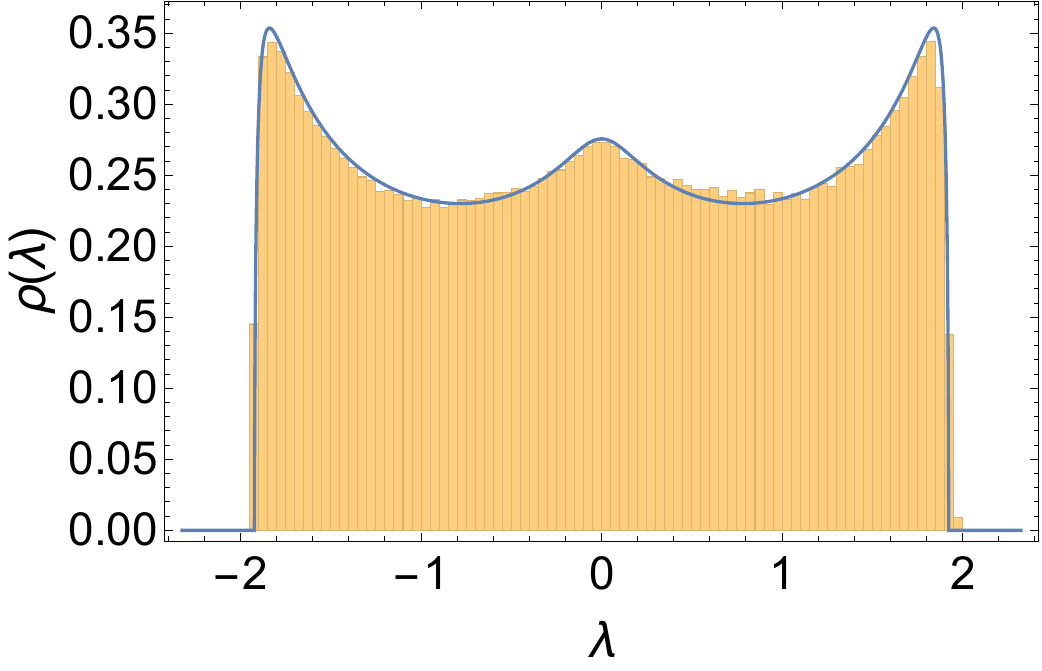}&
    \includegraphics[width=0.45\linewidth]{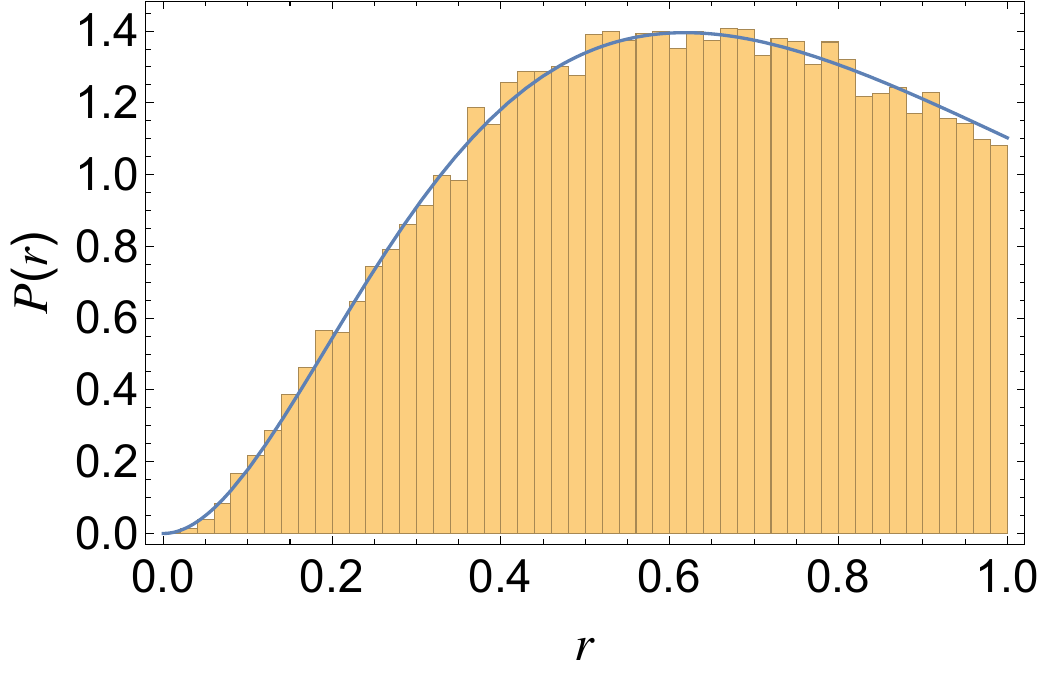}\\
    (a) Eigenvalue density & (b) $r$-parameter distribution
    \end{tabular}
    \caption{Spectral statistics of the operator $\sqrt{2/3}(Y_1^{(s)} + Y_{6}^{(s)}(t))$ based on 100 realizations of the Hamiltonian \eqref{eq-HSIsing} at the chaotic point \( (h_x, h_z, g) = (-1.05, 0.5, 0.2) \), with \( s = 1 \), \( L = 6 \) and $t=50$. The solid lines represent (a) the free probability prediction and (b) the Wigner $r$-parameter distribution for $\nu=2$. The $r$-parameter distribution was fitted using the form \eqref{eq: r-dis-N=3}, yielding the Dyson index \( \nu = 2.000 \pm 0.004 \). The average $r$-parameter was found to be $\langle \tilde{r} \rangle = 0.60$.  The analysis focuses on the central 50\% of the spectrum, excluding the outermost 25\% on each edge.}

    \label{fig:fluctuationsHSY}
\end{figure}

\begin{figure}[h!]
    \centering
    \begin{tabular}{cc}
    \includegraphics[width=0.45\linewidth]{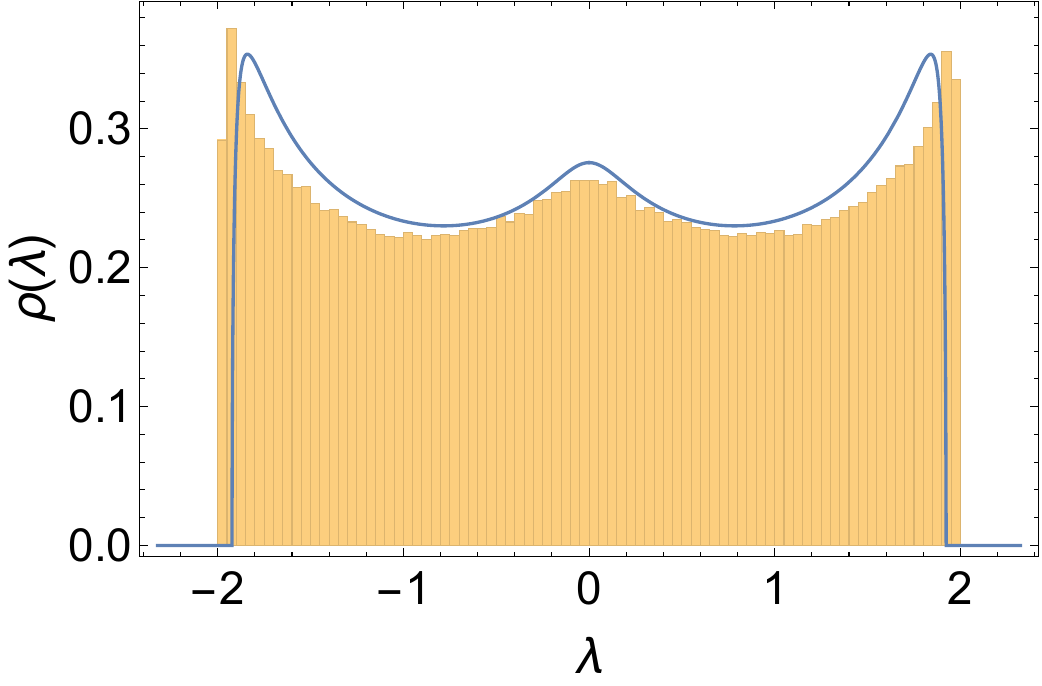}&
    \includegraphics[width=0.45\linewidth]{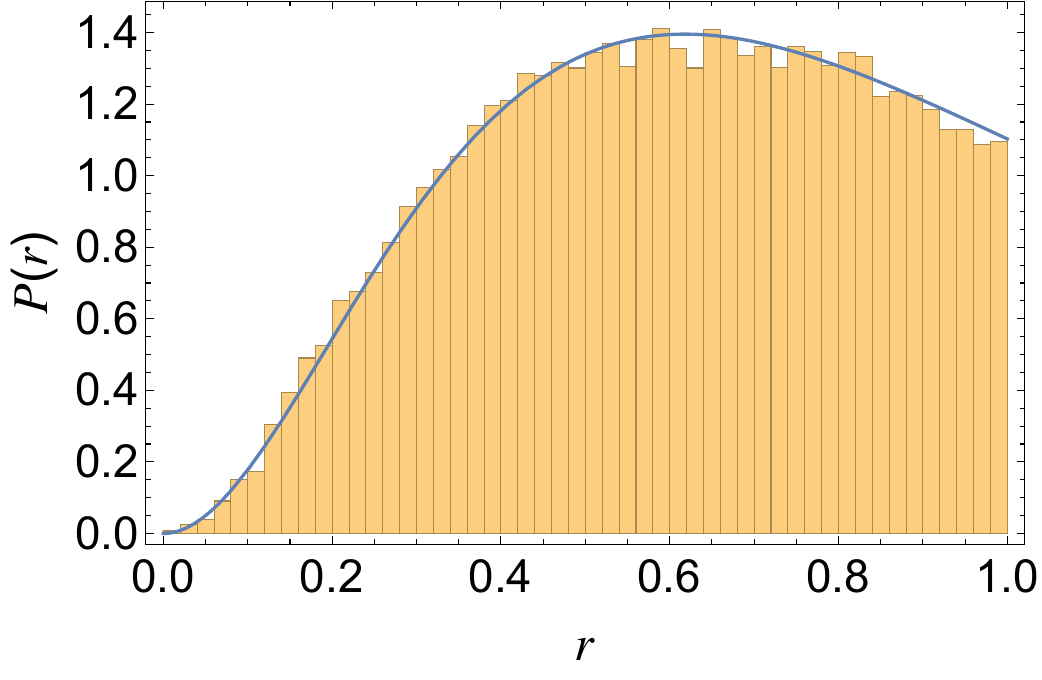}\\
    (a) Eigenvalue density & (b) $r$-parameter distribution
    \end{tabular}
    \caption{Spectral statistics of the operator $\sqrt{2/3}(X_1^{(s)} + X_{6}^{(s)}(t))$ based on 100 realizations of the Hamiltonian \eqref{eq-HSIsing} at the chaotic point \( (h_x, h_z, g) = (-1.05, 0.5, 0.2) \), with \( s = 1 \), \( L = 6 \) and $t=50$. The solid lines represent (a) the free probability prediction and (b) the Wigner $r$-parameter distribution for $\nu=2$. The $r$-parameter distribution was fitted using the form \eqref{eq: r-dis-N=3}, yielding the Dyson index \( \nu = 2.000 \pm 0.004 \). The average $r$-parameter was found to be $\langle \tilde{r} \rangle = 0.60$.  The analysis focuses on the central 50\% of the spectrum, excluding the outermost 25\% on each edge.}

    \label{fig:fluctuationsHSX}
\end{figure}

In this subsection, we investigate the fluctuations in the spectrum of the sum of spin-1 operators of the form $A + B(t)$. In the presence of chaotic dynamics, the operators $A$ and $B(t)$ are expected to become asymptotically free, with the distribution of their sum given by Eq.~(\ref{eq: sumdis_spin1}). The results are presented in Figures~\ref{fig:fluctuationsHS}, \ref{fig:fluctuationsHSY}, and \ref{fig:fluctuationsHSX}. In all cases, we consider 100 realizations of the Hamiltonian \eqref{eq-HSIsing} at the chaotic point \( (h_x, h_z, g) = (-1.05, 0.5, 0.2) \), with \( s = 1 \), \( L = 6 \), and \( t = 50 \).

\paragraph{$Z$-operators.} 
The results for the operator $\sqrt{2/3}(Z_1^{(s)} + Z_{6}^{(s)}(t))$ are shown in Figure~\ref{fig:fluctuationsHS}. The left panel illustrates that the free probability prediction agrees well with the data, except for a deviation near the right edge of the spectrum. The right panel shows that the corresponding $r$-parameter distribution is well described by a Wigner $r$-parameter distribution with \( \nu = 2 \). 

\paragraph{$Y$-operators.} 
Figure~\ref{fig:fluctuationsHSY} presents the results for the operator $\sqrt{2/3}(Y_1^{(s)} + Y_{6}^{(s)}(t))$. In this case, the free probability prediction provides an excellent fit across the entire spectrum, and the fluctuations are also well captured by random matrix theory with \( \nu = 2 \). 

\paragraph{$X$-operators.} 
The results for the operator $\sqrt{2/3}(X_1^{(s)} + X_{6}^{(s)}(t))$ are displayed in Figure~\ref{fig:fluctuationsHSX}. Here, while the overall shape of the curve follows the free probability prediction, quantitative discrepancies are observed. Nonetheless, the fluctuations remain well described by random matrix theory with \( \nu = 2 \).

The key takeaway from this section is that, although the level spacing statistics of the model at the chaotic point are well described by random matrix theory, the free probability prediction for the sum of operators at different times holds only approximately, with deviations that depend on the specific choice of operators. In particular, the above results suggest that finite-size effects on the mixed cumulants in the model \eqref{eq-HSIsing} are operator-dependent, with larger residual values observed for cumulants involving the $X$-operators, similar to what occurs in the spin chain model with $s=1/2$.

%%%%%%%%%%%%%%%%%%%%%%%%%%%%%%%%%%%%%%%%%%%%%%%%%%%%%%%%%
%%%%%%%%%%%%%%% SYK Analysis %%%%%%%%%%%%%%%%%%%%%%%%%%%%
%%%%%%%%%%%%%%%%%%%%%%%%%%%%%%%%%%%%%%%%%%%%%%%%%%%%%%%%%

\section{Asymptotic Freeness in the Sachdev--Ye--Kitaev Model } \label{sec-SYK}
Finally, we discuss the emergence of asymptotic freeness in the SYK model.
We consider the Hamiltonian of the standard $q=4$ SYK model with \( N \) Majorana fermions, given by \cite{SachdevYeModel, Kitaev2015}
\begin{equation} \label{eq-SYK4}
    H_{\text{SYK}_4} = \sqrt{\frac{6}{N^3}} \sum_{j<k<l<m} J_{jklm} \, \psi_j \psi_k \psi_l \psi_m
\end{equation}
where \( \{\psi_j, \psi_k \} = \delta_{jk} \), and \( J_{jklm} \) is a random Gaussian variable with zero mean and standard deviation \( J \).

The Hamiltonian \eqref{eq-SYK4} conserves charge parity, defined as \( Q \mod 2 \), where the charge operator \( Q \) is expressed in terms of Dirac fermions \( c_i \) as:  
\begin{equation}
    Q = \sum_{i=1}^{\left\lfloor N/2 \right\rfloor} \bar{c}_i c_i\,.
\end{equation}  
where$\left\lfloor N/2 \right\rfloor$ denotes the largest integer less than or equal to $N/2$.
The Dirac fermions are related to the Majorana fermions by the transformations:  
\begin{equation}
    \psi_{2i} = \frac{c_i + \bar{c}_i}{\sqrt{2}}\,, \quad \psi_{2i-1} = \frac{i(c_i - \bar{c}_i)}{\sqrt{2}}\,.
\end{equation} 
As a result, the Hamiltonian \eqref{eq-SYK4} can be cast into a block diagonal form, consisting of two blocks corresponding to positive and negative charge parity sectors \cite{Cotler:2016fpe}. By representing the Majorana fermions in the chiral basis~\cite{Pais1962}:
\begin{equation}
    \psi_i = \frac{1}{\sqrt{2}}
\begin{cases} 
\bigotimes_{k=1}^{\nu} \sigma_1 & \text{if } i = 1, \\[10pt]
\left( \bigotimes_{k=1}^{\nu - \left\lfloor \frac{i}{2} \right\rfloor} \sigma_1 \right) \otimes \sigma_2 \otimes \left( \bigotimes_{k=1}^{\left\lfloor \frac{i}{2} \right\rfloor - 1} \sigma_0 \right) & \text{if } 2 \leq i \leq 2\nu \text{ and } i \text{ is even}, \\[10pt]
\left( \bigotimes_{k=1}^{\nu - \left\lfloor \frac{i-1}{2} \right\rfloor} \sigma_1 \right) \otimes \sigma_3 \otimes \left( \bigotimes_{k=1}^{\left\lfloor \frac{i-1}{2} \right\rfloor - 1} \sigma_0 \right) & \text{if } 3 \leq i \leq 2\nu \text{ and } i \text{ is odd}, \\[10pt]
\sigma_2 \otimes \left( \bigotimes_{k=1}^{\nu - 1} \sigma_0 \right) & \text{if } i = 2\nu.
\end{cases}
\end{equation}
the Hamiltonian \eqref{eq-SYK4} naturally assumes this block diagonal structure.

\begin{figure}[h!]
    \centering
    \begin{tabular}{cc}
    \includegraphics[width=0.45\linewidth]{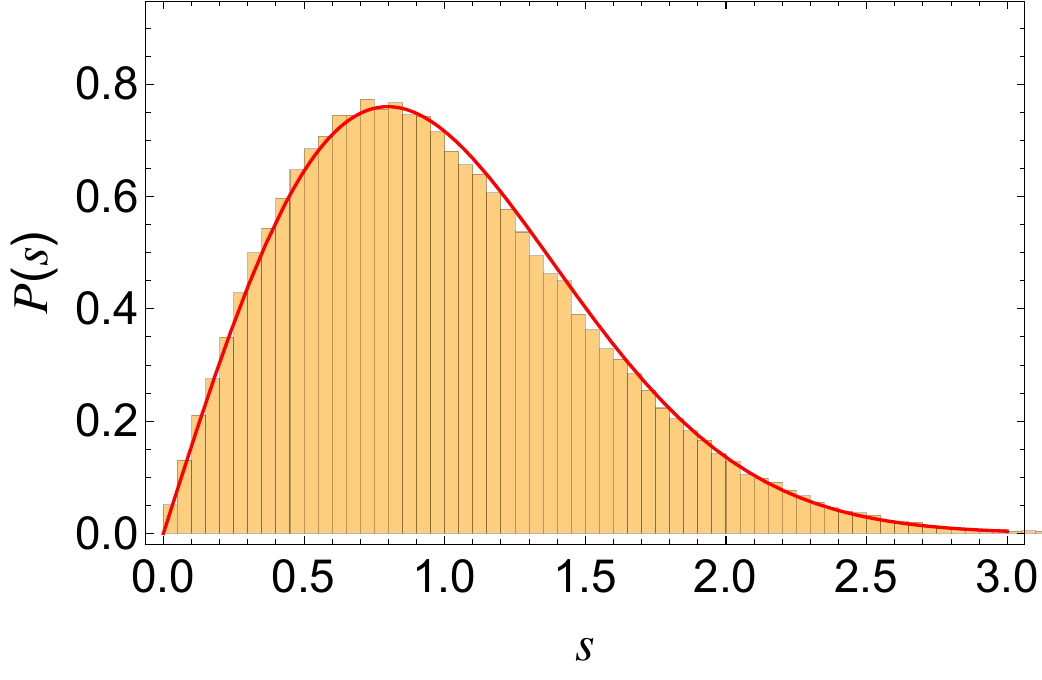}&
    \includegraphics[width=0.45\linewidth]{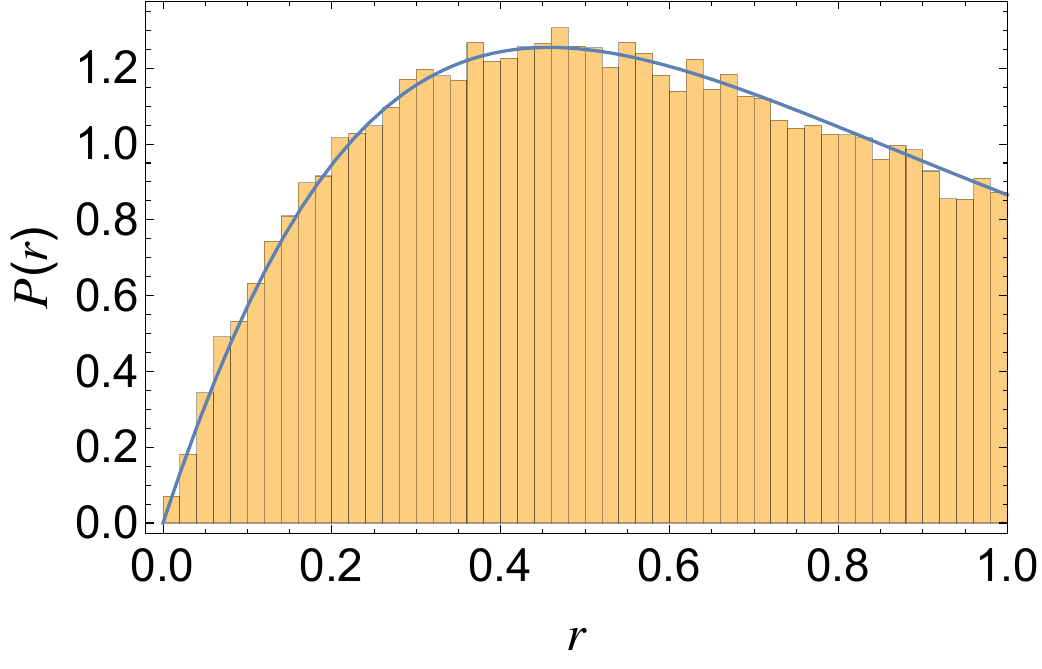}
    \end{tabular}
    \caption{Level spacing statistics (left panel) and $r$-parameter statistics (right panel) for 1000 realizations of the Hamiltonian~(\ref{eq-SYK4}) with parameters $J=1$ and $N=16$. The average $r$-parameter obtained is \( \langle \tilde{r} \rangle = 0.53 \). The red and blue curves represent the Wigner-Dyson distributions for level spacing and $r$-parameter, respectively, corresponding to $\nu=1$ (GOE). The analysis considers the full spectrum.}
    \label{fig:SpectralStatisticsSYK4}
\end{figure}

It is well known that the Hamiltonian \eqref{eq-SYK4} has level spacing statistics described by random matrix theory~\cite{Garcia-Garcia:2016mno, Garcia-Garcia:2017pzl}. Figure \ref{fig:SpectralStatisticsSYK4} shows the level spacing statistics and the $r$-parameter statistics for this model for $N=16$, in which case the spectral statistics follow that of GOE of random matrices~\cite{Cotler:2016fpe}.

\begin{figure}[h!]
    \centering
    \begin{tabular}{cc}
        \includegraphics[width=0.45\textwidth]{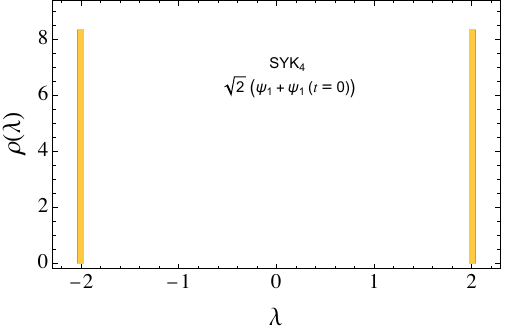} &
        \includegraphics[width=0.45\textwidth]{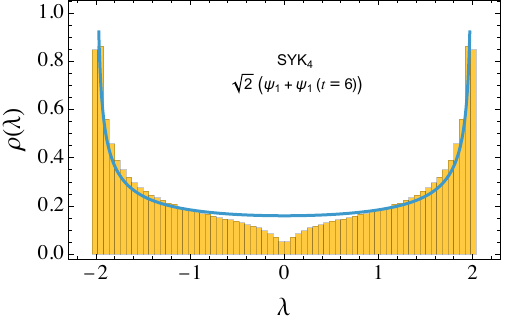}\\
        %(a) Caption for figure 1 & (b) Caption for figure 2 \\
        \includegraphics[width=0.45\textwidth]{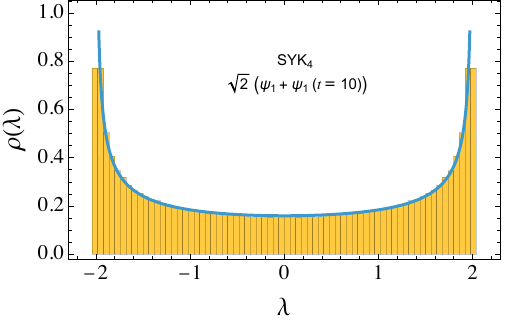}&
        \includegraphics[width=0.45\textwidth]{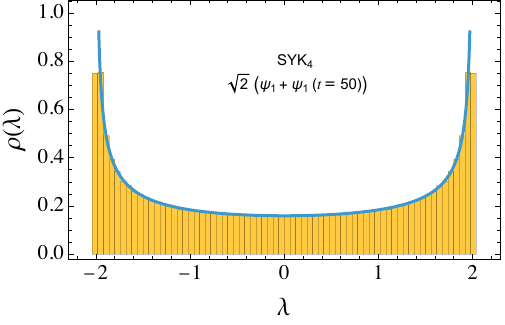}\\
        %(c) Caption for figure 3 & (d) Caption for figure 4 \\
    \end{tabular}
    \caption{Spectral statistics (eigenvalue density) of the operator $\sqrt{2}\left(\psi_{1}+\psi_{1}(t)\right)$ for $1000$ realizations of the SYK$_{4}$ Hamiltonian~\eqref{eq-SYK4}, without fixing a chirality sector, with $N=16$ fermions for different times, indicated in each individual figure. The bin size in each histogram is 0.06. 
 }
    \label{fig:OpStatSYK4Psi1Psi1} 
\end{figure}

\begin{figure}[h!]
    \centering
    \begin{tabular}{cc}
        \includegraphics[width=0.45\textwidth]{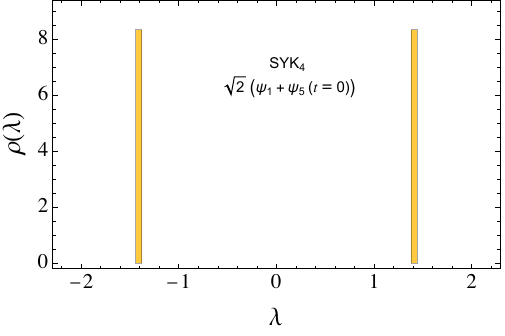} &
        \includegraphics[width=0.45\textwidth]{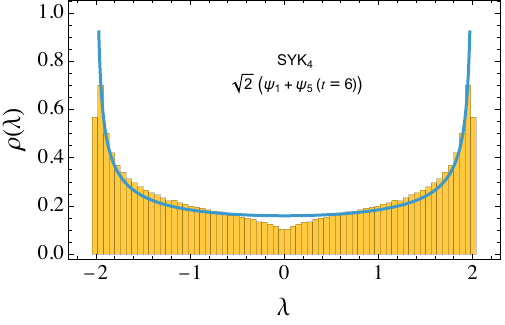}\\
        %(a) Caption for figure 1 & (b) Caption for figure 2 \\
        \includegraphics[width=0.45\textwidth]{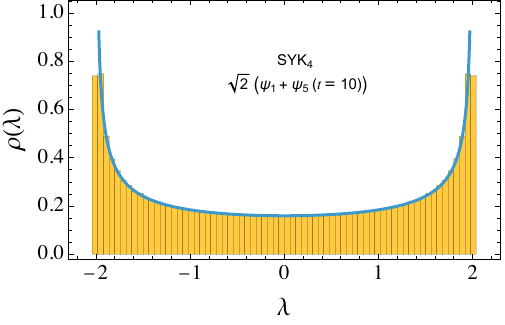}&
        \includegraphics[width=0.45\textwidth]{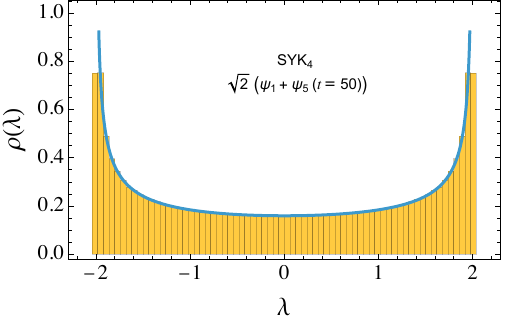}\\
        %(c) Caption for figure 3 & (d) Caption for figure 4 \\
    \end{tabular}
    \caption{Spectral statistics (eigenvalue density) of the operator $\sqrt{2}\left(\psi_{1}+\psi_{5}(t)\right)$ for $1000$ realizations of the SYK$_{4}$ Hamiltonian~\eqref{eq-SYK4}, without fixing a chirality sector, with $N=16$ fermions for different times, indicated in each individual figure. The bin size in each histogram is 0.06. 
 }
    \label{fig:OpStatSYK4Psi1Psi5} 
\end{figure}

Figures~\ref{fig:OpStatSYK4Psi1Psi1} and~\ref{fig:OpStatSYK4Psi1Psi5} show the behavior of the spectral statistics for two choices of operator, $\sqrt{2}\left(\psi_{1}+\psi_{1}(t)\right)$ and $\sqrt{2}\left(\psi_{1}+\psi_{5}(t)\right)$ for different times. In both cases, the arcsine law emerges for timescales $t\sim O(10^{1})$.

To illustrate integrable dynamics, we also consider the integrable $q=2$ SYK Hamiltonian:
\begin{equation}\label{eq-SYK2}
    H_{\text{SYK}_2} = \frac{i}{\sqrt{N}} \sum_{j<k} K_{jk} \, \psi_j \psi_k
\end{equation}
where \( K_{jk} \) is drawn from a Gaussian distribution with zero mean and standard deviation \( K \). In the chiral basis, the Hamiltonian \eqref{eq-SYK2} also takes a block diagonal form.

\begin{figure}[h!]
    \centering
    \begin{tabular}{cc}
    \includegraphics[width=0.45\linewidth]{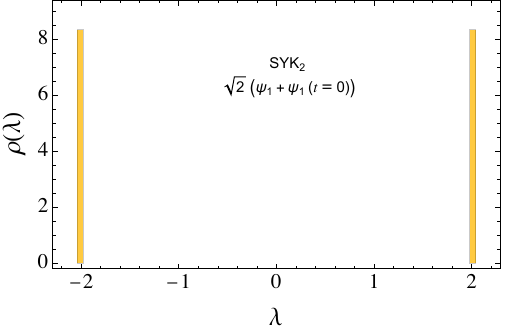}&
    \includegraphics[width=0.45\linewidth]{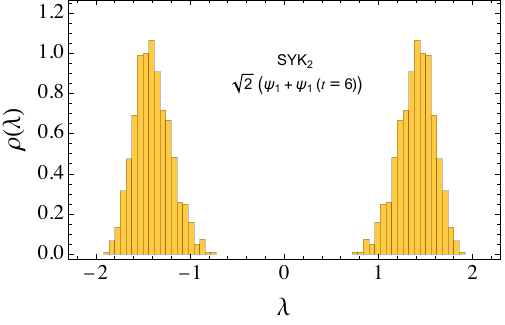}\\
      \includegraphics[width=0.45\linewidth]{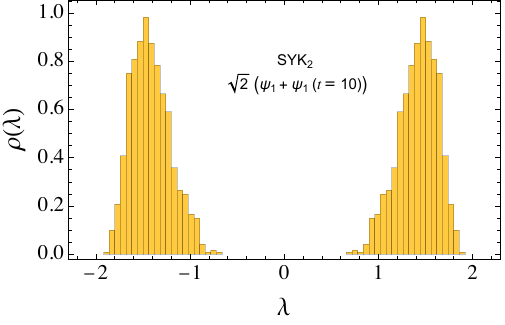}&
    \includegraphics[width=0.45\linewidth]{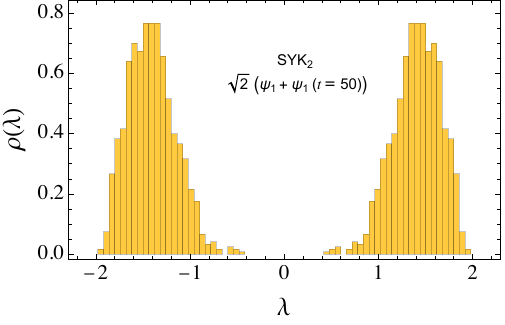}\\
    \end{tabular}
    \caption{Spectral statistics (eigenvalue density) of the operator $\sqrt{2}\left(\psi_{1}+\psi_{1}(t)\right)$ for $1000$ realizations of the SYK$_{2}$ Hamiltonian~\eqref{eq-SYK2}, without fixing a chirality sector, with $N=16$ fermions for different times, indicated in each individual figure. The bin size in each histogram is 0.06. The acrsine distribution does not form, at least within timescales up to order $t\sim O(10^{6})$.}
    \label{fig:OpStatSYK2Psi1Psi1}
\end{figure}

\begin{figure}[h!]
    \centering
    \begin{tabular}{cc}
    \includegraphics[width=0.45\linewidth]{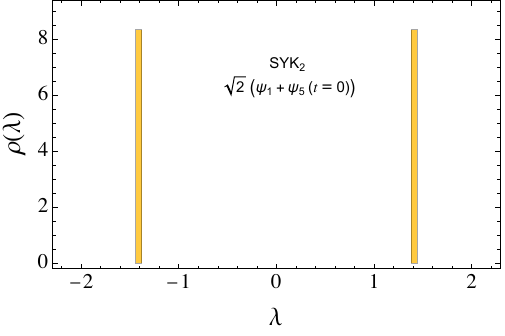}&
    \includegraphics[width=0.45\linewidth]{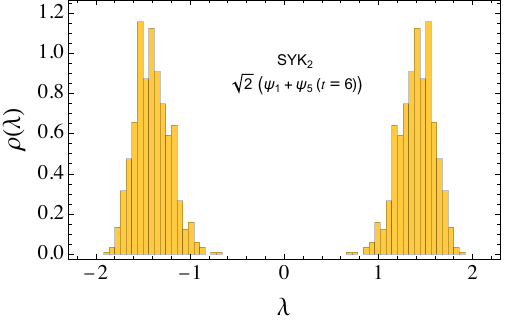}\\
    \includegraphics[width=0.45\linewidth]{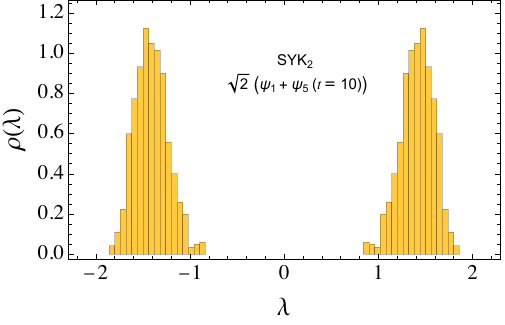}&
    \includegraphics[width=0.45\linewidth]{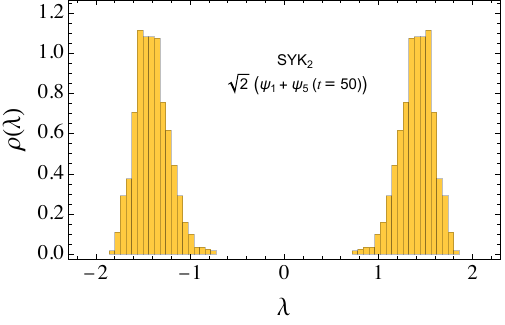}\\
    \end{tabular}
    \caption{Spectral statistics (eigenvalue density) of the operator $\sqrt{2}\left(\psi_{1}+\psi_{5}(t)\right)$ for $1000$ realizations of the SYK$_{2}$ Hamiltonian~\eqref{eq-SYK2}, without fixing a chirality sector, with $N=16$ fermions for different times, indicated in each individual figure.The bin size in each histogram is 0.06. The arcsine distribution does not form, at least within timescales up to order $t\sim O(10^{6})$.}
    \label{fig:OpStatSYK2Psi1Psi5}
\end{figure}

Figures~\ref{fig:OpStatSYK2Psi1Psi1} and~\ref{fig:OpStatSYK2Psi1Psi5} show the behavior of the spectral statistics for the two choices of the operator as in the SYK$_{4}$ case above, $\sqrt{2}\left(\psi_{1}+\psi_{1}(t)\right)$ and $\sqrt{2}\left(\psi_{1}+\psi_{5}(t)\right)$, for different times. In both cases, the arcsine law never emerges, at least within timescales up to order $t\sim O(10^{6})$.

%%%%%%%%%%%%%%%%%%%%%%%%%%%%%%%%%%%%%%%%%%%%%%%%%%%%%%%%%%%
%%%%%%%%%%%%%%%%%%%%%%%%%%%%%%%%%%%%%%%%%%%%%%%%%%%%%%%%%%%
%%%%%%%%%%%%%%%%%%%%%%%%%%%%%%%%%%%%%%%%%%%%%%%%%%%%%%%%%%%

\section{Conclusions and discussions } \label{sec-discussion}

In this work, we argue that free probability theory provides the correct mathematical framework to describe quantum chaos. Beyond offering new tools to characterize quantum chaotic behavior, it also serves as a bridge connecting different manifestations of chaos in quantum systems.

We began by reviewing classical and quantum ergodic theories, emphasizing the probability spaces on which these theories are based. We then explored how different degrees of chaoticity can be quantified through the extent to which future operators, denoted by $B(t)$, become independent of past operators, $A(0)$. In free probability theory, statistical independence between two operators is encapsulated by the concept of freeness, which is defined by the vanishing of all mixed free cumulants involving $A(0)$ and $B(t)$ (see Eq. \eqref{eq-DefFreeness}). This property implies that mixed correlators involving $A(0)$ and $B(t)$ can be factorized in terms of products of their individual moments (see Sec. \ref{subsec-freeness}).

Strictly speaking, freeness can only be realized for operators belonging to type II or type III von Neumann algebras, which may emerge in the thermodynamic limit, $\mathcal{D} \to \infty$, where $\mathcal{D}$ denotes the matrix rank of the operators in the algebra or, equivalently, the dimension of the Hilbert space. For finite-dimensional systems, the algebra of operators is always type I, which does not allow strict mixing. In such cases, mixed free cumulants in chaotic systems do not vanish at large times but instead approach a small non-zero value that decreases with increasing $\mathcal{D}$. This leads to the notion of asymptotic freeness, where the free cumulants vanish only in the limit $\mathcal{D} \to \infty$.

The spectral properties of asymptotically free operators provide a useful probe for quantum chaos. In particular, free probability theory predicts the eigenvalue distributions of combinations of asymptotically free operators. This prediction serves as a benchmark for testing the emergence of approximate asymptotic freeness in finite-dimensional quantum many-body systems under time evolution. Consequently, the spectral statistics of operators can be leveraged to test for asymptotic freeness.

Given two deterministic matrices $A$ and $B$, and a Haar-random unitary matrix $U$, it can be shown that $A$ and $U^\dagger B U$ are asymptotically free in the large-$\mathcal{D}$ limit with respect to the map $\varphi_{\mathcal{D}}(\cdot)$ in \eqref{eq-expectationValueRM}, where $\mathbb{E}$ denotes the appropriate ensemble average over the unitary $U$. Since a strongly chaotic dynamics can often be approximated by replacing the time evolution operator with a Haar-random unitary, it is natural to expect that, in realistic Hamiltonians appearing in quantum many-body physics, strong chaos would lead to $A(0)$ and $B(t)$ becoming asymptotically free with respect to the state $\varphi_{\mathcal{D}}(\cdot)$, with $\mathbb{E}$ now denoting an average over possibles random parameters in the Hamiltonian; or $\varphi_{\mathcal{D}}(\cdot)= \Tr(\cdot)/\mathcal{D}$ if the Hamiltonian does not contain any random parameters. In this case, a smoking gun of asymptotic freeness is whether the spectral statistics of $A(0)+B(t)$ follow the free convolution prediction.

As can be seen from the Eq. \eqref{eq-DefFreeness}, the condition that defines two free operators depends on the map $\varphi(\cdot):\mathcal{A} \rightarrow \mathbb{C}$. Since, for a quantum system, this map essentially represents the state of the system, whether two quantum operators are free with respect to each other depends on this state. 
The primary quantity we have used in this paper as a signature of asymptotic freeness of two deterministic operators ($A$ and $B$) under time evolution, namely the spectral statistics of an operator of the form $A(0) + B(t)$, refers to a specific state of the quantum system for which the map acts as $\varphi(\cdot)=\varphi_{\mathcal{D}}(\cdot)$.\footnote{Note that the expansion 
of $G(z)$ (the Cauchy transformation of the eigenvalue density) in terms of the moments uses this map explicitly (see the discussion in footnote \ref{moments_gen}). In the context of random matrix ensembles, $\mathbb{E}$ refers to an average over the probability weight of the ensemble, whereas, in the present context of realistic quantum many-body systems, this is essentially an average over random parameters that might be present in the Hamiltonian. } For a generic quantum system, this map would essentially represent an expectation value taken with respect to the infinite temperature thermal density matrix. Therefore, the observation that for chaotic systems, the eigenvalue density of $A(0) + B(t)$ starts to closely mimic the free convolution prediction around scrambling time is a strong indication that $A(0)$ and $B(t)$ indeed become asymptotically free under time evolution with respect to the above map.\footnote{ We emphasize that, for a finite $\mathcal{D}$, the mixed free cumulants of $A(0)$ and $B(t)$ would not vanish at late times, but they are expected to reach a stationary residual value whose amplitude decreases as $\mathcal{D}$ is increased, indicating that $A(0)$ and $B(t)$ become free in the limit $\mathcal{D} \rightarrow \infty$.}

The above discussion raises the question of whether we can test 
asymptotic freeness with respect to some other quantum state, say a finite temperature density matrix, in a similar way. As argued above, since the free probability prediction of the eigenvalue distribution of an operator of the form $A(0) + B(t)$ makes use of a specific instance of the map $\varphi(\cdot)$, in our opinion the eigenvalue distribution of $A(0) + B(t)$ is not a suitable quantity for that purpose. However, it would be interesting to look for other measures that might be useful in that context. In appendix \ref{app:finiteTemperature}, we elaborate on possible approaches to extend our results to finite temperature.

In the rest of this work, we performed a systematic study of the spectral statistics of the eigenvalues of operators of the form $A(0) + B(t)$ in various quantum many-body systems, including the mixed-field Ising model with a random magnetic field, a higher-spin generalization of this model, and the SYK model. We analyzed the timescale at which $A(0)$ and $B(t)$ become approximately free by comparing the eigenvalue distribution of $A(0) + B(t)$ with the predictions of free probability theory, which we also obtain in closed analytic form. Additionally, we investigated fluctuations on top of the free probability prediction by studying the level-spacing statistics of the eigenvalues of $A(0) + B(t)$. We now present a discussion of our results, contrasting the behavior in chaotic, near-integrable, and integrable regimes in the sense of nearest-neighbor level spacing statistics.

\paragraph{Chaotic cases.}  
For both the spin-1/2 and spin-1 Ising models with a random magnetic field, the eigenvalue density of operators of the form \( A(0) + B(t) \) follows the free convolution prediction at times around the scrambling time. While this prediction accurately describes the eigenvalue density for some operators, others exhibit small deviations, particularly near the edges of the spectrum. A possible explanation for these deviations lies in the nature of spectral correlations in these models. They exhibit random matrix behavior in terms of nearest-neighbor level spacing statistics, which primarily capture short-range correlations between adjacent energy levels. However, full random matrix behavior, which ensures the emergence of free probability spectral prediction, requires more than just nearest-neighbor statistics. If the system deviates from random matrix behavior at longer energy level correlations, this could account for the observed discrepancies. To quantify these deviations and their impact on the emergence of free probability predictions, it would be interesting to analyze the \( k \)-th level spacing statistics~\cite{Shir:2023olc} in these models and compare them with the corresponding random matrix expectations.

In the SYK\(_4\) model, the spectra of operators of the form \(\psi_i(0) + \psi_j(t)\) are presented in Figures \ref{fig:OpStatSYK4Psi1Psi1} and \ref{fig:OpStatSYK4Psi1Psi5} for the cases \(i = j = 1\) and \(i = 1, j = 5\), respectively. In both cases, the eigenvalue density converges to the arcsine law, the free probability prediction in this context, for times around \(t \approx 10\). This indicates that the operators \(\psi_i(0)\) and \(\psi_j(t)\) become asymptotically free on this timescale. Notably, unlike the spin chain cases -- where certain operators exhibit slight deviations from the free probability prediction -- the SYK\(_4\) model shows no such discrepancies. This confirms the expectation that the SYK\(_4\) model is more chaotic than the spin chain models considered in this work.

The emergence of the free probability prediction in the SYK\(_4\) model at finite \(N\) suggests that the model satisfies the F-property in the limit \(N \to \infty\). Since the F-property implies mixing, which typically requires a type II or III algebra, this provides strong evidence that the corresponding algebra of observables becomes type II or III in this limit, in agreement with expectations from the literature \cite{Leutheusser:2021frk}.

\paragraph{Near-integrable cases.}  
Previous works suggest that the free convolution prediction for the spectrum of operators of the form \( A + B(t) \) holds even in systems near integrability~\cite{Chen:2024zfj}, albeit emerging at significantly longer times compared to fully chaotic cases. However, increasing the statistical sampling reveals deviations from the free convolution prediction for the \( X \) Pauli operators, whereas no such deviations are observed for the \( Z \) Pauli operators (see Sec.~\ref{sec-fluctuationsSpinHalf}). This suggests that, near the integrable point, \( Z_i(0) \) and \( Z_j(t) \) remain asymptotically free at sufficiently large times, while \( X_i(0) \) and \( X_j(t) \) do not. Such an operator-dependent asymptotic freeness could be a distinctive feature of near-integrable systems. In the next subsection, we outline a strategy to investigate whether this is indeed the case.

For the mixed-field Ising model, we also investigate how the timescale at which the arcsine law emerges varies as the system is driven from the chaotic regime toward integrability. More specifically, we analyze how the arcsine time depends on the average \( r \)-parameter, with the results presented in Figure~\ref{fig:arcsinetimeversusrparameter}. We observe that the arcsine time increases as \( r \) decreases from its GOE value of approximately 0.53, eventually diverging as the system approaches integrability. Indeed, we verified that at the precisely integrable point, the arcsine law never emerges, regardless of how large we set the time. 

\paragraph{Integrable cases.}
In SYK$_2$, the spectra of operators of the form $\psi_i(0) + \psi_j(t)$ are shown in Figures \ref{fig:OpStatSYK2Psi1Psi1} and \ref{fig:OpStatSYK2Psi1Psi5} for the cases $i = j = 1$ and $i = 1, j = 5$, respectively. We observe that, regardless of how large the time parameter is, the arcsine law never emerges.  As observed before, the same happens at the integrable point of the mixed-field Ising model.

\paragraph{Fluctuations and symmetry resolution.} In the spin-chain models we considered, the random magnetic field breaks parity symmetry, leaving energy as the only conserved quantity. Consequently, there is no need to account for distinct symmetry sectors. Symmetry resolution techniques are typically employed when analyzing spectral fluctuations, which suggests that in systems with conserved charges, symmetry resolution plays a crucial role in accurately capturing fluctuations beyond the predictions of free probability. However, it seems to be less relevant for determining whether the eigenvalue density follows free probability predictions. 

For spin chains, we analyzed the \( r \)-parameter statistics associated with spectral fluctuations and found that they are well described by the Wigner-Dyson distribution with either \(\nu = 1\) or \(\nu = 2\). Specifically, for the sum of operators at different sites, we observe a GUE-type distribution (\(\nu = 2\)), whereas for operators acting on the same site, the statistics follow a GOE-type distribution (\(\nu = 1\)). The emergence of GUE-type statistics in the former case can be attributed to the breaking of time-reversal symmetry\footnote{For operators acting on the same site, the time-reversal symmetry operator \(\mathcal{T}\) acts as \(\mathcal{T}(X_i(0) + X_i(t)) = X_i(0) + X_i(-t)\). Moreover, invariance under time translations allows us to shift the time argument in both terms by \(t\), leading to  
\(X_i(0) + X_i(-t) = X_i(t) + X_i(0) = \mathcal{T}(X_i(0) + X_i(t))\).  
Since this relation does not hold for operators at different sites, time-reversal symmetry is effectively broken in that case, resulting in the observed GUE statistics. We thank Pratik Nandy for insightful discussions on this matter.}.

For the SYK model in the chiral basis, the Hamiltonian takes a block-diagonal form, corresponding to the two charge-parity sectors of the model. The analysis of fluctuations in this case should account for these two symmetry sectors. Operators of the form $\psi_i(0) + \psi_j(t)$ take an off-diagonal block form, with each block presumably corresponding to one of the charge-parity sectors. However, each block results in a matrix that is not Hermitian. It is not yet entirely clear to us how to perform the level spacing or $r$-parameter statistics in this case, so we leave this for future work. 

\paragraph{Relation between freeness and random matrix universality}

Our analysis of decorrelated ensembles in subsection~\ref{sec:decorrelatedEnsembles}, along with the emergence of freeness near the integrable point in spin chains, suggests that freeness can arise even without random matrix universality. In such cases, however, it appears in an operator-dependent manner -- some operators become free from their past versions or other past operators, while others, particularly those linked to conserved charges, do not. Since freeness essentially captures the notion of strong operator growth or scrambling, this suggests that while random matrix universality implies scrambling, the converse is not true: scrambling can occur without random matrix behavior, albeit in an operator-dependent way. This is consistent with the findings of \cite{Dowling:2023hqc}, which state that scrambling is a necessary but not sufficient condition for random matrix chaos. Here we emphasize that by scrambling we mean the vanishing of OTOCs and their higher-order counterparts, not any possible exponential behavior. 

\paragraph{Previous Works Connecting Different Notions of Quantum Chaos.}  
Finally, we comment on previous works that connect different notions of quantum chaos. In~\cite{Cotler:2017jue}, the authors establish a connection between correlators involving Haar random unitary operators and a generalized version of the spectral form factor (see Appendix \ref{app-averageOTOC} for a brief review). Specifically, they show that higher-order OTOCs involving Haar random unitaries are related to the $k$-th power of the spectral form factor, thus linking the vanishing of OTOCs with spectral statistics exhibiting random matrix behavior. In contrast, the concept of freeness between operators \( A(0) \) and \( B(t) \) encompasses the vanishing of all mixed free cumulants, implying the vanishing of two-point functions, OTOCs, and higher-order out-of-time-ordered correlators for traceless operators. If the time evolution operator is modeled by a Haar random unitary \( U \), one can show that \( A(0) \) and \( U^\dagger B U \) are asymptotically free, implying the vanishing of all mixed free cumulants. In this case, an ensemble average is taken in the definition of the map $\varphi(A) = \lim_{\mathcal{D} \rightarrow \infty} \mathbb{E}\left( \frac{\Tr A}{\mathcal{D}} \right)$.

Another relevant work connecting OTOCs with operator spectral statistics is~\cite{rozenbaum2019universal}, which argues that in chaotic systems, out-of-time-order operators (OTOOs) of the form \( A(0)B(t)A(0)B(t) + B(t)A(0)B(t)A(0) \) display universal level spacing statistics. In free probability, the fact that \( A(0) \) and \( B(t) \) become asymptotically free under chaotic dynamics should allow one to obtain the free probability prediction for the eigenvalue density of the OTOO mentioned above. In the next section, we elaborate on how to proceed in this direction of research.

\subsection{Future directions}
This work can be extended in several interesting directions, some of which we now briefly discuss.

\begin{itemize}
    \item In this work, we have examined the free convolution prediction for the sum of two free operators, focusing on spin-1/2 and spin-1 operators. While these results, specially, the spin-1/2 case, apply to a broad class of quantum mechanical systems, including spin chains and SYK-like models, it is crucial to extend the free convolution prediction to sums of other types of operators that naturally arise in quantum mechanics. In Sec.~\ref{sec-numFreeConvolution}, we numerically determine the free additive convolution for the sum of two free spin-3/2, spin-2, spin-5/2, as well as large spin operators as a first step in this direction. However, many important cases remain unexplored, such as bosonic operators in, say, the Bose-Hubbard model, and the position $x$ and momentum $p$ operators in quantum billiard systems, and numerous others. A systematic classification of free convolution predictions for different classes of operators would be highly valuable for future studies on quantum chaos and asymptotic freeness in quantum mechanics. 

    \item The approximate validity of the free convolution prediction for systems near integrability, when considering the \( Z \) Pauli operators, suggests that even a slight breaking of integrability could lead to mixing in the thermodynamic limit for cumulants involving these operators. A natural way to investigate this would be to analyze the scaling of the residual value of mixed free cumulants with \( N \) in systems with an intermediate degree of chaos. This would extend the numerical studies of \cite{Huang:2017fng}, which focused on strongly chaotic systems, to such intermediate cases.

    \item In this paper we have considered spectral properties of the sum of two non-commutative operators (say $A$ and $B$). However, one can similarly consider the spectral properties of the products of two such operators, as well as their different linear combinations. If these variables are traceless, i.e., $\varphi(A)=\varphi(B)=0$, then their free product is trivial since the trace of a combination of them vanishes from the definition freeness, $\varphi \big((AB)^k\big)=\varphi\big(ABAB\cdots AB\big)=0$. For variables with non-zero trace, one can obtain the free multiplicative convolution of these variables by using an analogue of the $R$-transform, known as the $S$-transform, which has the property that for a pair of free variables, it is multiplicative \cite{voiculescu1987multiplication}. This allows us to compute the free multiplicative convolution of the eigenvalue density of these two operators and obtain the eigenvalue density of the product by using the rule $S_{\rho_{A} \boxtimes \rho_{B}}(x)=S_{\rho_A}(x) S_{\rho_B}(x) $. A particular set of combinations of addition and multiplication of two non-commutative variables that are relevant for our purposes is the squared commutator  $(-i)^2 \big[A(t), B(0)\big]^2$ and its higher-order generalizations. If $A(t)$ and $B(0)$ are asymptotically free, a prediction for the spectral statistics of this operator 
would help us to understand, in precise terms, the connection between previous proposals regarding the spectral statistics of out-of-time-order operators and the Lyapunovian operator~\cite{rozenbaum2019universal}\footnote{We have reviewed this proposal briefly in appendix \ref{app-Lyapunovian}.}, and the emergence of asymptotic freeness in chaotic quantum many-body dynamics that we have advocated here.\footnote{We also note here that, it is known in the free probability literature how to obtain the eigenvalue density of the commutator of two operators from the 
individual eigenvalue density \cite{nica1998commutators}. One can follow such an approach to derive the free probability prediction for the convolved distribution of the commutator of, say, two spin-$1/2$ operators and check whether it matches the results from numerical simulations for interacting quantum many-body systems.  }

\item It would be interesting to investigate the role of freeness in quantum field theories (QFTs). Notably, freeness imposes nontrivial constraints on these theories. In two-dimensional integrable conformal field theories (CFTs), for instance, two-point functions exhibit universal behavior, always decaying to zero as \( t \to \infty \). However, four-point OTOCs depend on the choice of operators: they remain constant when involving only energy operators or a combination of energy and spin operators. In contrast, chaotic CFTs display a more uniform decay, where both two-point functions and OTOCs tend to zero for almost any choice of operators, and higher-order OTOCs are also expected to vanish at late times (see Appendix B of~\cite{Roberts:2014ifa}).  

Moreover, it would be valuable to study operator spectral statistics in interacting QFTs using light-cone truncation methods~\cite{Delacretaz:2022ojg}.

   \item Exploring the emergence of stronger notions of freeness, such as second-order freeness~\cite{collins2024second}, under chaotic dynamics is also an interesting direction for future research.

 \item Recently, a new manifestation of quantum chaotic behavior was identified in the BPS sectors of supersymmetric theories, a phenomenon referred to as ``BPS chaos"~\cite{Chen:2024oqv}. The procedure begins with an operator \( \mathcal{O} \) in the UV theory, which is then projected into a BPS sector via \( \mathcal{O}_\text{BPS} = P_\text{BPS} \mathcal{O} P_\text{BPS} \), where \( P_\text{BPS} \) is a projection operator, morally taking the form \( P_\text{BPS} = \lim_{\beta \rightarrow \infty} e^{-\beta H} \). For BPS sectors dual to black hole geometries, one expects to observe random matrix behavior upon studying the spectral properties of \( \mathcal{O}_\text{BPS} \). It would be interesting to investigate whether this new manifestation of quantum chaos can be derived from free probability, as a consequence of the free independence between the operator \( \mathcal{O} \) and the projector \( P_\text{BPS} \).

\

\end{itemize}

We look forward to reporting on these developments in future work.

\begin{comment}

\end{comment}

\acknowledgments
We would like to thank Antonio Garcia-Garcia, Ping Gao, Zhenbin Yang, Neil Dowling, and Silvia Pappalardi for valuable discussions. We are also grateful to Pratik Nandy, Tanay Pathak, Rathindra Nath Das, Sergio E. Aguilar-Gutierrez, and Leo Shaposhnik for useful discussions and comments on the draft. In particular, we thank Pratik Nandy for carefully checking many of the numerical calculations. We also thank Heide Narnhofer for helpful correspondence. The authors thank the Yukawa Institute for Theoretical Physics at Kyoto University, where this work was presented and discussed during the YITP workshop {\it Quantum Gravity and Information in an Expanding Universe} (YITP-W-24-19).
This work was supported by the Basic Science Research Program through the National Research Foundation of Korea (NRF) funded by the Ministry of Science, ICT $\&$ Future Planning (NRF-2021R1A2C1006791),  by the Ministry of Education (NRF-2020R1I1A2054376) and the AI-based GIST Research Scientist Project grant funded by the GIST in 2025.
This work was also supported by Creation of the Quantum Information Science R$\&$D Ecosystem (Grant No. 2022M3H3A106307411) through the National Research Foundation of Korea (NRF) funded by the Korean government (Ministry of Science and ICT). H.~A. Camargo, and V.~Jahnke were supported by the Basic Science Research Program through the National Research Foundation of Korea (NRF) funded by the Ministry of Education (NRF-2022R1I1A1A01070589, and RS-2023-00248186). H. Camargo, Y. Fu, V. Jahnke, and K. Pal should be recognized as co-first authors.

\appendix

\section{Cumulants from partitions} \label{sec-appA}

In this appendix we provide a brief introduction to the  free cumulants associated with non-commutating random variables, starting from the standard
definition of the classical cumulants of commutating random variables. We discuss the diagrammatic method based on crossing and non-crossing partitions of a set to obtain the classical and free cumulants, which has been
used recently to connect the eigenstate thermalization hypothesis and the dynamics of fluctuations of coherence in noisy mesoscopic systems with the free probability theory \cite{Pappalardi:2022aaz, Hruza:2022uqj}. For more detail on the combinatorics in free probabilities, we refer to  \cite{speicher2009free, speicher2019lecture, mingo2017free, nica2006lectures, Potters}. 

Consider an algebra of random variables $\mathcal{R}$, and a positive faithful linear functional $\varphi(\cdot)$ which maps $\mathcal{R} \rightarrow \mathbb{C}$. 
Usual examples of this functional include the expectation operator ($\mathbb{E}[\cdot]$) in  standard probability theory, or the (normalized) trace operator $\Tr(\cdot)/N$
for an algebra of matrices, or it could be a combination of these two operations, $\mathbb{E}[\text{Tr}(\cdot)]/N$.

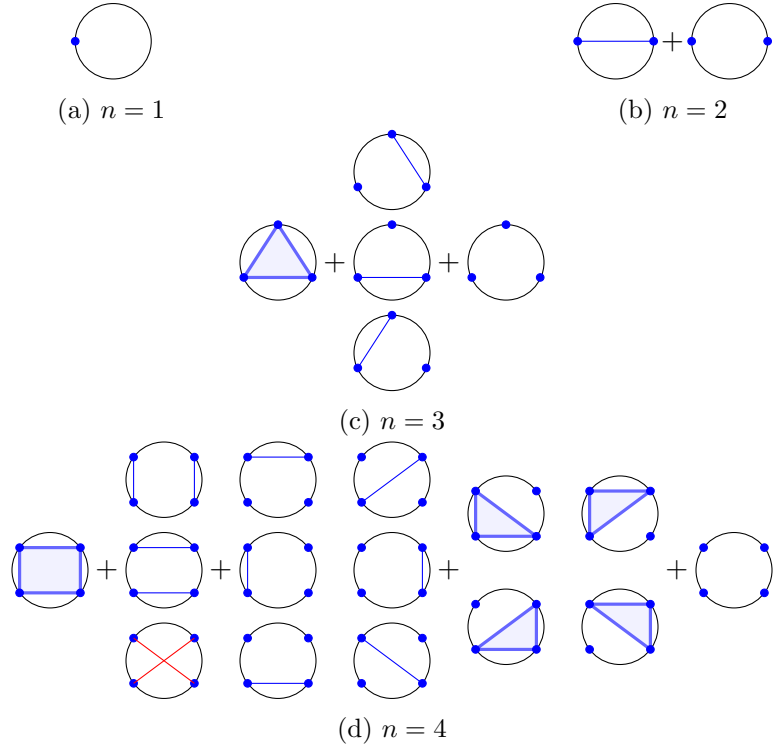
\begin{figure}
    \centering
    \begin{subfigure}[b]{0.48\textwidth}
    \centering
    \begin{tikzpicture}
        \draw (0,0) circle(0.5);
        \draw[fill=blue,blue](-0.5,0) circle [radius=0.05];
    \end{tikzpicture}
    \caption{$n=1$}
    \label{fig: partition_diag_n1}
    \end{subfigure}
\begin{subfigure}[b]{0.48\textwidth}
    \centering
    \begin{tikzpicture}
        \draw (-0.5,0) circle(0.5);
        \draw[fill=blue,blue](-1,0) circle [radius=0.05];
        \draw[fill=blue,blue](0,0) circle [radius=0.05];
        \draw[blue] (-1,0)--(0,0);
        \node at (0.25,0){$+$};
        \draw (1,0) circle(0.5);
        \draw[fill=blue,blue](0.5,0) circle [radius=0.05];
        \draw[fill=blue,blue](1.5,0) circle [radius=0.05];
    \end{tikzpicture}
    \caption{$n=2$}
    \label{fig: partition_diag_n2}
    \end{subfigure}
    \begin{subfigure}[b]{\textwidth}
    \centering
    \begin{tikzpicture}
        \draw (-0.5,0) circle(0.5);
        \filldraw[color=blue!60, fill=blue!5, very thick](-0.95,-0.2) -- (-0.05,-0.2) -- (-0.5,0.5) -- cycle;
        \draw[fill=blue,blue](-0.95,-0.2) circle [radius=0.05];
        \draw[fill=blue,blue](-0.05,-0.2) circle [radius=0.05];
        \draw[fill=blue,blue](-0.5,0.5) circle [radius=0.05];

        \node at (0.25,0){$+$};
        
        \draw (1,0) circle(0.5);
        \draw[fill=blue,blue](1.45,-0.2) circle [radius=0.05];
        \draw[fill=blue,blue](0.55,-0.2) circle [radius=0.05];
        \draw[fill=blue,blue](1,0.5) circle [radius=0.05];
        \draw[blue] (1.45,-0.2)--(0.55,-0.2);

        \draw (1,1.2) circle(0.5);
        \draw[fill=blue,blue](1.45,1) circle [radius=0.05];
        \draw[fill=blue,blue](0.55,1) circle [radius=0.05];
        \draw[fill=blue,blue](1,1.7) circle [radius=0.05];
        \draw[blue] (1.45,1)--(1,1.7);

        \draw (1,-1.2) circle(0.5);
        \draw[fill=blue,blue](1.45,-1.4) circle [radius=0.05];
        \draw[fill=blue,blue](0.55,-1.4) circle [radius=0.05];
        \draw[fill=blue,blue](1,-0.7) circle [radius=0.05];
         \draw[blue] (0.55,-1.4)--(1,-0.7);
         
         \node at (1.75,0){$+$};

         \draw (2.5,0) circle(0.5);
        \draw[fill=blue,blue](2.95,-0.2) circle [radius=0.05];
        \draw[fill=blue,blue](2.05,-0.2) circle [radius=0.05];
        \draw[fill=blue,blue](2.5,0.5) circle [radius=0.05];
    \end{tikzpicture}
    \caption{$n=3$}
    \label{fig: partition_diag_n3}
    \end{subfigure}
    
    \begin{subfigure}[b]{\textwidth}
    \centering
        \begin{tikzpicture}
        \centering
        \draw (-0.5,0) circle(0.5);
        \filldraw[color=blue!60, fill=blue!5, very thick](-0.9,-0.3) rectangle (-0.1,0.3);
        \draw[fill=blue,blue](-0.9,-0.3) circle [radius=0.05];
        \draw[fill=blue,blue](-0.1,-0.3) circle [radius=0.05];
        \draw[fill=blue,blue](-0.9,0.3) circle [radius=0.05];
        \draw[fill=blue,blue](-0.1,0.3) circle [radius=0.05];

        \node at (0.25,0){$+$};
        
        \draw (1,0) circle(0.5);
        \draw[fill=blue,blue](0.6,-0.3) circle [radius=0.05];
        \draw[fill=blue,blue](1.4,-0.3) circle [radius=0.05];
        \draw[fill=blue,blue](0.6,0.3) circle [radius=0.05];
        \draw[fill=blue,blue](1.4,0.3) circle [radius=0.05];
        \draw[blue] (0.6,-0.3)--(1.4,-0.3);
        \draw[blue] (0.6,0.3)--(1.4,0.3);

        \draw (1,1.2) circle(0.5);
        \draw[fill=blue,blue](0.6,0.9) circle [radius=0.05];
        \draw[fill=blue,blue](1.4,0.9) circle [radius=0.05];
        \draw[fill=blue,blue](0.6,1.5) circle [radius=0.05];
        \draw[fill=blue,blue](1.4,1.5) circle [radius=0.05];
        \draw[blue] (0.6,1.5)--(0.6,0.9);
        \draw[blue] (1.4,0.9)--(1.4,1.5);

        \draw (1,-1.2) circle(0.5);
        \draw[fill=blue,blue](0.6,-1.5) circle [radius=0.05];
        \draw[fill=blue,blue](1.4,-1.5) circle [radius=0.05];
        \draw[fill=blue,blue](0.6,-0.9) circle [radius=0.05];
        \draw[fill=blue,blue](1.4,-0.9) circle [radius=0.05];
         \draw[red] (0.6,-1.5)--(1.4,-0.9);
         \draw[red] (1.4,-1.5)--(0.6,-0.9);
         
         \node at (1.75,0){$+$};

         \draw (2.5,0) circle(0.5);
        \draw[fill=blue,blue](2.1,-0.3) circle [radius=0.05];
        \draw[fill=blue,blue](2.9,-0.3) circle [radius=0.05];
        \draw[fill=blue,blue](2.1,0.3) circle [radius=0.05];
        \draw[fill=blue,blue](2.9,0.3) circle [radius=0.05];
        \draw[blue] (2.1,0.3)--(2.1,-0.3);

        \draw (2.5,1.2) circle(0.5);
        \draw[fill=blue,blue](2.1,0.9) circle [radius=0.05];
        \draw[fill=blue,blue](2.9,0.9) circle [radius=0.05];
        \draw[fill=blue,blue](2.1,1.5) circle [radius=0.05];
        \draw[fill=blue,blue](2.9,1.5) circle [radius=0.05];
        \draw[blue] (2.9,1.5)--(2.1,1.5);

        \draw (2.5,-1.2) circle(0.5);
        \draw[fill=blue,blue](2.1,-1.5) circle [radius=0.05];
        \draw[fill=blue,blue](2.9,-1.5) circle [radius=0.05];
        \draw[fill=blue,blue](2.1,-0.9) circle [radius=0.05];
        \draw[fill=blue,blue](2.9,-0.9) circle [radius=0.05];
        \draw[blue] (2.9,-1.5)--(2.1,-1.5);

        \draw (4,0) circle(0.5);
        \draw[fill=blue,blue](3.6,-0.3) circle [radius=0.05];
        \draw[fill=blue,blue](4.4,-0.3) circle [radius=0.05];
        \draw[fill=blue,blue](3.6,0.3) circle [radius=0.05];
        \draw[fill=blue,blue](4.4,0.3) circle [radius=0.05];
        \draw[blue] (4.4,-0.3)--(4.4,0.3);

        \draw (4,1.2) circle(0.5);
        \draw[fill=blue,blue](3.6,0.9) circle [radius=0.05];
        \draw[fill=blue,blue](4.4,0.9) circle [radius=0.05];
        \draw[fill=blue,blue](3.6,1.5) circle [radius=0.05];
        \draw[fill=blue,blue](4.4,1.5) circle [radius=0.05];
        \draw[blue] (4.4,1.5)--(3.6,0.9);

        \draw (4,-1.2) circle(0.5);
        \draw[fill=blue,blue](3.6,-1.5) circle [radius=0.05];
        \draw[fill=blue,blue](4.4,-1.5) circle [radius=0.05];
        \draw[fill=blue,blue](3.6,-0.9) circle [radius=0.05];
        \draw[fill=blue,blue](4.4,-0.9) circle [radius=0.05];
        \draw[blue] (3.6,-0.9)--(4.4,-1.5);

         \node at (4.75,0){$+$};

        \draw (5.5,0.75) circle(0.5);
        \filldraw[color=blue!60, fill=blue!5, very thick](5.1,1.05) -- (5.1,0.45) -- (5.9,0.45) -- cycle;
        \draw[fill=blue,blue](5.1,0.45) circle [radius=0.05];
        \draw[fill=blue,blue](5.9,0.45) circle [radius=0.05];
        \draw[fill=blue,blue](5.1,1.05) circle [radius=0.05];
        \draw[fill=blue,blue](5.9,1.05) circle [radius=0.05];

        \draw (5.5,-0.75) circle(0.5);
        \filldraw[color=blue!60, fill=blue!5, very thick](5.9,-0.45) -- (5.9,-1.05) -- (5.1,-1.05) -- cycle;
        \draw[fill=blue,blue](5.1,-1.05) circle [radius=0.05];
        \draw[fill=blue,blue](5.9,-1.05) circle [radius=0.05];
        \draw[fill=blue,blue](5.1,-0.45) circle [radius=0.05];
        \draw[fill=blue,blue](5.9,-0.45) circle [radius=0.05];

        \draw (7,0.75) circle(0.5);
        \filldraw[color=blue!60, fill=blue!5, very thick](6.6,0.45) -- (7.4,1.05) -- (6.6,1.05) -- cycle;
        \draw[fill=blue,blue](6.6,0.45) circle [radius=0.05];
        \draw[fill=blue,blue](7.4,0.45) circle [radius=0.05];
        \draw[fill=blue,blue](6.6,1.05) circle [radius=0.05];
        \draw[fill=blue,blue](7.4,1.05) circle [radius=0.05];

        \draw (7,-0.75) circle(0.5);
        \filldraw[color=blue!60, fill=blue!5, very thick](6.6,-0.45) -- (7.4,-0.45) -- (7.4,-1.05) -- cycle;
        \draw[fill=blue,blue](6.6,-1.05) circle [radius=0.05];
        \draw[fill=blue,blue](7.4,-1.05) circle [radius=0.05];
        \draw[fill=blue,blue](6.6,-0.45) circle [radius=0.05];
        \draw[fill=blue,blue](7.4,-0.45) circle [radius=0.05];

         \node at (7.75,0){$+$};

         \draw (8.5,0) circle(0.5);
        \draw[fill=blue,blue](8.1,-0.3) circle [radius=0.05];
        \draw[fill=blue,blue](8.9,-0.3) circle [radius=0.05];
        \draw[fill=blue,blue](8.1,0.3) circle [radius=0.05];
        \draw[fill=blue,blue](8.9,0.3) circle [radius=0.05];
    
    \end{tikzpicture}
    \caption{$n=4$}
    \label{fig: partition_diag_n4}
    \end{subfigure}
    \caption{Diagrams for partitions of a set used to obtain the classical and free cumulants. All the partitions contribute to classical cumulants, while only the non-crossing partitions contribute to free cumulants. Crossing partitions start to appear from $n=4$ onward (distinguished above from the non-crossing ones by red lines).}
    \label{fig:partition_diag}
\end{figure}

\textbf{Commutating random variables: Classical cumulants.} We start by considering the cumulants of commutating random variables, known as the classical 
cumulants $c_n(x)$. These are essentially the connected correlation functions of the random variables of different orders and can be conveniently written as combinations
of  $n$-th order and lower order moments $m_n = \varphi(x^n)$.\footnote{Here the linear functional $\varphi$ is the usual expectation value with respect to a 
probability measure $\varphi(x^n)=\mathbb{E}(x^n)=\int x^n p(x) \text{d}x$.} One can obtain the relation between the first few moments and cumulants straightforwardly from their respective generating functions.  We start from the cumulant generating function $K(x)=\sum_n c_n x^n/n!$, which is related with the moment generating function $G(x)$ as 
\begin{equation}
    G(x)=e^{K(x)}=\sum_n m_n x^n/n!~.
\end{equation}
Taking successive derivatives of  the relation between the first few moments and cumulants
are given by 
\begin{align}\label{mn_cn}
    m_1 &= \frac{d}{dx}\big(e^{K(x)}\big)\Big|_{x=0}=c_1 \nonumber\\
    m_2 &= \frac{d^2}{dx^2}\big(e^{K(x)}\big)\Big|_{x=0}=e^{K(x)}\Big(K'(x)^2+K''(x)\Big)\Big|_{x=0}=c_1^2+c_2\nonumber\\
    m_3 &= \frac{d^3}{dx^3}\big(e^{K(x)}\big)\Big|_{x=0}=e^{K(x)}\Big(K'(x)^3+3 K'(x)K''(x)+K'''(x)\Big)\Big|_{x=0}=c_1^3+3c_1 c_2+c_3\nonumber\\
    m_4 &= \frac{d^4}{dx^4}\big(e^{K(x)}\big)\Big|_{x=0}=e^{K(x)}\Big(K'(x)^4+6 K'(x)^2K''(x)+4 K'(x)K'''(x)+3K''(x)^2 \nonumber\\
    &+K^{(4)}(x)\Big)\Big|_{x=0}=c_1^4+6c_1^2 c_2+4c_1c_3+3c_2^2+c_4~.
\end{align}
Inverting these relations we obtain the expressions for the cumulants 
in terms of the moments
\begin{align}
    c_1 &= m_1 \nonumber\\
    c_2 &=m_2-m_1^2 \nonumber\\
    c_3 &=m_3-3 m_2 m_1 + 2 m_1^3 \nonumber\\
    c_4 &=m_4- 4 m_1 m_3 + 12 m_2 m_1^2 -3 m_2^2 -6 m_1^4~.
\end{align}

These relations between the classical cumulants and the moments can be written
compactly using the concept of partitions. For a set $\{1, \cdots n\}$, a partition $\pi$ represents a decomposition of the set in terms of blocks that do not overlap 
while the union of all these blocks gives back the full set.
 The set of all partitions of  $\{1, 2, \cdots n\}$ is denoted as $P(n)$. We have shown the set of all partitions for $n=1,2,3,4$ in Figure~\ref{fig:partition_diag}. Note that for $n=1,2,3$, all the partitions are non-crossing, while for $n=4$, there is one crossing partition (shown in red). 
The total number of partitions (crossing as well as non-crossing) of a set with $n$ elements is known as the Bell number ($B_n$), which satisfies the recurrence relations $B_{n+1} = \sum_{k=0}^{n}  \Big( \begin{array}{c} n \\ k \end{array}\Big) B_k$. Then, the moments are related to the classical cumulants
through the formula
\begin{equation}\label{mn_to_cn}
	m_n(x) = \sum_{ \pi \in P(n)} c_\pi(x), ~~\text{where}~~c_\pi (x)= \prod_{b\in \pi} c_{|b|}(x)~. 
\end{equation}
According to this formula, the quantity $c_\pi (x)$ is defined as the product of the cumulants over the number of the elements of the block $b$ of a certain partition $\pi$. 
As an example, consulting the partitions for $n=4$ shown in Fig. \ref{fig:partition_diag} it can be deduced that the fourth order moment is given in terms of the classical cumulants as
$m_4(x)\equiv \mathbb{E}(x^4)= c_4(x)+ 3 c_2^2 (x) + 6 c_2(x) c_1^2(x)+ 4 c_3(x)c_1(x) + c_1^4(x)$, which is the same as 
the expression for $m_4$ we obtained in \eqref{mn_cn} by taking successive derivatives of the cumulant generating function. The formula in eq. \eqref{mn_to_cn} can be inverted for each $n$ to obtain the 
classical cumulants in terms of the moments.

Finally, for more than one random variable $(x_1 x_2 \cdots x_n)$, the formula connecting the moments and the cumulants can be generalized to obtain
\begin{equation}
	\mathbb{E}(x_1x_2\cdots x_n) = \sum_{ \pi \in P(n)} c_\pi(x_1x_2\cdots x_n)~, ~~\text{where}~~~c_\pi (x_1x_2\cdots x_n)
	= \prod_{b\in \pi} c_{|b|}\Big(x_{b(1)}x_{b(2)}\cdots x_{b(n)}\Big)~.
\end{equation}

\textbf{Non-commutating random variables: Free cumulants.}
Now we consider the cumulants for non-commutating random variables ($A_1, A_2, \cdots A_n$), known as the free cumulants ($\kappa_n$). To relate these with $\varphi(A^n)$, we need to consider the partitions of the set $\{1, 2, \cdots n\}$ which do not cross. While the total number of partitions of this set is given by the Bell
numbers, the number of non-crossing partitions ($NC(n)$) is given by the Catalan numbers $C_n= \frac{1}{n+1} \Big( \begin{array}{c} 2n \\ n \end{array}\Big)$,
 and the relation  between   $\varphi(A^n)$ and $\kappa_n$ of different order can be written as a sum over the non-crossing partitions in the following way,
 \begin{equation}\label{tau_to_kappa}
 	\varphi(A^n) =  \sum_{ \pi \in NC(n)} \kappa_\pi(A), ~~\text{where}~~~~\kappa_\pi (A)= \prod_{b\in \pi} \kappa_{|b|}(A)~.
 \end{equation}
 The expressions for the first four free cumulants are, therefore,
\begin{align}
    \kappa_1 &= \varphi(A) \nonumber\\
    \kappa_2 &=\varphi(A^2)-\varphi(A)^2 \nonumber\\
    \kappa_3 &=\varphi(A^3)-3 \varphi(A^2)\varphi(A) + 2 \varphi(A)^{3} \nonumber\\
    \kappa_4 &=\varphi(A^4) - 4 \varphi(A) \varphi(A^3)+ 10 \varphi(A^2) \varphi(A)^2 -2 \varphi(A^2)^2 -5 \varphi(A)^4~.
\end{align} 
 As can be expected from the partitions shown in Figure \ref{fig:partition_diag},   the difference between the classical and free cumulants starts to appear for $n=4$, so that we have 
the relation $\varphi(A^4)= \kappa_4(A)+ 2 \kappa_2^2 (A) + 6 \kappa_2(A) \kappa_1^2(A)+ 4 \kappa_3(A)\kappa_1(A) + \kappa_1^4(A)$.
For more than one random variable, we can extend the relation in eq. \eqref{tau_to_kappa} to 
\begin{equation}
\begin{split}
	\varphi(A_1 A_2 \cdots A_n) =  \sum_{ \pi \in NC(n)} \kappa_\pi(A_1, A_2, \cdots A_n)~,\\ \text{where}~~\kappa_\pi (A_1, A_2, \cdots A_n)= \prod_{b\in \pi} \kappa_{|b|}\Big(
	A_{b(1)}, A_{b(2)}, \cdots A_{b(n)}\Big)~.
\end{split}
\end{equation}
Here $b(I)$ denotes the different elements of the blocks of the partitions, $b=(b(1), b(2), \cdots b(n))$  and $|b|$ indicates its length.

\section{Details about the Level Spacing Analysis} 

The average density of states in a chaotic spectrum is typically dependent on the specific model and is non-universal. In contrast, it is conjectured that fluctuations around the average density of states can be accurately described by the random matrix theory corresponding to the appropriate symmetry class~\cite{BGS, BGS2}. Thus, it is crucial to isolate the spectral fluctuations from the average density of states. The standard method to achieve this is known as {\it unfolding}, which involves mapping the original energy levels $E_n$ to new levels $\epsilon_n$ that preserve the fluctuation properties while normalizing the average density of states to one. In this appendix, we provide a review of the unfolding process, using a chaotic spin chain as an illustrative example.

Consider the scenario where we directly diagonalize the Hamiltonian of a given system and obtain the energy spectrum $E_n$ within a particular symmetry sector. The initial step involves determining the average density of states, $\rho_\text{avg}(E)$.

After obtaining the average density of states, we compute the corresponding cumulative distribution function:
\begin{equation} 
    N_\text{avg}(E) = \int_{-\infty}^{E} \textrm{d}E' \rho_\text{avg}(E')\,. 
\end{equation}
The unfolding procedure is then defined by the mapping:
\begin{equation} 
    \varepsilon_n = N_\text{avg}(E_n)\,. 
\end{equation}
The average density of states of the unfolded spectrum is very close to one. The nearest-neighbor level spacing distribution of the unfolded spectrum can then be obtained by computing the spacings:
\begin{equation}
    s_i = \varepsilon_{i+1}-\varepsilon_{i} \approx \rho(E_i)(E_{i+1}-E_i)\,,
\end{equation}
and constructing a histogram of $s_i$. For a sufficiently large number of spacings, one can obtain a good estimate of the level spacing distribution, $P(s)$. When this distribution resembles those observed in random matrix ensembles, the system is considered 'quantum chaotic'. In contrast, for many integrable systems, the energy levels are uncorrelated and follow Poisson statistics~\cite{BerryTabor1977}.

The Wigner-Dyson distributions can be compactly written as:
\begin{equation}
    P_\nu(s) = a(\nu) s^\nu e^{-b(\nu) s^2}\,,
\end{equation}
where the constants $a(\nu)$ and $b(\nu)$ are determined by the conditions:
\begin{eqnarray}
    \int_{0}^{\infty}P_\nu(s)\textrm{d}s = 1\,,\,\,\,\,\,\, \int_{0}^{\infty}s\, P_\nu(s)\textrm{d}s = 1\,.
\end{eqnarray}
The index $\nu$ specifies the symmetry class of the random matrix ensemble, with $\nu=1$ for GOE, $\nu=2$ for GUE, and $\nu=4$ for GSE. In these cases, the formula takes the following form:
\begin{equation} \label{eq-P(s)wigner}
    P_\nu (s) = 
    \begin{cases}
        \frac{\pi}{2} s \,e^{-\pi s^2/4}, & \nu=1\,\,\text{(GOE)} \\
        \frac{32}{\pi^2} s^2 \, e^{-4 s^2/\pi}, &\nu=2\,\, \text{(GUE)} \\
        \frac{2^{18}}{3^6 \pi^3} s^4\, e^{-64 s^2/(9\pi)}, &\nu=4\,\, \text{(GSE)}
    \end{cases}  
\end{equation}
For a spectrum with uncorrelated energy levels, the level spacing distribution follows Poisson statistics:
\begin{equation}\label{eq-P(s)poisson}
    P_\text{Poisson}(s) = e^{-s}\,.
\end{equation}
One limitation of using the level spacing distribution to characterize random matrix behavior is the need to unfold the spectrum, which requires knowledge of the system's average density of states. In the cases considered in this paper, the average density of states follows either a Wigner semicircle distribution (for SYK$_4$) or a Gaussian distribution (for spin chains), making the unfolding procedure straightforward. However, for models without an analytic formula for the average density of states (SYK$_2$ for example), the unfolding procedure becomes more complex and may introduce errors. Therefore, it is important to use alternative methods that do not require unfolding, reducing the risk of such errors. One such method is the $r$-parameter statistics, which we discuss in Appendix \ref{app:r-parameter}.

\section{$r$-parameter statistics } \label{app:r-parameter}

In this Appendix, we discuss the statistics of the so-called $r$-parameter, which is essentially the distribution of the ratio of nearest-neighbour spacings,
and therefore serves as an important tool for diagnosing the correlation in the energy spectrum.
Within a symmetry sector of the system, consider an ordered spectrum ${E_n }$, and define the nearest-neighbour level 
spacings $s_n = E_{n+1} - E_n$. 
From these level spacings, we define the $r_n \equiv s_n/s_{n-1}$,  and consider the distribution of the following quantity, known as the $r$-parameter \cite{Oganesyan:2007wpd}
\begin{equation}\label{rtilde}
	\tilde{r}_n \equiv \frac{\text{min} (s_n, s_{n-1})}{\text{max} (s_n, s_{n-1})} = \text{min}~ \Big(r_n, \frac{1}{r_n}\Big)~.
\end{equation}
One of the most important advantages of studying the $r$-parameter is that one does not need to unfold the energy spectrum, and hence 
does not need to have knowledge of the local density of states, since the ratios of successive
level spacings are independent of the local density of states. 
For an $N \times N$ random matrix Hamiltonian belonging to the classical Gaussian ensembles, the distribution of ratio $r$ is given by 
\begin{equation}
   P(r_1, \dots, r_{N-2}) = \left(\prod_{i=1}^N \int \textrm{d}E_i \right) P(E_1, \dots, E_N) \prod_{i=1}^{N-2}\delta\left(r_i-\frac{E_{i+2}-E_{i+1}}{E_{i+1}-E_{i}}\right)~,
\end{equation}
where $P(E_1, \dots, E_N)$ denotes the joint probability distribution of the eigenvalues, which for the Gaussian ensembles under consideration is given by \cite{mehta1991random}:
\begin{equation}
    P(E_1, \dots, E_N)=C_{\nu,N} e^{-\frac{\nu N}{4} \sum_{k=1}^N  E_k^2} \prod_{i<j} |E_i-E_j|^\nu~,  
\end{equation}
$C_{\nu,N}$ being a constant which normalizes the distribution.

An important result on the statistics of the $r$-parameter was obtained 
in \cite{Atas_2013}, where a Wigner-like surmise for the distribution of $r_n$ was obtained analytically for all three classical random matrix ensemble
by considering a model of $3 \times 3 $ matrices. This expression, which already provides a very good approximation for the distribution of $P(r)$
obtained from exact numerical analysis of matrices of large sizes, is given by
\begin{equation} 
	P_{W}(r) = \frac{1}{Z_{\nu}} \frac{(r+r^2)^\nu}{(1+r+r^2)^{1+\frac{3}{2}\nu}}~, 
    \label{eq: r-dis-N=3}
\end{equation}
where $Z_\nu$ is a normalization constant that depends on the Dyson index $\nu$. 
This distribution of $r$ has the similar level repulsion as that of the distribution of the level spacing for small values of $r$, i.e., $P_{W}(r) \sim r^\nu$.
On the other hand, for large values $r$, unlike the distribution of level spacings, the $P_{W}(r)$ does not decay exponentially, rather it has slower polynomial decay, $P_{W}(r) \sim r^{-(2+\nu)}$. 

From the above expression for the distribution of $r$, one can analytically calculate the average value of the $r$ parameter to be 
\begin{equation}
	\langle\tilde{r}\rangle_W = \int_0^1 \textrm{d}r~ r P_{W}(r) + \int_{1}^{\infty} \textrm{d}r~r^{-1} P_{W}(r)~.
\end{equation}
We also note that, for an uncorrelated spectrum,  with level spacing statistics following a Poisson distribution, the probability distribution of the ratio $r$ is given by:
\begin{equation}
    P_{\text{Poisson}}(r)=\frac{2}{(1+r)^2}.
\end{equation}
Numerically, one can calculate each of the individual 
$\tilde{r}$ parameter from eq. \eqref{rtilde}, and obtain the average using 
\begin{equation} 
\langle \tilde{r} \rangle = \frac{1}{N-1} \sum_{n=1}^{N-1} \tilde{r}_n\,. 
\label{eq: averaged r}
\end{equation}
Table \ref{tab:r_parameter} presents the average $r$-parameter for various random matrix ensembles, including the GOE, GUE, and Gaussian Symplectic Ensemble (GSE), as well as for a spectrum that follows Poisson statistics.
\begin{table}[h]
    \centering
    \caption{Average \( r \)-parameter for different ensembles}
    \begin{tabular}{lc}
        \hline
        Ensemble & \(\langle r \rangle\) \\
        \hline
        Poisson  & \( \approx 0.386 \) \\
        GOE      & \( \approx 0.536 \) \\
        GUE      & \( \approx 0.602 \) \\
        GSE      & \( \approx 0.676 \) \\
        \hline
    \end{tabular}
    \label{tab:r_parameter}
\end{table}

\section{Average correlators and spectral form factor} \label{app-averageOTOC}
In this appendix we review some relations between the averages of time-ordered and out-of-time-ordered correlation functions and 
the spectral form factor~\cite{Cotler:2017jue}.  Specifically, the connections between the OTOC and the spectral form factor (SFF) we discuss can be thought of as 
a bridge between two different popular ways of quantifying quantum chaos.
	
We start by considering a generic system described by a Hamiltonian $H$,  having Hilbert space dimension $L=2^n$, and a unitary operator 
$\mcu$.  Using the formula for the first moment of the Haar ensemble associated with this operator, the average of the corresponding 
2-point function over the Haar measure defined on $U(2^n)$ can be related to the 2-point spectral form factor as,\footnote{The subscript $U$ denotes an average over the Haar measure defined on $U(2^n)$. }  
\begin{equation}\label{SFF_2point}
	  \big<\mcu(t=0) \mcu^{\dagger}(t) \big>_U  \equiv  \frac{1}{L}  \int \textrm{d}U ~\text{Tr} \big[\mcu e^{- i H t} \mcu ^{\dagger}  e^{- i H t} \big]\\
	=   \frac{1}{L^2} \big|\text{Tr} \big[e^{- i H t} \big]\big|^2~=    \frac{1}{L^2}  \mathcal{R}^H_2(t)~,
\end{equation}
	where $ \mathcal{R}^H_2(t)$ denotes the 2-point form factor associated with the Hamiltonian $H$. This formula provides a direct way of estimating the
spectral form factor from the knowledge of the 2-point function, since, the variance is suppressed by $1/L$, i.e.  \cite{Cotler:2017jue},
\begin{equation}
	\Delta  \big<\mcu(t) \mcu^{\dagger}(t) \big>^2=   \frac{1}{L^2}\int \textrm{d}U \big| \text{Tr}\big[\mcu(0) \mcu^{\dagger}(t) \big]\big|^2- \frac{1}{L^2}\bigg|\int \textrm{d}U ~ \text{Tr}\big[\mcu(0) \mcu^{\dagger}(t) \big]\bigg|^2 \sim \mco (1/L^2)~.
\end{equation}
That the formula in eq. \eqref{SFF_2point}  is true, and the error is small, were verified in \cite{Cotler:2017jue} by numerical simulation of a random non-local spin system.  

For an OTOC, defined for  two unitary operators, an analogous computation using the first moment of the Haar ensemble gives a different type of form factor
\begin{equation}
 \big<\mcu_1(0) \mcu_1^{\dagger}(t) \mcu_2(0) \mcu_2^\dagger(t)\big>_{U}=	\frac{1}{L}\int \textrm{d}U_1  \textrm{d}U_2  ~\text{Tr}\big[\mcu_1(0) \mcu_1^{\dagger}(t) \mcu_2(0) \mcu_2^\dagger(t)\big] =  \frac{1}{L^3}  \text{Tr} \big[e^{- i H t} \big] ^2 
	\text{Tr} \big[e^{-2 i H t} \big] ~.
\end{equation}
This identity can be easily generalized for $k$ different operators to get, 
\begin{align}
	\big<\mcu_1(0) \mcu_1^{\dagger}(t)\cdots  \mcu_k(0) \mcu_k^\dagger(t)\big>_U&=\frac{1}{L} \int \textrm{d}U_1  \textrm{d}U_2 \cdots \textrm{d}U_k  ~\text{Tr}\big[\mcu_1(0) \mcu_1^{\dagger}(t)\cdots  \mcu_k(0) \mcu_k^\dagger(t)\big] 
	\\&=  \frac{1}{L^{k+1}}  \text{Tr} \big[e^{- i H t} \big] ^k \text{Tr} \big[e^{-k i H t} \big] ~.
\end{align}
To obtain the $2k$-point form factor from an OTOC average, one can consider the following average of a $2k$-point OTOC, and using the first moment 
of the Haar ensemble $2k-1$ times obtain \cite{Cotler:2017jue}, 
\begin{align}\label{SFF_kpoint}
	\big<\mcu_1(0) \mcv_1(t)\cdots  \mcu_k(0) \mcv_k(t)\big>_{U,V}&=\frac{1}{L}\int \textrm{d}U_1  \textrm{d}V_1 \cdots \textrm{d}V_{k-1} \textrm{d}U_k  ~\text{Tr}\big[\mcu_1(0) \mcv_1(t)\cdots  \mcu_k(0) \mcv_k(t)\big] 
	\\&=     \frac{1}{L^{2k}} \big|\text{Tr} \big[e^{- i H t} \big]\big|^{2k}~=    \frac{1}{L^{2k}}  \mathcal{R}^H_{2k}(t)~.
\end{align}
To summarize, these relations connecting the averages of the time-ordered correlators and OTOC with various SFFs indicate that the spectral statistics can not capture either of the spatial or temporal locality of the operators.

\textbf{Average correlation function for random matrix Hamiltonians}.
The relations between the averages of the correlation function and the SFFs discussed above are valid for 
any arbitrary Hamiltonian. One can also derive similar such relations when the Hamiltonian is drawn from some random matrix ensemble and the average is performed over the ensemble.  For instance, for a random matrix Hamiltonian taken from the GUE, utilizing the invariance of the integral measure under unitary transformations,  the following relation can be established
between the ensemble-averaged 2-point correlation function of two self-adjoint operators and the 2-point SFF of the ensemble \cite{Cotler:2017jue}
\begin{equation}\label{2pt_SFF}
	\big<A(0) B(t) \big>_{\text{GUE}} \equiv \frac{1}{L} \int \textrm{d}H ~ \text{Tr}\big[A(0) B(t) \big] = \big<A B \big> \frac{\mathcal{R}_2(t)-1}{L^2-1}
	+ \big<A \big> \big<B \big> \frac{L^2-\mathcal{R}_2(t)}{L^2-1}~.
\end{equation}
Here $\mathcal{R}_2(t)$ is the spectral form factor defined as 
\begin{equation}
	\mathcal{R}_2(t) = \int \td H~ \big|\text{Tr}\big(e^{-i t H}\big)\big|^2~,
\end{equation}
and $\brac{A}=\text{Tr}(A)/L$.
If $A$ is a non-identity Pauli operator, and $\mathcal{R}_2(t) \gg1$, the above formula implies a relation between the GUE averaged
2-point function and 2-point form factor for the ensemble, which is very similar to the one in eq. \eqref{SFF_2point},
\begin{equation}
	\big<A(0) A^{\dagger}(t) \big>_{\text{GUE}}  \simeq \frac{1}{L^2}\mathcal{R}_2(t)~.
\end{equation}
Computing the $2k$-point OTOC for the GUE, one can similarly show that
\begin{equation}
	\big<A_1(0) B_1(t)\cdots  A_k(0) B_k(t)\big> _{\text{GUE}} \simeq \big<A_1(0) B_1(0)\cdots  A_k (0)B_k(0)\big>\frac{\mathcal{R}_{2k}(t)}{L^{2k}} ~.
\end{equation}
This is non-zero only when $A_1 B_1\cdots  A_k B_k=I$, and in that case, this equation resembles the one in eq. \eqref{SFF_kpoint}. 
Performing a similar analysis, one can show that 
\[\big<A_1(0) A_2(t) A_3(2t) A_4(t)\big> _{\text{GUE}} \simeq \big<A_1(0) A_2(t) A_3(0) A_4(t)\big> _{\text{GUE}}\,~, \] i.e., the GUE does not care about the spatial and temporal locality of the operators in the correlation functions. 

It is possible to extend the relation between the average OTOC and the SFF for theories with infinite-dimensional Hilbert space, e.g., to Bosonic quantum mechanics and QFTs \cite{deMelloKoch:2019rxr}. For such theories, to establish a relation between the OTOC and SFF, one only needs to average over the Lie group generated by the canonical position and momentum operators, i.e., over the so-called Heisenberg group.

%\paragraph{Mixed free cumulants and the SFF time scales.}

\paragraph{Finite temperature correlators.}
So far, we have only considered the infinite temperature correlation functions. However, we can follow the strategy described above to compute the finite-temperature correlation function as well, and see whether they can be related to the finite-temperature SFF. To begin with, consider the correlation function defined through the standard thermal 
expectation value, $\brac{\mathcal{O}}_\beta= Z^{-1}(\beta) \text{Tr} (e^{-\beta H} \mathcal{O})$, $Z(\beta)$ being the partition function.  For a random matrix belonging to the GUE, using the unitary invariance of the ensemble, we have the following expression for the two-point correlation function with respect to the thermal inner product%\footnote{In order to avoid cluttering the notation, we have removed the subscript GUE in the correlation functions used previously. }
\begin{align}
	\brac{AB(t)}^{\text{GUE}}_\beta&=\int \text{d}U~\text{d}H~\brac{A_U e^{i H t} B_U e^{-i H t}}_\beta \\
	&=\frac{\brac{AB}}{(L^2-1)} \big(L~r_2(t, \beta)-1\big)+\frac{L\brac{A}\brac{B}}{L^2-1}\big(L-r_2(t, \beta)\big) ~.
\end{align}
Here we have defined, the function $ r_2(t, \beta)$, and the GUE-averaged partition function, respectively, as 
\begin{equation}
    r_2(t, \beta) = \frac{1}{\brac{Z(\beta)}_{\text{GUE}} }\int \text{d} H~ \text{Tr}\big(e^{-i t H}\big) \text{Tr}\big(e^{i H(t+i \beta)}\big)~,~~\brac{Z(\beta)}_{\text{GUE}} = \int \td E \sum_j e^{- \beta E_j} ~.
\end{equation}
On the other hand, using the Wightman inner product instead of the standard thermal inner product, we have the following relation for the two-point correlation function
\begin{equation}
    \brac{AB(t)}^{\text{GUE}}_\text{W} = \brac{e^{\beta H/2}A e^{-\beta H/2}e^{i H t} B e^{-i H t}}^{\text{GUE}}_\beta\\
    =\frac{L}{\brac{Z(\beta)}_{\text{GUE}} }\int \text{d}U~~\brac{A_U B_U\Big(t-\frac{i\beta}{2}\Big)}_{\text{GUE}}~.
\end{equation}
Following a procedure similar to the one to obtain \eqref{2pt_GUE_SFF}, we derive the expression for the Wightman two-point correlation function at an inverse temperature $\beta$ to be
\begin{equation}
     \brac{AB(t)}^{\text{GUE}}_\text{W} =\frac{L}{\brac{Z(\beta)}_{\text{GUE}} } \Bigg[\brac{AB}\bigg(\frac{L\mathcal{R}_2(t, \beta/2)-Z(\beta)}{L(L^2-1)}\bigg)+ \brac{A}\brac{B}\bigg(\frac{L Z(\beta)-\mathcal{R}_2(t, \beta/2)}{(L^2-1)}\bigg)\Bigg]~,
\end{equation}
where we have defined analytically continued partition function $Z(t, \beta/2)=\text{Tr} \big(e^{- i H (t-i\frac{\beta}{2}) }\big)$ and the associated ensemble-averaged SFF, $\mathcal{R}_2(t, \beta/2)=\brac{Z(t, \beta/2)Z^*(t, \beta/2)}_\text{GUE}$. 

\section{Towards operator statistics at finite temperature}\label{app:finiteTemperature}
As we have discussed in section \ref{sec-discussion}, the predictions for the operator statistics obtained from the free probability theory that we have used in the main text are applicable for a quantum system at infinite temperature. Here we provide a brief discussion on how one can proceed in obtaining the statistics of the sum of free operators for finite temperature states. 

We begin by noticing that, in general, if two operators $A(0)$ and $B(t)$ are free with respect to an infinite-temperature state, we then expect them to also be approximately free with respect to finite-temperature states, as long as the temperature is sufficiently high. However, as the temperature is lowered, we approach the regime where expectation values are computed near the quantum system’s ground state. In this case, we do not generally expect freeness, except perhaps in special systems that remain highly chaotic even near the ground state.

The free probability prediction for the free convolution of two operators is obtained by computing the Cauchy transform $G(z)$, whose definition is appropriate for an infinite-temperature state (see eq. \eqref{Cauchy_tra} and the discussion in footnote \ref{moments_gen}). One approach to obtaining the free convolution prediction for a finite temperature state with the map, taken e.g., the standard thermal inner product  $\varphi_\beta(\mathcal{O})=\brac{\mathcal{O}}_\beta=\text{Tr}( \rho_\beta \mathcal{O} )/\text{Tr}(\rho_\beta)$, $\rho_\beta=e^{-\beta H}$ being the thermal density matrix, would be to use an appropriate generalization of the Cauchy transform which takes into account the finite temperature of the quantum system. 

However, in principle, one can choose a different definition of the thermal inner product as well. In particular, choosing the Wightman inner product to define the thermal two-point function, we see that for a generic Hamiltonian $H$, it can be  rewritten in terms of an infinite temperature state as
\begin{equation}
    \brac{A(0)B(t)}_\text{W} = \brac{\rho_\beta^{1/2} A \rho_\beta^{-1/2} e^{i H t} B e^{-i H t}}_\beta = \frac{L}{Z(\beta)} \brac{A(0) B\Big(t-\frac{i\beta}{2}\Big)}~.
\end{equation}
Therefore, this two-point function also captures thermal effects and is related to the spectral properties of $A(0)+\rho^{1/2} B(t) \rho^{1/2}$. By considering the composite operator $\rho^{1/2} B \rho^{1/2}$ (instead of just $B$), we can incorporate thermal effects into the operator $B$ while still using the (infinite temperature) map $\varphi(\cdot)=\text{Tr}( \cdot )/N$. This allows us to apply the formalism of the standard Cauchy transform and obtain the free convolution prediction, while still incorporating thermal effects. The price to pay is that the spectrum of $\rho^{1/2} B \rho^{1/2}$ differs from that of $B$ and needs to be determined.

To this end, one could proceed as follows. If the dynamics of a quantum system is chaotic and Hamiltonian ($H$) behaves like a random matrix, we expect $\rho^{1/2}=e^{-\beta H/2}$ to also approximately behave like a random matrix and to be free from the deterministic operator $ B$. This allows us to use free multiplicative convolution, based on the S-transform, to find the spectrum of $\rho^{1/2} B \rho^{1/2}$. Knowing the spectra of $A(0)$ and $\rho^{1/2} B \rho^{1/2}$, we can compute the corresponding R-transforms and thus determine the spectral properties of $A(0)+\rho^{1/2} B \rho^{1/2}$. 
However, the details of this procedure are beyond the scope of the present paper, and we plan to report on finite-temperature generalizations of our results in the near future.

\section{Universal statistics of the out-of-time order operators} \label{app-Lyapunovian}
In this appendix, we review the proposal introduced in~\cite{rozenbaum2019universal}, which, to some extent, bridges two different notions of quantum chaos: level spacing statistics and scrambling quantified by the OTOCs. Another notable work in a similar direction is~\cite{Gharibyan:2018fax}.

Since the dependence of the OTOC on the state with respect to which it is computed is not universal, \cite{rozenbaum2019universal} proposed to instead look at statistics of the so-called Lyapunovian operator (which
is a Hermitian operator obtained by taking the logarithm of the squared commutator of two operators) to search for random matrix universality in these correlators. Below, we first review this proposal more precisely. Consider the following correlators between two Hermitian operators, and define %positive definite 
the operators $\mathbf{\Lambda}_k$, 
\begin{equation}
	\mcc^k (t) \equiv  \exp \big[i k t \mathbf{\Lambda}_k(t)\big] = (-i)^k \big[A(t), B(0)\big]^k~,
\end{equation}
where $k \in \mathbb{N}$, and when $k$ is odd the operator $\mathbf{\Lambda}_k(t)$ is defined only within the subspace where the eigenvalues of $\mcc^{(2n-1)} (t)$ are positive. For our purposes, we assume the operators $A$ and $B$ are conjugate to each other. When $k=2$,
the corresponding operator $\mathbf{\Lambda}_2(t)$ is called the Lyapunovian \cite{rozenbaum2019universal}. Apart from $\mcc^k (t)$, we can also define another Hermitian operator from the OTOC as, 
\begin{equation}
	\mathcal{F}(t) = A(t)B(0)A(t)B(0) + \text{H.c.} = \exp[\Gamma(t)]~. 
\end{equation}
As an example, if one considers the (two-dimensional) stadium billiard system, one of the paradigmatic models of classical chaotic systems that can be 
shown to follow the Wigner-Dyson energy level spacing statistics, then the operators $A$ and $B$ can be taken, respectively, the position coordinate ($x$) and its conjugate momenta ($p_x$). 

According to the proposal advocated in \cite{rozenbaum2019universal}, one first constructs the $N \times N$ matrices  $\mcc^{k}_{nm}(t) = \bra{E_n} 	\mcc^k (t) \ket{E_m}$ and 
$\mathcal{F}_{nm}(t) = \bra{E_n} 	\mathcal{F}(t) \ket{E_m}$, where $\ket{E_m}$ denotes the energy eigenstates. Subsequently, these matrices are diagonalized by taking $t$ to be a fixed value, and the statistics of the spacing between the eigenvalues (as well as the logarithm of the eigenvalues) are obtained to look for random 
matrix universality in the correlators. One can then repeat this procedure for different values of time to compare the statistics as time evolves. 
This proposal was studied in detail in \cite{rozenbaum2019universal}  by taking the operators $A=x$, and $B=p_x$, and   the Hamiltonian that
 of a stadium billiard
\begin{equation}
	H= \frac{1}{2} \big(p_x^2 + p_y^2 \big) + V_{\text{walls}}(x,y)~,
\end{equation}
so that (in units $\hbar=1$)  $\mathcal{F}(t=0) = - \Big(x \big[H,x\big]\Big)^2+ \text{H.c}$. At the initial times, the bulk level spacing statistics of the operators 
$\mathcal{F}(0)$ and $\Lambda(0)$ are found to be the same as that of the Hamiltonian itself, i.e., they are closely approximated by the GOE Wigner-Dyson
distribution.\footnote{There are subtleties from different parity sectors of the eigenstates and the so-called bouncing  ball modes that
one needs to take care of before obtaining the level spacing statistics. See \cite{rozenbaum2019universal} for 
more details.}  On the other hand, the level spacing distribution of these operators is different when the time takes a non-zero value. For finite values of time 
and odd values of $k=2n-1$, the bulk level spacing statistics of $\mcc^{k=2n-1}(t \neq0)$ follow GUE Wigner-Dyson distribution, while for even values of $k=2n$, the bulk level spacing of statistics of every second eigenvalue of the matrix $\mcc^{k=2n}_{ij}(t \neq0)$ matches with the level spacing distribution of a Gaussian ensemble with the Dyson index $\nu=3$.  

One can perform similar studies of the level spacing distribution of the Lyapunovian and other
such operators defined above for integrable systems. It can be seen that there are multiple degeneracies in the spectrum of these operators, and hence 
the level spacing distribution of $\mathbf{\Lambda}_2$ is not well defined -  a property of this kind of system that is drastically different from the chaotic ones discussed above.

\section{Example of subordination function notebook } \label{app-SubFunctMathematica}

In this section, we provide a Mathematica code example that implements the subordination function method~\cite{speicher2019lecture}, as reviewed in Section~\ref{sec-numFreeConvolution}, to compute the free additive convolution of two random variables whose auxiliary functions \eqref{auxi_fn} are identical $H_a(z)=H_b(z)$. This code is used to generate the free probability predictions shown in Figures~\ref{fig:FreeConvolutionRM} and~\ref{fig:freeConvolutionRM2}.

\begin{table}[h!]
\centering
\caption{Mathematica code for free additive convolution based on the subordination function method}
\setlength{\extrarowheight}{2pt}
\renewcommand{\arraystretch}{1.2}
\begin{tabular}{>{\ttfamily}l}
\hline
\textbf{(* Solves fixed-point equation for $\omega(z)$ using 20,000 iterations *)} \\
Clear[MapF] \\
MapF[z\_] := Module[\{HH, MAP, w0, wf\}, \\
\quad HH[x\_] := Ha[x]; (* auxiliary function *) \\
\quad MAP[ww\_] := z + HH[z + HH[ww]]; \\
\quad w0 = 1. + 1.1 I; \\
\quad wf = Nest[MAP, w0, 20000]; \\
\quad wf \\
] \\
\\
\textbf{(* Inverse Stieltjes transform *)} \\
Clear[G$\mu$] \\
G$\mu$[z\_] := G[z] \\
Density[x\_] := -1/Pi * Im[G$\mu$[MapF[x + 0.001 I]]] \\

\hline
\end{tabular}
\end{table}

\section{Asymptotic Freeness under Time evolution of Decorrelated Gaussian Ensembles} \label{app-decorrensembles}

Here we collect plots for operator spectra comparing between the true time evolution and time evolution under decorrelated ensembles (Sec.~\ref{sec:decorrelatedEnsembles}) introduced in ~\cite{Balasubramanian:2022tpr} and also discussed in~\cite{Magan:2024nkr}.

\begin{figure}[ht]
    \centering
    \begin{tabular}{cc}
        \includegraphics[width=0.45\textwidth]{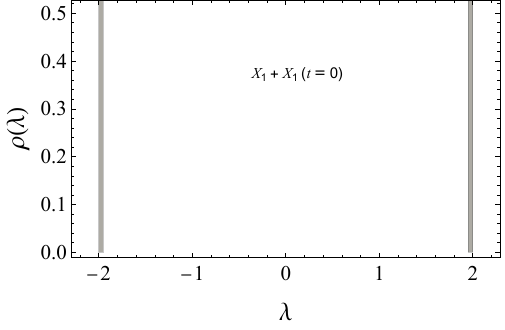}  & \includegraphics[width=0.45\textwidth]{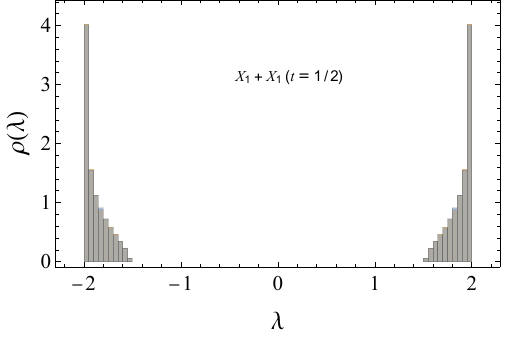} \\
        %(a) Caption for figure 1 & (b) Caption for figure 2 \\
        \includegraphics[width=0.45\textwidth]{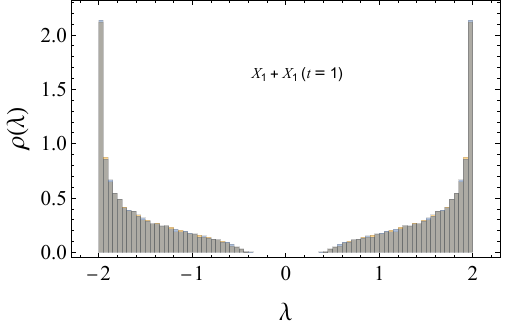} &
        \includegraphics[width=0.45\textwidth]{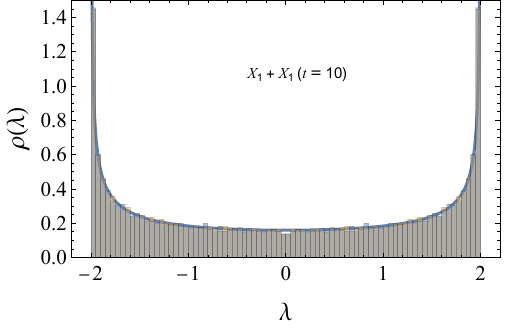} \\
        %(c) Caption for figure 3 & (d) Caption for figure 4 \\
    \end{tabular}
    \caption{Spectrum of eigenvalues of the spin-$1/2$ operator $X_1(0)+X_{1}(t)$ for increasing values of time under time evolution by a Hamiltonian drawn from the GOE (yellow) and under time evolution by a Hamiltonian drawn from a decorrelated GOE ensemble (blue). The overlap between these two distributions is shown in grey. The blue curve is the arcsine distribution~\eqref{eq:arcsine}. The bin size in each histogram is $\lbrace 0.05\rbrace$, $L=8$ and the number of realizations is $50$.}
    \label{fig:SpinHalfXi1j1Poissonian} 
\end{figure}

\begin{figure}[ht]
    \centering
    \begin{tabular}{cc}
        \includegraphics[width=0.45\textwidth]{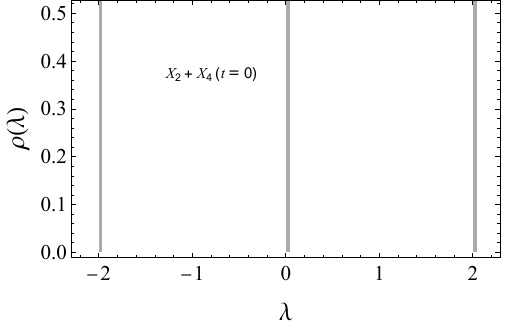}  & \includegraphics[width=0.45\textwidth]{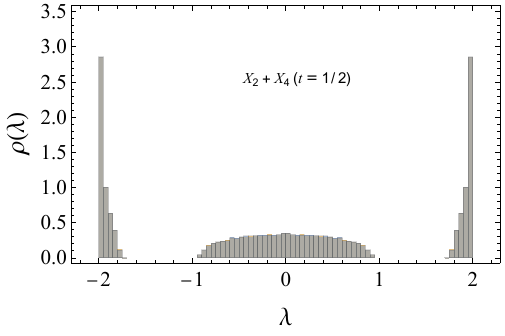} \\
        %(a) Caption for figure 1 & (b) Caption for figure 2 \\
        \includegraphics[width=0.45\textwidth]{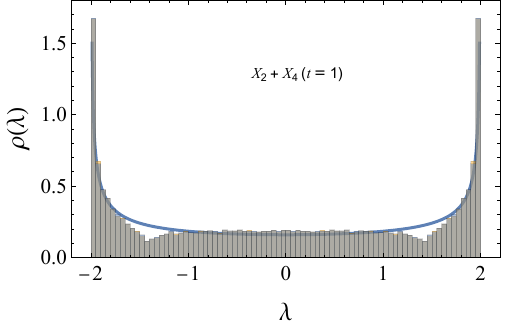} &
        \includegraphics[width=0.45\textwidth]{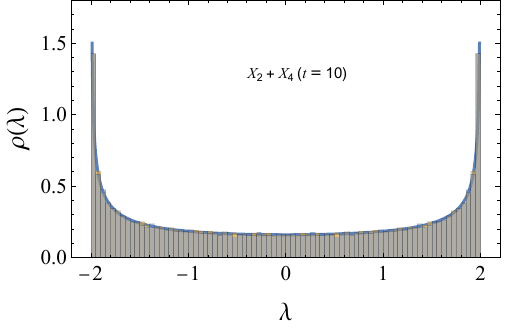} \\
        %(c) Caption for figure 3 & (d) Caption for figure 4 \\
    \end{tabular}
    \caption{Spectrum of eigenvalues of the spin-$1/2$ operator $X_2(0)+X_{4}(t)$ for increasing values of time under time evolution by a Hamiltonian drawn from the GOE (yellow) and under time evolution by a Hamiltonian drawn from a decorrelated GOE ensemble (blue). The overlap between these two distributions is shown in grey. The blue curve is the arcsine distribution~\eqref{eq:arcsine}. The bin size in each histogram is $\lbrace 0.05\rbrace$, $L=8$ and the number of realizations is $50$.}
    \label{fig:SpinHalfXi2j4Poissonian} 
\end{figure}

\begin{figure}[ht]
    \centering
    \begin{tabular}{cc}
        \includegraphics[width=0.45\textwidth]{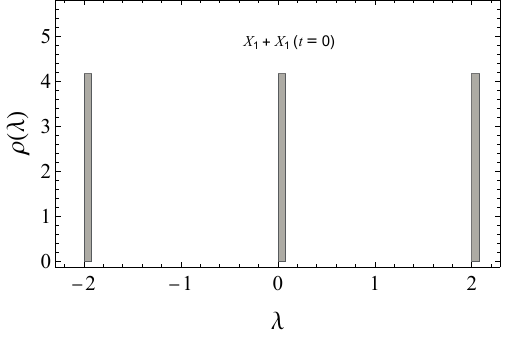}  & \includegraphics[width=0.45\textwidth]{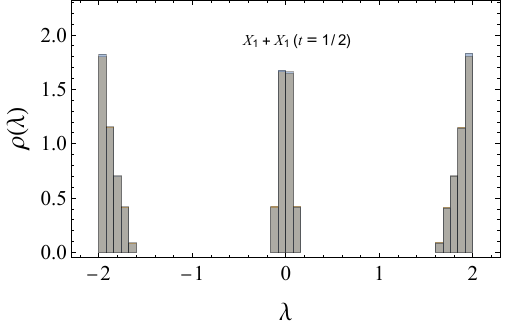} \\
        %(a) Caption for figure 1 & (b) Caption for figure 2 \\
        \includegraphics[width=0.45\textwidth]{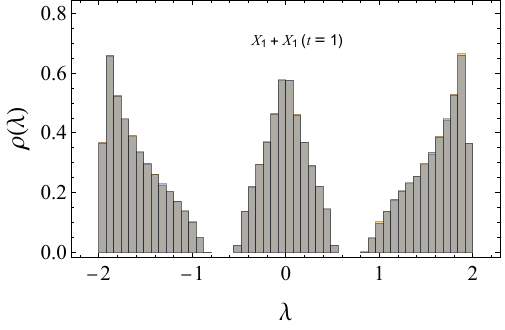} &
        \includegraphics[width=0.45\textwidth]{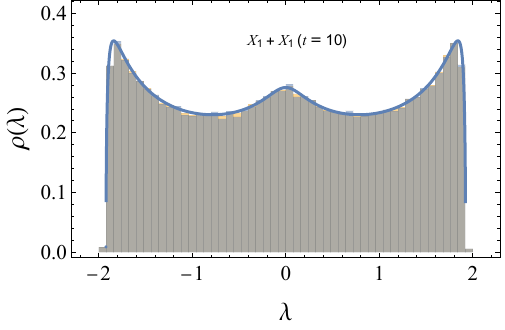} \\
        %(c) Caption for figure 3 & (d) Caption for figure 4 \\
    \end{tabular}
    \caption{Spectrum of eigenvalues of the spin-$1$ operator $X_1(0)+X_{1}(t)$ for increasing values of time under time evolution by a Hamiltonian drawn from the GOE (yellow) and under time evolution by a Hamiltonian drawn from a decorrelated GOE ensemble (blue). The overlap between these two distributions is shown in grey. The blue curve in the bottom right panel is the distribution~\eqref{eq: sumdis_spin1}. The bin size in each histogram is $\lbrace 0.08\rbrace$, $L=6$ and the number of realizations is $20$.}
    \label{fig:SpinOneXi1j1Poissonian} 
\end{figure}

\begin{figure}[ht]
    \centering
    \begin{tabular}{cc}
        \includegraphics[width=0.45\textwidth]{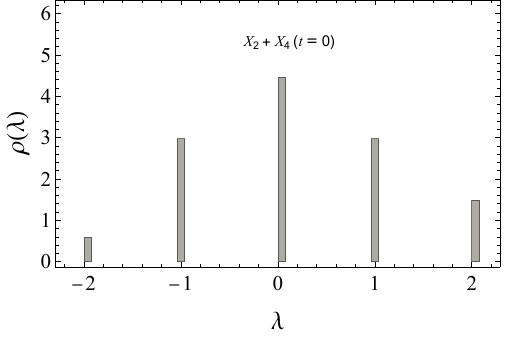}  & \includegraphics[width=0.45\textwidth]{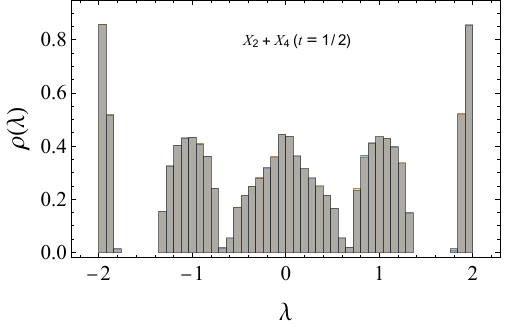} \\
        %(a) Caption for figure 1 & (b) Caption for figure 2 \\
        \includegraphics[width=0.45\textwidth]{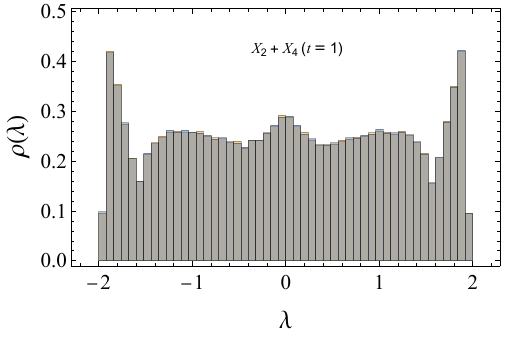} &
        \includegraphics[width=0.45\textwidth]{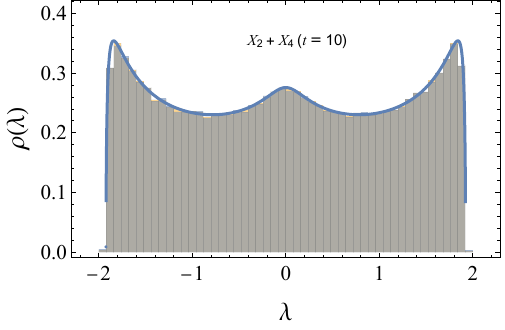} \\
        %(c) Caption for figure 3 & (d) Caption for figure 4 \\
    \end{tabular}
    \caption{Spectrum of eigenvalues of the spin-$1$ operator $X_2(0)+X_{4}(t)$ for increasing values of time under time evolution by a Hamiltonian drawn from the GOE (yellow) and under time evolution by a Hamiltonian drawn from a decorrelated GOE ensemble (blue). The overlap between these two distributions is shown in grey. The blue curve in the bottom right panel is the distribution~\eqref{eq: sumdis_spin1}. The bin size in each histogram is $\lbrace 0.08\rbrace$, $L=6$ and the number of realizations is $20$.}
    \label{fig:SpinOneXi2j4Poissonian} 
\end{figure}

Figures~\ref{fig:SpinHalfXi1j1Poissonian} and~\ref{fig:SpinHalfXi2j4Poissonian} focus on the spin-$1/2$ case, for the operator sums $X_1(0)+X_{1}(t)$ and $X_{2}(0)+X_{4}(t)$ respectively. Figures~\ref{fig:SpinOneXi1j1Poissonian} and~\ref{fig:SpinOneXi2j4Poissonian} focus on the spin-$1$ case for the same choice of operators. For both choices of spins the operator statistics is found to be consistent for time evolutions generated by the true GOE and decorrelated GOE Hamiltonians.

%%%%%%%%%%%%%%%%%%%%%%%%%%%%
%%%%%%%%%%%%%%%%%%%%%%%%%%%%
\sloppy
\bibliography{references_JHEP}
\bibliographystyle{JHEP}

\end{document}